\makeatletter
\newcommand{\dontusepackage}[2][]{%
  \@namedef{ver@#2.sty}{9999/12/31}%
  \@namedef{opt@#2.sty}{#1}}
\makeatother
\dontusepackage{subfigure}

\documentclass[]{article}

\usepackage{lmodern}
\usepackage{amssymb,amsmath}
\usepackage{ifxetex,ifluatex}
\usepackage[usenames,dvipsnames]{color}
\usepackage{fixltx2e} %
\ifnum 0\ifxetex 1\fi\ifluatex 1\fi=0 %
  \usepackage[T1]{fontenc}
  \usepackage[utf8]{inputenc}
\else %
  \ifxetex
    \usepackage{mathspec}
    \usepackage{xltxtra,xunicode}
  \else
    \usepackage{fontspec}
  \fi
  \defaultfontfeatures{Mapping=tex-text,Scale=MatchLowercase}
  
\fi
\IfFileExists{upquote.sty}{\usepackage{upquote}}{}
\IfFileExists{microtype.sty}{%
\usepackage{microtype}
\UseMicrotypeSet[protrusion]{basicmath} %
}{}
\usepackage[margin=1.0in,bottom=1.5in]{geometry}
\usepackage[]{natbib}
\usepackage{listings}
\usepackage{color}
\usepackage{fancyvrb}

\DefineVerbatimEnvironment{Highlighting}{Verbatim}{commandchars=\\\{\}}
\newenvironment{Shaded}{}{}
\newcommand{\KeywordTok}[1]{\textcolor[rgb]{0.00,0.44,0.13}{\textbf{{#1}}}}
\newcommand{\DataTypeTok}[1]{\textcolor[rgb]{0.56,0.13,0.00}{{#1}}}

\newcommand{\FloatTok}[1]{\textcolor[rgb]{0.25,0.63,0.44}{{#1}}}

\newcommand{\StringTok}[1]{\textcolor[rgb]{0.25,0.44,0.63}{{#1}}}
\newcommand{\CommentTok}[1]{\textcolor[rgb]{0.38,0.63,0.69}{\textit{{#1}}}}

\newcommand{\NormalTok}[1]{{#1}}
\usepackage{longtable,booktabs}
\usepackage{graphicx}
\makeatletter
\def\maxwidth{\ifdim\Gin@nat@width>\linewidth\linewidth\else\Gin@nat@width\fi}
\def\maxheight{\ifdim\Gin@nat@height>\textheight\textheight\else\Gin@nat@height\fi}
\makeatother
\setkeys{Gin}{width=\maxwidth,height=\maxheight,keepaspectratio}
\usepackage{caption}
\usepackage{float}
\usepackage{subfig}
\usepackage{algorithm} %
\let\scholmdAlgorithm\algorithm
\let\endscholmdAlgorithm\endalgorithm
\let\algorithm\relax \let\endalgorithm\relax

{
 \catcode`\*=11\relax
 \global\let\scholmdAlgorithm*\algorithm*
 \global\let\endscholmdAlgorithm*\endalgorithm*
 \global\let\algorithm*\relax 
 \global\let\endalgorithm*\relax
}
\ifxetex
  \usepackage[setpagesize=false, %
              unicode=false, %
              xetex]{hyperref}
\else
  \usepackage[unicode=true]{hyperref}
\fi
\hypersetup{breaklinks=true,
            bookmarks=true,
            pdfauthor={},
            pdftitle={Algorithms and software for projections onto intersections of convex and non-convex sets with applications to inverse problems.},
            colorlinks=true,
            citecolor=black,
            urlcolor=blue,
            linkcolor=black,
            pdfborder={0 0 0}}
\def\argmin{\mathop{\rm arg\min}}

\title{Algorithms and software for projections onto intersections of convex and
non-convex sets with applications to inverse problems.}
\author{Bas Peters\textsuperscript{*}, Felix J.
Herrmann\textsuperscript{\#}\\\textsuperscript{*}Seismic Laboratory for
Imaging and Modeling (SLIM), University of British
Columbia\\\textsuperscript{\#}Georgia Institute of Technology}
\date{}

\begin{document}
\maketitle

\section{Abstract}\label{abstract}

We propose algorithms and software for computing projections onto the
intersection of multiple convex and non-convex constraint sets. The
software package, called SetIntersectionProjection, is intended for the
regularization of inverse problems in physical parameter estimation and
image processing. The primary design criterion is working with multiple
sets, which allows us to solve inverse problems with multiple pieces of
prior knowledge. Our algorithms outperform the well known Dykstra's
algorithm when individual sets are not easy to project onto because we
exploit similarities between constraint sets. Other design choices that
make the software fast and practical to use, include recently developed
automatic selection methods for auxiliary algorithm parameters, fine and
coarse grained parallelism, and a multilevel acceleration scheme. We
provide implementation details and examples that show how the software
can be used to regularize inverse problems. Results show that we benefit
from working with all available prior information and are not limited to
one or two regularizers because of algorithmic, computational, or
hyper-parameter selection issues.

\section{Keywords}\label{keywords}

inverse problems, alternating direction method of multipliers, parallel
computing, software, projection methods

\section{Introduction}\label{introduction}

We consider problems of the form
\begin{equation}
\mathcal{P}_{\mathcal{V}} (m) \in \argmin_{x} \frac{1}{2} \| x - m \|_2^2 \quad \text{subject to} \quad x \in \bigcap_{i=1}^p \mathcal{V}_i,
\label{proj_intersect}
\end{equation}
 which is the projection of a vector $m \in \mathbb{R}^N$ onto the
intersection of $p$ convex and possibly non-convex sets $\mathcal{V}_i$.
The projection in equation~\eqref{proj_intersect} is unique if all sets
are closed and convex. The projection operation is a common tool used
for solving constrained optimization problems of the form
\begin{equation}
\min_{m} f(m) \quad \text{subject to} \quad m \in \bigcap_{i=1}^p \mathcal{V}_i.
\label{prob_constr}
\end{equation}
 Examples of algorithms that use projections include spectral projected
gradient descent \citep[SPG,][]{Birgin:1999:NSP:588891.589081},
projected quasi-Newton \citep{schmidt2009optimizing}, and projected
Newton-type methods \citep{doi:10.1137/0320018, pnmethods}. In the above
optimization problem, the function
$f(m) : \mathbb{R}^N \rightarrow \mathbb{R}$ is at least twice
differentiable and may also be non-convex. Alternatively, proximal
algorithms solve
\begin{equation}
\min_{m} f(m) + \iota_{\mathcal{V}}(m),
\label{proximal_min}
\end{equation}
 which is equivalent to~\eqref{prob_constr} and where
$\iota_\mathcal{V}(m)$ is the indicator function of the set
$\mathcal{V} \equiv \bigcap_{i=1}^p \mathcal{V}_i$, which returns zero
if $m$ is an element of the intersection and infinity otherwise. Because
applications may benefit from using non-convex sets $\mathcal{V}_i$, we
also consider those sets in the numerical examples. While we do not
provide convergence guarantees for this case, we will work with some
useful/practical heuristics.

The main applications of interest in this work are inverse problems for
the estimation of physical (model) parameters ($m \in \mathbb{R}^N$)
from observed data ($d_\text{obs} \in \mathbb{C}^s$). Notable examples
are geophysical imaging problems with seismic waves \citep[full-waveform
inversion, see, e.g.,][]{TarantolaA, Pratt98, virieux09} for acoustic
velocity estimation and direct-current resistivity problems
\citep[DC-resistivity, see, e.g.,][]{haber2014computational} to obtain
electrical conductivity information. These problems have `expensive'
forward operators, i.e., evaluating the objective (data-misfit) $f(m)$
requires solutions of many partial-differential-equations (PDEs) if the
PDE constraints are implicit in $f(m)$, which corresponds to a reduced
data-misfit \citep{Haber2000}. In our context, each set $\mathcal{V}_i$
describes a different type of prior information on the model $m$.
Examples of prior knowledge as convex sets are bounds on parameter
values, smoothness, matrix properties such as the nuclear norm, and
whether or not the model is blocky with sharp edges (total-variation
like constraints via the $\ell_1$ norm). Non-convex sets that we use in
the numerical examples include the annulus (minimum and maximum $\ell_2$
norm), limited matrix rank, and vector cardinality.

Aside from the constrained minimization as in
problem~\eqref{prob_constr}, we consider feasibility (also known as
set-theoretic estimation) problem formulations
\citep[e.g.,][]{STE_2, STE_1, STEstimation, COMBETTES1996155}.
Feasibility only formulations accept any point in the intersection of
sets $\mathcal{V}_i$ that describe constraints on model parameter
properties, and a data-fit constraint $\mathcal{V}_p^{\text{data}}$ that
ties the unknown model vector $x$ to the observed data
$d_\text{obs} \in \mathbb{R}^M$ via a forward operator
$F \in \mathbb{R}^{M \times N}$. Examples of data-constraint sets are
$\mathcal{V}^{\text{data}} = \{ x \: | \: l \leq (F x - d_\text{obs}) \leq u\}$
and
$\mathcal{V}^{\text{data}} = \{ x \: | \: \| F x - d_\text{obs} \|_2 \leq \sigma \}$.
The upper and lower bounds are vectors $l$ and $u$ and $\sigma > 0$ is a
scalar that depends on the noise level. The forward operators are linear
and often computationally `cheap' to apply. Examples include masks and
blurring kernels. In case there is a good initial guess available, we
can choose to solve a projection rather than feasibility problem by
adding the squared $\ell_2$ distance term as follows:
\begin{equation}
\min_{x} \frac{1}{2} \| x - m \|_2^2 \quad \text{s.t.} \quad \begin{cases}
x \in \mathcal{V}_p^{\text{data}} \\ x \in \bigcap_{i=1}^{p-1} \mathcal{V}_i
\end{cases}.
\label{proj_intersect_lininvprob}
\end{equation}
 To demonstrate the benefits of this constrained formulation, we recast
joint denoising-deblurring-inpainting and image desaturation problems
as~\eqref{proj_intersect_lininvprob}. Especially when we have a few
training examples from which we can learn constraint set parameters, the
feasibility and projection approaches conveniently add many pieces of
prior knowledge in the form of multiple constraint sets, but without any
penalty or trade-off parameters. For instance, \citep{TVLearn} show that
we can observe `good' constraint sets, such as the average of the total
variation of a few training images. We address increasing computational
demand that comes with additional constraint sets with a reformulation
of problem~\eqref{proj_intersect_lininvprob}, such that we take into
account similarity between sets, and split the problem up into simple
parallel computations where possible.

Projected gradient and similar algorithms naturally split
problem~\eqref{prob_constr} into a projection and data-fitting part. In
this setting, software for computing projections onto the intersection
of sets can work together with codes for physical simulations that
compute $f(m)$ and $\nabla_m f(m)$, as we show in one of the numerical
examples. See \texttt{dolfin-adjoint} \citep{doi:10.1137/120873558},
\texttt{Devito} \citep{DevitoDSL, doi:10.1190/tle37010069.1} in
\texttt{Python} and \texttt{WAVEFORM} \citep{waveformCurt},
\texttt{jInv} \citep{doi:10.1137/16M1081063}, and \texttt{JUDI}
\citep{doi:10.1190/tle37020142.1} in \texttt{Julia} for examples of
recent packages.

Compared to regularization via penalty functions (that are not an
indicator function), constrained problem formulations (\ref{prob_constr}
and~\ref{proj_intersect_lininvprob}) have several advantages when
solving physical parameter estimation problems. Penalty methods
\begin{equation}
\min_m f(m) + \sum_i^p \alpha_i R_i(m)
\label{eq:penalty}
\end{equation}
 add prior knowledge through $p\geq 1$ penalty functions
$R_i(m) : \mathbb{R}^N \rightarrow \mathbb{R}$ with scalar weights
$\alpha_i > 0$ to the data-misfit term $f(m)$. Alternatively, we can add
penalties to the objective and work with a data constraint
instead---i.e., we have
\begin{equation}
\min_m \sum_{i=1}^p \alpha_i R_i(m) \quad \text{s.t.} \quad f(m) \leq \sigma,
\label{BPdenoise}
\end{equation}
 generally referred to as Basis Pursuit Denoise
\citep{MallatZhang, doi:10.1137/S003614450037906X, doi:10.1137/080714488, doi:10.1137/130919210},
Morozov/residual regularization \citep{ivanov2013theory}, or Occam's
inversion \citep{doi:10.1190/1.1442303}. The scalar $\sigma$ relates to
the noise level in the data. For convex
constraints/objectives/penalties, constrained, penalty and
data-constrained problems are equivalent under certain conditions and
for specific $\alpha$ - $\sigma$ pairs
\citep{Vasin1970, Gander1980, Golub1991, doi:10.1137/080714488, aravkin2016level, NIPS2017_6655},
but differ in algorithmic implementation and in their ability to handle
multiple pieces of prior information ($p>1$). In that case, the
simplicity of adding penalties is negated by the challenge of selecting
multiple trade-off parameters ($\alpha_i$). For this, and for reasons we
list below, we prefer constrained formulations that involve projections
onto the intersection of constraint sets (problem~\ref{proj_intersect}).
Constrained formulations

\begin{itemize}
\item
  \textbf{satisfy prior information at every iteration} PDE-based
  inverse problems require model parameters that are in an interval for
  which the mesh (PDE discretization) is suitable, i.e., we have to use
  bound constraints. Projection-based algorithms satisfy all constraints
  at every iteration and give the user precise control of the model
  properties. This allows us to start solving a non-convex inverse
  problem with certain constraints, followed by a solution stage with
  `looser' constraints.
  \citep{smithyman2015constrained, doi:10.1190/tle35030235.1, Esser2016arch, doi:10.1190/tle36010094.1, ournewpreprint}
  apply this strategy to seismic full-waveform inversion to avoid local
  minimizers that correspond to geologically unrealistic models.
\item
  \textbf{require a minimum number of manual tuning parameters for
  multiple constraints} We want to avoid the time-consuming and possibly
  computationally costly procedure of manually tuning numerous nuisance
  parameters. Constraint sets have the advantage that their definitions
  are independent of all other constraint definitions. For penalty
  functions, the effect of the weights $\alpha_i$ associated with each
  $R_i$ on the solutions of an inverse problem depends on all other
  $\alpha_i$ and $R_i$. For this reason, selecting multiple scalar
  weights to balance multiple penalty functions becomes increasingly
  difficult as we increase the number of penalties.
\item
  \textbf{make direct use of prior knowledge} We can observe model
  properties from training examples and use this information directly as
  constraints \citep[see also numerical examples in this work]{TVLearn}.
  Penalty and basis-pursuit type methods first need to translate this
  information into penalty functions and scalar weights.
\end{itemize}

Most classical and recently proposed methods to project onto an
intersection of multiple (convex) sets, such as Dykstra's algorithm and
variants
\citep{doi:10.1080/01621459.1983.10477029, Dyk, CENSOR2006111, bauschke2015projection, Lopez2016, AragonArtacho2018},
(see also Appendix A), use projections onto each set separately, denoted
as
$\mathcal{P}_{\mathcal{V}_i}(\cdot) \: : \: \mathbb{R}^n \rightarrow \mathbb{R}^N$.
The projection is a black box, and this may create difficulties if the
projection onto one or more sets has no known closed-form solution. We
then need another iterative algorithm to solve the sub-problems. This
nesting of algorithms may lead to problems with the selection of
appropriate stopping criteria for the sub-problem solver. In that case,
we need two sets of stopping criteria: one for Dykstra's algorithm
itself and one for the iterative algorithm that computes the individual
projections. Projections need to be sufficiently accurate such that
Dykstra's algorithm converges. At the same time, we do not want to waste
computational resources by solving sub-problems more accurately than
necessary. A second characteristic of the black-box projection
algorithms is that they treat every set individually and do not attempt
to exploit similarities between the sets. If we work with multiple
constraint sets, some of the set definitions may include the same or
similar linear operators in terms of sparsity (non zero) patterns.

Besides algorithms that are designed specifically to compute projections
onto the intersection of multiple sets, there exist software packages
capable of solving a range of generic optimization problems. However,
many of the current software packages are not designed to compute
projections onto intersections of multiple constraint sets where we
usually do not know the projection onto each set in closed form. This
happens, for instance, when the set definitions include linear operators
$A$ that satisfy the relation $AA^\top\neq\alpha I$ for $\alpha>0$. A
package such as \texttt{Convex} for Julia \citep{ConvexJL}, an example
of disciplined convex programming (DCP), does not handle non-convex sets
and requires lots of memory even for large and sparse linear operators
from problems on 2D grids. The high memory demands are a result of the
packages that \texttt{Convex} can call as the back-end, for example,
\texttt{SCS} \citep{ODonoghue2016} or \texttt{ECOS}
\citep{domahidi2013ecos}. These solvers work with matrices that possess
a structure similar to
\begin{equation}
\begin{aligned}
\begin{pmatrix}
\star & \begin{pmatrix} A_1^\top & \dots & A_p^\top \end{pmatrix}\\
\begin{pmatrix} A_1 \\ \vdots \\ A_p \end{pmatrix}     & \star
\end{pmatrix},
\end{aligned}
\label{saddle}
\end{equation}
 This block-structured system becomes prohibitively large in case we
work with multiple constraint sets that include a linear operator in
their definitions. The software that comes closer to our implementation
is \texttt{Epsilon} \citep{epsilonsoftware}, which is written in Python.
Like our proposed algorithms, \texttt{Epsilon} also employs the
alternating direction method of multipliers (ADMM), but reformulates
optimization problems by emphasizing generalized proximal mappings as in
equation (\ref{prox_affine}, see below). Linear equality constraints
then appear as indicator functions, which leads to different linear
operators ending up in different sub-problems. Contrary, we work with a
single ADMM sub-problem that includes all linear operators. The
\texttt{ProxImaL} software \citep{Heide2016PEI28978242925875} for Python
is designed for linear inverse problems in imaging using ADMM with a
similar problem reformulation. However, \texttt{ProxImaL} differs
fundamentally since it applies regularization with a relatively small
number of penalty functions. While in principle it should be possible to
adapt that package to constrained problem formulations by replacing
penalties with indicator functions, \texttt{ProxImaL} is in its current
form not set up for that purpose. Finally there is
\texttt{StructuredOptimization} \citep{arXiv180301621A} in
\texttt{Julia}. This package also targets inverse problems by
smooth+non-smooth function formulations. Different from the goal of this
work, \texttt{StructuredOptimization} focusses on problems with easy to
compute generalized proximal mappings~\eqref{prox_affine}, i.e., penalty
functions or constraints that are composed with linear operators that
satisfy $AA^\top=\alpha I$. Contrary, we focus on the situation where we
have many constraints with operators ($AA^\top \neq \alpha I$) that make
generalized proximal mappings~\eqref{prox_affine} difficult to compute.
Below, we list additional benefits of our approach compared to existing
packages that can solve intersection projection problems.

\subsection{Contributions}\label{contributions}

Our aim is to design and implement parallel computational optimization
algorithms for solving projection problems onto intersections of
multiple constraint sets in the context of inverse problems. To arrive
at this optimization framework, \texttt{SetIntersectionProjection}, we
propose

\begin{itemize}
\item
  an implementation that avoids nesting of algorithms and exploits
  similarities between constraint sets, unlike black-box alternating
  projection methods such as Dykstra's algorithm. Taking similarities
  between sets into account allows us to work with many sets at a
  relatively small increase in computational cost.
\item
  algorithms that are based on a relaxed variant of the simultaneous
  direction method of multipliers
  \citep[SDMM,][]{SalsaPaper, prox_split, CoSpSDMM}. By merging SDMM
  with recently developed schemes for automatically adapting the
  augmented-Lagrangian penalty and relaxation parameters
  \citep{Xu_2017_CVPR, pmlr-v54-xu17a}, we achieve speedups when solving
  problem~\eqref{proj_intersect} compared to the straightforward
  application of operator splitting methods.
\item
  a software design specifically for set intersection projection
  problems. Our specializations enhance computational performance and
  include \emph{(i)} a relatively simple multilevel strategy for
  ADMM-based algorithms that does part of the computations on coarser
  grids; \emph{(ii)} solutions of banded linear systems in compressed
  diagonal format (CDS) with multi-threaded matrix-vector products
  (MVP). These MVPs are faster than general purpose storage formats like
  compressed sparse column storage (CSC) and support linear operators
  with spatially varying (blurring) kernels; \emph{(iii)} more intuitive
  stopping criteria based on set feasibility.
\item
  to make our work available as a software package in \texttt{Julia}
  \citep{doi:10.1137/141000671}. Besides the algorithms, we also provide
  scripts for setting up the constraints, projectors and linear
  operators, as well as various examples. All presented timings,
  comparisons, and examples are reproducible.
\item
  an implementation that is suitable for small matrices (2D) up to
  larger tensors (3D models, at least
  $m \in \mathbb{R}^{300 \times 300 \times 300}$). Because we solve
  simple-to-compute sub-problems in closed form and independently in
  parallel, the proposed algorithms work with large models and many
  constraints. We achieve this because there is only a single inexact
  linear-system solve that does not become much more computationally
  expensive as we add more constraint sets.
\end{itemize}

To demonstrate the capabilities of our optimization framework and
implementation, we provide examples how projections onto an intersection
of multiple constraint sets can be used to solve linear image processing
tasks such as denoising an deconvolution and more complicated inverse
problems including nonlinear parameters estimation problems with PDEs.

\subsection{Notation, assumptions, and
definitions}\label{notation-assumptions-and-definitions}

Our goal is to estimate the model vector (e.g., discretized medium
parameters such as the acoustic wave speed) $m \in \mathbb{R}^N$, which
in 2D corresponds to a vectorized (lexicographically ordered) matrix of
size $n_z \times n_x$. Coordinate $z$ is the vertical direction and $x$
the horizontal direction. There are $N = n_x \times n_z$ elements in a
2D model. Our work applies to 2D and 3D models but to keep the
derivations simpler we limit the descriptions to 2D models discretized
on a regular grid. We use the following discretization for the vertical
derivative in our constraint definitions
\begin{equation}
\begin{aligned}
D_z = \frac{1}{h_z} \begin{pmatrix}
 -1 & 1  &        &        & \\
    & -1 & 1      &        & \\
    &    & \ddots & \ddots & \\
    &    &        &  -1    & 1
\end{pmatrix},
\end{aligned}
\label{FD_mat}
\end{equation}
 where $h_z$ is the vertical grid size. We define the discretized
vertical derivative for the 2D model as the Kronecker product of $D_z$
and the identity matrix corresponding to the x-dimension:
$D_z \otimes I_x$.

The indicator function of a convex or non-convex set $\mathcal{C}$ is
defined as
\begin{equation}
    \iota_{\mathcal{C}}(m) =
    \begin{cases}
      0 & \text{if } m \in \mathcal{C},\\
      +\infty & \text{if } m \notin \mathcal{C}.\\
\end{cases}
\label{indicator}
\end{equation}
 We define the Euclidean projection onto a set $\mathcal{C}$ as
\begin{equation}
\mathcal{P_{C}}({m})=\operatorname*{arg\,min}_{x} \| x-m \|^2_2 \quad \text{s.t.} \quad m \in \mathcal{C}.
\label{proj_prob}
\end{equation}
 This projection is unique if $\mathcal{C}$ is a closed and convex set.
If $\mathcal{C}$ is a non-convex set, the projection may not be unique
so the result is any vector in the set of minimizers of the projection
problem. The proximal map of a function
$g(m) : \mathbb{R}^N \rightarrow \mathbb{R} \cup \{+\infty\}$ is defined
as
\begin{equation}
\operatorname{prox}_{\gamma,g}(m) = \argmin_x g(x) + \frac{\gamma}{2}\|x-m\|_2^2,
\label{prox_scaled}
\end{equation}
 so
$\operatorname{prox}_{\gamma,g}(m) : \mathbb{R}^N \rightarrow \mathbb{R}^N$,
where $\gamma >0$ is a scalar. The case when $g(x)$ includes a linear
operator $A \in \mathbb{R}^{M \times N}$ is of particular interest to us
and we make it explicit with the definition
\begin{equation}
\operatorname{prox}_{\gamma, g \circ A}(m) = \argmin_x g(A x) + \frac{\gamma}{2}\|x-m\|_2^2.
\label{prox_affine}
\end{equation}
 Even though $\operatorname{prox}_{\gamma,g}(m)$ is often available in
closed-form solution, or cheap to compute \citep[ Chapter 6 \&
7]{prox_split, OPT-003, doi:10.1137/1.9781611974997},
$\operatorname{prox}_{\gamma, g \circ A}(m)$ is usually not available in
closed form if $A A^\top \neq \alpha I, \: \alpha > 0$ and more
expensive to compute. Here, the symbol $^\top$ refers to (Hermitian)
transpose. The proximal map for the indicator function is the
projection:
\begin{equation*}
\operatorname{prox}_{\gamma,\iota_\mathcal{C}}(m)= \mathcal{P}_{\iota_{\mathcal{C}}}(m)
\end{equation*}
 with $\mathcal{P}_{\iota_{\mathcal{C}}}(m)$ defined as in
(\ref{proj_prob}). The intersection of an arbitrary number of convex
sets, $\bigcap_{i=1}^p \mathcal{C}_i$, is also convex. We assume that
all constraints are chosen consistently, such that the intersection of
all selected constraint sets is nonempty:
\begin{equation}
\bigcap_{i=1}^p \mathcal{C}_i \neq \emptyset.
\label{nonemptyset}
\end{equation}
 This assumption is not restrictive in practice because apparently
contradicting constraint sets often have a non-empty intersection. For
example, $\ell_1$-norm based total-variation constraints and smoothness
promoting constraints have at least one model in their intersection: a
homogeneous model has a total-variation equal to $0$ and maximal
smoothness.

We use $m[i]$ to indicate entries of the vector $m$. Subscripts like
$y_i$ refer to one of the sub-vectors that are part of
$\tilde{y} = ( y_1^\top \: y_2^\top \: \dots \: y_p^\top)^\top$.

The Euclidean inner product of two vectors is denoted as $a^\top b$, and
$\| a \|_2^2 = a^\top a$.

\section{PARSDMM: Exploiting similarity between constraint
sets}\label{parsdmm-exploiting-similarity-between-constraint-sets}

As we briefly mentioned in the introduction, we want to construct an
algorithm to compute projections onto the intersection of multiple sets
that \emph{(i)} avoids nesting multiple algorithms if we do not know a
projection onto one of the sets in closed-form; \emph{(ii)} explicitly
exploit similarities between the $i=1,2,\dots,p$ linear operators
$A_i \in \mathbb{R}^{M_i \times N}$; \emph{(iii)} do most computational
work in parallel. The first step to accomplish this is writing each
constraint set $\mathcal{V}_i$ in problem~\eqref{proj_intersect} as the
indicator function of a `simple' set ($\iota_{\mathcal{C}_i}$) and a
possibly non-orthogonal linear operator:
$x \in \mathcal{V}_i \Leftrightarrow A_i x \in \mathcal{C}_i$. We
formulate projection of $m \in \mathbb{R}^N$ onto the intersection of
$p$ sets as
\begin{equation}
\min_{x} \frac{1}{2} \| x - m \|_2^2 + \sum_{i=1}^{p} \iota_{\mathcal{C}_i}(A_i x).
\label{proj_intersect_indicator}
\end{equation}
 In this section, we introduce our main algorithm, Projection Adaptive
Relaxed Simultaneous Method of Multipliers (PARSDMM). The name derives
from the relation to existing works about adaptive relaxation and
ADMM-variants for minimizing sums of functions. It is designed to solve
inverse problems that call for multiple pieces of prior knowledge in the
form of constraints. Each piece of prior knowledge corresponds to a
single set. We focus on intersections up to about $16$ sets, which we
found adequate to regularize inverse problems. To avoid technical issues
with non-convexity, we, for now, assume all sets to be closed and
convex.

We use ADMM as a starting point. ADMM is known to solve intersection
projection (and feasibility) problems
\citep{Boyd:2011:DOS:2185815.2185816, doi:10.1080/10556788.2014.902056, bauschke2015projection, JIA2017320, NIPS2017_6655, arXiv171006465K}.
However, it remains a black-box algorithm and struggles with projections
that do not have closed-form solutions. For completeness and to
highlight the differences with the algorithm we propose below, we
describe in Appendix A a black-box algorithm for the projection onto the
intersection of sets based on ADMM.

\subsection{The augmented Lagrangian}\label{the-augmented-lagrangian}

To start the derivation of PARSDMM, we introduce separate vectors
$y_i \in \mathbb{R}^{M_i}$ for each of the $i=1,\dots,p$ constraint sets
of problem~\eqref{proj_intersect_indicator} and we add linear equality
constraints as follows:
\begin{equation}
\min_{x,\{y_i\}} \frac{1}{2}\| x - m \|_2^2 + \sum_{i=1}^{p} \iota_{\mathcal{C}_i}(y_i) \quad \text{s.t.} \quad A_i x = y_i.
\label{PARSDMM_1}
\end{equation}
 The augmented Lagrangian \citep[e.g.,][ Chapter 17]{Nocedal:2000} of
problem~\eqref{PARSDMM_1} is a basis for ADMM (see~\eqref{aug_lag}
below). To ensure that the $x$-minimization of the augmented Lagrangian
remains quadratic, we make this minimization problem independent of the
distance term $\frac{1}{2}\| x - m \|_2^2$. This choice has the
additional benefit of allowing for other functions that measure distance
from $m$. We remove the direct coupling of the distance term by
\emph{(i)} introducing the additional variables and constraints
$y_{p+1}=A_{p+1}x=I_N x$; \emph{(ii)} defining
$\frac{1}{2}\| x - m \|_2^2 \equiv f(y_{p+1})$; \emph{(iii)} creating
the function
\begin{equation}
\tilde{f}(\tilde{y}) = f(y_{p+1}) + \sum_{i=1}^{p} \iota_{\mathcal{C}_i}(y_i),
\label{PARSDMM_2}
\end{equation}
 where we use the $\tilde{}$ symbol to indicate concatenated matrices
and vectors, as well as functions that are the sum of multiple functions
to simplify notation. The concatenated matrices and vectors read
\begin{equation}
\begin{split}
\tilde{A} = \begin{pmatrix} A_1 \\ \vdots \\ A_{p+1}=I_N \end{pmatrix}, \quad \tilde{y} = \begin{pmatrix} y_1 \\ \vdots \\ y_{p+1} \end{pmatrix}, \quad \tilde{v} = \begin{pmatrix} v_1 \\ \vdots \\ v_{p+1} \end{pmatrix}.
\end{split}
\label{conc_variables}
\end{equation}
 The vectors $v_i \in \mathbb{R}^{M_i}$ are the Lagrangian multipliers
that occur in the augmented Lagrangian for the projection problem. We
always have $A_{p+1} x = I_N x = y_{p+1}$ for the Euclidean projection
that uses the squared $\ell_2$-distance $\frac{1}{2}\| x - m \|_2^2$.
With these new definitions, problem~\eqref{PARSDMM_1} becomes
\begin{equation}
\min_{x,\tilde{y}} \tilde{f}(\tilde{y}) \quad \text{s.t.} \quad \tilde{A} x = \tilde{y}.
\label{proj_intersect_final}
\end{equation}
 This formulation has the same form as problems that regular ADMM
solves---i.e.,
$\min_{x,y} f(x) + g(y) \:\: \text{s.t.} \:\: A x + B y = c$. It follows
that we can guarantee convergence under the same conditions as for ADMM.
According to
\citep{Boyd:2011:DOS:2185815.2185816, eckstein2015understanding}, ADMM
converges when
$f(x) : \mathbb{R}^{N_1} \rightarrow \mathbb{R} \cup \{+\infty\}$ and
$g(y) : \mathbb{R}^{N_2} \rightarrow \mathbb{R} \cup \{+\infty\}$ are
proper and convex. The linear equality constraints involve matrices
$A \in \mathbb{R}^{M \times N_1}$ and $B \in \mathbb{R}^{M \times N_2}$
and vectors $x \in \mathbb{R}^{N_1}$, $y \in \mathbb{R}^{N_2}$ and
$c \in \mathbb{R}^M$.

To arrive at the main iterations of PARSDMM, we continue based on the
augmented Lagrangian for~\eqref{proj_intersect_final}, which reads
\begin{equation}
L_{\rho_1, \dots, \rho_{p+1}} (x, y_1, \dots , y_{p+1}, v_1, \dots, v_{p+1}) = \sum_{i=1}^{p+1} \bigg[ f_i(y_i) + v_i^\top (y_i - A_i x ) + \frac{\rho_i}{2} \| y_i - A_i x \|^2_2 \bigg].
\label{aug_lag}
\end{equation}
 As we can see, this expression has a separable structure with respect
to the Lagrangian multipliers $v_i$, and the auxiliary vectors $y_i$.
Following the ADMM variants for multiple functions, as formulated by
\citep{Song:2016:FAA:3015812.3015924, CoSpSDMM, pmlr-v70-xu17c}, we use
a different penalty parameter $\rho_i > 0$ for each index $i$. In this
way, we make sure all linear equality constraints $A_i x = y_i$ are
satisfied sufficiently when running a limited number of iterations.
Because the different matrices $A_i$ may have widely varying scalings
and sizes, a fixed penalty for all $i$ could cause slow convergence of
$x$ towards one of the constraint sets. To further accelerate the
algorithm we also introduce a different relaxation parameter
($\gamma_i$) for each index $i$. After we derive the main steps of our
proposed algorithm, we describe the automatic selection of the scalar
parameters.

\subsection{The iterations}\label{the-iterations}

With the above definitions, iteration counter $k$, and inclusion of
relaxation parameters, which we assume to be limited to the interval
$\gamma_i \in [1,2)$ \citep[see][]{Xu_2017_CVPR}, the iterations can be
written as
\begin{align}\label{SDMM_iters}
x^{k+1} &= \arg\min_{x} \sum_{i=1}^{p+1} \Big( \frac{\rho_i^k}{2} \| y_i^{k} - A_i x + \frac{v_i^k}{\rho_i^k} \|^2_2 \Big) \nonumber\\
\bar{x}_i^{k+1} &= \gamma_i^k A_i x_i^{k+1} + ( 1-\gamma_i^k ) y_i^{k} \\
y_i^{k+1} &= \arg\min_{y_i} \Big[ f_i(y_i) + \frac{\rho_i^k}{2} \| y_i^{k} - \bar{x}_i^{k+1} + \frac{v_i^k}{\rho_i^k} \|^2_2 \Big] \nonumber\\
v_i^{k+1} &= v_i^k + \rho_i^k (y_i^{k+1} - \bar{x}_i^{k+1}). \nonumber
\end{align}
To arrive at our final algorithm, we rewrite these iterations in a more
explicit form as
\begin{align}\label{SDMM_adaptive}
x^{k+1} &= \Big[ \sum_{i=1}^{p} (\rho_i^k A_i^\top A_i ) + \rho_{p+1}^k I_N \big]^{-1} \sum_{i=1}^{p+1} \Big[A_i^\top(\rho_i^k y_i^{k} + v_i^k) \Big]\nonumber\\ 
\bar{x}_i^{k+1} &= \gamma_i^k A_i x_i^{k+1} + ( 1-\gamma_i^k ) y_i^{k} \\ 
y_i^{k+1} &= \operatorname{prox}_{f_i,\rho_i^k}(\bar{x}_i^{k+1} - \frac{v_i^k}{\rho_i^k} )\nonumber\\ 
v_i^{k+1} &= v_i^k + \rho_i^k (y_i^{k+1} - \bar{x}_i^{k+1}). \nonumber
\end{align}
 In this expression, we used the fact that $A_{p+1}$ is always the
identity matrix of size $N$ for projection problems. Without over/under
relaxation \citep[$\bar{x}_i^{k+1}$
computation,][]{Eckstein1992, doi:10.1080/10556788.2017.1396601, Xu_2017_CVPR},
these iterations are known as SALSA \citep{SalsaPaper} or the
simultaneous direction method of multipliers
\citep[SDMM,][]{prox_split, CoSpSDMM}. The derivation in this section
shows that ADMM/SDMM solve the projection onto an intersection of
multiple closed and convex sets. However, the basic iterations
from~\eqref{SDMM_adaptive} are not yet a practical and fast algorithm,
because there are scalar parameters that need to be selected, no
stopping conditions, and no specializations to constraints typically
found in the imaging sciences. We address these issues in the following
sections.

\subsection{Computing the proximal
maps}\label{computing-the-proximal-maps}

The proximal maps in the iterations~\eqref{SDMM_adaptive} become
projections onto simple sets (e.g., bounds/$\ell_1$ and $\ell_2$
norm-ball/cardinality/rank), which permit closed-form solutions that do
not depend on the $\rho_i$. When $f_{p+1}(w)=1/2\|w-m\|_2^2$, (squared
$\ell_2$ distance of $w$ to the reference vector $m$) the proximal map
is also available in closed form:
\begin{equation}
\begin{aligned}
\operatorname{prox}_{f_{p+1},\rho_{p+1}}(w)&=\argmin_{z} 1/2\|z-m\|_2^2 + \rho_{p+1}/2\|z-w\|_2^2\\ &=(m+\rho_{p+1} w)/(1+\rho_{p+1}),
\end{aligned}
\label{prox_sq_dist}
\end{equation}
 where we understand the division in a point-wise sense. We thus avoided
other convex optimization algorithms for computations of the proximal
maps of interest.

\subsection{Solving the linear system and automatic parameter
selection}\label{solving-the-linear-system-and-automatic-parameter-selection}

We can see from~\eqref{SDMM_adaptive} that the computation of $x^{k+1}$
involves the solution of a single system of normal equations that
contains all linear operators. The system matrix equals
\begin{equation}
C \equiv \sum_{i=1}^{p+1} (\rho_i A_i^\top A_i ) = \sum_{i=1}^{p} (\rho_i A_i^\top A_i ) + \rho_{p+1} I_N
\label{sys_mat}
\end{equation}
 and is by construction always positive-definite because $\rho_i > 0$
for all $i$. The minimization over $x$ is therefore uniquely defined. As
suggested by \citet{pmlr-v54-xu17a}, we adapt the $\rho_i$'s every two
iterations using the scheme we discuss below.

While we could use direct matrix factorizations of $C$, we would need to
refactorize every time we update any of the $\rho_i$'s. This makes
computing $x^{k+1}$ too costly. Instead, we rely on warm-started
iterative solvers with $x^k$ used as the initial guess for $x^{k+1}$.
There exist several alternatives including LSQR
\citep{Paige:1982:LAS:355984.355989} to solve the above linear system
($x^{k+1}$ computation in~\ref{SDMM_adaptive}) iteratively. We choose to
apply the conjugate-gradient (CG) method to the normal equations for the
following reasons:

\begin{enumerate}
\def\labelenumi{\arabic{enumi}.}
\item
  Contrary to LSQR, transforms that satisfy $A_i^\top A_i= \alpha I_N$
  are free in CG because we explicitly form the sparse system matrix
  $C$, which already includes the identity matrix.
\item
  By limiting the relative difference between the $\rho_i$ and
  $\rho_{p+1}$, where the latter corresponds to the identity matrix
  in~\eqref{sys_mat}, we ensure $C$ is sufficiently well conditioned so
  squaring the condition number does not become a numerical problem.
\item
  For many transforms, the matrices $A_i^\top A_i$ are sparse and have
  at least partially overlapping sparsity patterns (discrete derivative
  matrices for one or more directions, orthogonal transforms).
  Multiplication with $\sum_{i=1}^{p+1} (\rho_i A_i^\top A_i )$ is
  therefore not much more expensive than multiplication with a single
  $A_i^\top A_i$. However, LSQR requires matrix-vector products with all
  $A_i$ and $A_i^\top$ at every iteration.
\item
  Full reassembly of $C$ at iteration $k$ is not required. Every time we
  update any of the $\rho_i$, we update $C$ by subtracting and adding
  the block corresponding to the updated $\rho_i$. If the index that
  changed is indicated by $i=u$, the system matrix for the next
  $x^{k+1}$ computation becomes
  \begin{equation}
  \begin{aligned}
  C^{k+1} = \sum_{i=1}^{p+1} (\rho_i^{k+1} A_i^\top A_i ) &= \sum_{i=1}^{p+1} (\rho_i^k A_i^\top A_i ) - (\rho_u^k A_u^\top A_u ) + (\rho_u^{k+1} A_u^\top A_u )\\ & = C^k + A_u^\top A_u (\rho_u^{k+1} - \rho_u^{k}).
  \end{aligned}
  \label{C_update}
  \end{equation}
   For each $\rho_i$ update, forming the new system matrix involves a
  single addition of two sparse matrices (assuming all $A_i^\top A_i$'s
  are pre-computed).
\end{enumerate}

To further save computation time, we solve the minimization with respect
to $x$ inexactly. We select the stopping criterion for CG adaptively in
terms of the relative residual of the normal equations---i.e., we stop
CG if the relative residual drops below
\begin{equation}
0.1 \| \Big[ \sum_{i=1}^{p+1} (\rho_i^k A_i^\top A_i ) \big] x - \sum_{i=1}^{p+1} \Big[A_i^\top(\rho_i^k y_i^{k} + v_i^k) \Big]\|_2 / \|\sum_{i=1}^{p+1} \Big[A_i^\top(\rho_i^k y_i^{k} + v_i^k) \Big]\|_2.
\label{cg_stop}
\end{equation}
 Empirically, we found this choice robust and it also results in time
savings for solving problem~\eqref{proj_intersect_final} compared to a
fixed and accurate stopping criterion for the $x$-minimization step. The
stopping criterion for CG is relatively inexact during the first few
iterations from~\eqref{SDMM_adaptive} and requests more accurate
solutions later on, such that the conditions on inexact sub-problem
solutions from \citep{Eckstein1992} will be satisfied eventually.

Just like standard ADMM, our algorithm may also need a large number of
iterations~\eqref{SDMM_adaptive} for a fixed penalty parameter $\rho_i$
for all $i$ \citep[e.g.,][]{nishihara2015general, pmlr-v54-xu17a}. It is
better to update $\rho_i^k$ and $\gamma_i^k$ every couple of iterations
to ensure we reach a good solution in a relatively small number of
iterations. For this purpose, we use \citet{pmlr-v54-xu17a}'s automatic
selection of $\rho_i^k$ and $\gamma_i^k$ for ADMM. Numerical experiments
by \citet{empiricalncvxadmm} show that these updates also perform well
on various non-convex problems. The updates themselves are based on a
Barzilai-Borwein spectral step size \citep{BARZILAI01011988} for
Douglas-Rachford (DR) splitting applied to the dual of
$\min_{x,y} f(x) + g(y) \:\: \text{s.t.} \:\: A x + B y = c$ and derive
from equivalence between ADMM and DR
\citep{Eckstein1992, esser2009applications}.

\subsection{Exploiting parallelism}\label{exploiting-parallelism}

Given the grid size of 3D PDE-based parameter estimation problems and
our desire to work with multiple constraint sets, we seek a parallel
implementation that exploits multi-threading offered by programming
languages such as Julia \citep{doi:10.1137/141000671}. Since the
computational time for the x-minimization using the conjugate-gradient
algorithm is dominated by the matrix-vector products (MVP) with $C$, we
concentrate our efforts there by using compressed diagonal storage
(CDS), see, e.g., \citet{doi:10.1137/0910073};
\citet{CNM:CNM1630060505}; \citet{kotakemori2008performance}. This
format stores the non-zero bands of the matrix as a dense matrix, and we
compute MVPs directly in this storage sytem. These MVPs are faster than
the more general Compressed Sparse Column (CSC) format. CDS has the
additional benifit that it can efficiently handle matrices generated by
spatially varying (blurring, derivative) kernels. We can use CDS if all
matrices $A_i^\top A_i$ have a banded sparsity-pattern. Using Julia's
multi-threading, we compute the MVPs with $C$ in parallel. In cases
where the $A_i^\top A_i$'s do not have a banded structure we revert to
computations in the standard CSC format.

Aside from matrix-vector products during the inner iterations, most
calculation time in~\eqref{SDMM_adaptive} is used for $\bar{x}_i^{k+1}$,
$y_i^{k+1}$, $v_i^{k+1}$, $\rho_i^{k+1}$, and $\gamma_i^{k+1}$. To
reduce these computation times, we compute these quantities in parallel.
This is relatively straightforward to do because each problem is
independent so that the operations for the $p$ constraints can be
carried out by different Julia workers where each worker either uses
Julia threads, multi-threaded BLAS \citep[OpenBLAS,][]{openblas_paper},
or multi-threaded Fourier-transforms \citep[FFTW
library,][]{FFTW_paper}.

\subsection{Stopping conditions}\label{stopping-conditions}

So far, we focussed on reducing the time for each iteration
of~\eqref{SDMM_adaptive}. However, fast solutions to the full problem
are possible only if we know when to stop iterating. When working with a
single constraint set, stopping criteria based on a combination of the
primal $r^\text{pri} = \| \tilde{y} - \tilde{A} x^k \|$ and dual
residual
$r^\text{dual} = \| \tilde{\rho} \tilde{A}^\top (\tilde{y}^{k}-\tilde{y}^{k-1}) \|$
are adequate as long as both become sufficiently small
\citep[e.g.,][]{Boyd:2011:DOS:2185815.2185816, CoSpSDMM, pmlr-v54-xu17a}.
However, the situation is more complicated in situations where we work
with multiple constraint sets. In that case, the $\tilde{y}$ and
$\tilde{A}$ contain a variety of vectors and linear operators that
correspond to the different constraint sets. It then becomes more
difficult to determine the relationship between the size of the
residuals and the distance to each set. With other words, it becomes
challenging to decide at what primal and dual residual to stop such that
we are close to all constraint sets.

Instead, it may be more intuitive to look at the feasibility for each
set separately. This holds if $x$ is in the intersection of the
constraint sets but requires computationally costly projections onto
each $\mathcal{V}_i$ to verify, a situation we want to avoid in PARSDMM.
Instead, we rely on the transform-domain set feasibility error
\begin{equation}
r^{\text{feas}}_i = \frac{ \| A_i x - \mathcal{P}_{\mathcal{C}_i}(A_i x) \| }{ \| A_i x \| }, \, i=1\cdots p,
\label{feas_stop}
\end{equation}
 to which we have access at a relatively low cost since we already
computed $A_i x$ in the iterations from~\eqref{SDMM_adaptive}. Our first
stopping criterion thus corresponds to a normalized version of the
objective \citet{censor2005multiple} uses when solving convex multiple
set split-feasibility problems. We added this normalization
in~\eqref{feas_stop} to account for different types and sizes of the
linear operators $A_i$. The projections onto the constraint sets
$\mathcal{P}_{\mathcal{C}_i}(\cdot)$ themselves, are relatively cheap to
compute since they only include projections onto `simple' sets based on
norm-balls, bounds, cardinality, and matrix rank. By testing for
transform-domain feasibility every few iterations only (typically $5$),
we further reduce the computational costs for our stopping condition.

Satisfying constraints does not indicate whether or not $x^k$ is close
to the projection onto the intersection of the $p$ different constraint
sets or if it is just a feasible point, possibly `deep' inside the
intersection. If $x^k$ is indeed the projection of $m$, then
$\|x^k - x^{k-1}\|$ approaches a stationary point, assuming that $x^k$
converges to the projection. We make this property explicit by
considering the maximum relative change of $x^k$ over the $s$ previous
iterations: $j \in S \equiv \{ 1,2,\dots,s \}$. The relative evolution
of $x$ at the $k\mathrm{th}$ iteration thus becomes
\begin{equation}
r^\text{evol} = \frac{ \text{max}_{j \in S} \{ \|x^k - x^{k-j}\| \} }{ \|x^k\| }\,.
\label{obj_stop}
\end{equation}
 By considering the history (we use $s=5$ in our numerical examples),
our stopping criterion becomes more robust to oscillations in
$\|x^k - x^{k-1}\|$ as a function of $k.$ So we propose to stop PARSDMM
if
\begin{equation}
r^\text{evol} < \varepsilon^\text{evol} \quad \text{and} \quad r^{\text{feas}}_i < \varepsilon^{\text{feas}}_i \quad \forall \: i.
\label{stopping_conditions}
\end{equation}
 During our numerical experiments, we select
$\varepsilon^\text{evol}=10^{-2}$ and
$\varepsilon^{\text{feas}}_i = 10^{-3}$, which balance sufficiently
accurate solutions and short solution times. These are still two
constants to be chosen by the user, but we argue that
$r^{\text{feas}}_i$ may relate better to our intuition on feasibility
because it behaves like a distance to each set separately. The evolution
term $\|x^k - x^{k-1}\|$ is found in many optimization algorithms and is
especially informative for physical parameter estimation problems where
practitioners often have a good intuition to which $\|x^k - x^{k-1}\|$
the physical forward operator is sensitive.

\subsection{The PARSDMM algorithm}\label{the-parsdmm-algorithm}

We summarize our discussions from the previous sections in the following
Algorithms.

\begin{scholmdAlgorithm}
Algorithm~\texttt{PARSDMM}\\$\textbf{inputs:}$\\\hspace*{0.333em}\hspace*{0.333em}$m$~\texttt{//point\phantom{\ }to\phantom{\ }project}\\\hspace*{0.333em}\hspace*{0.333em}$A_1,A_2,\dots,A_{p},A_{p+1}=I_N$~\texttt{//linear\phantom{\ }operators}\\\hspace*{0.333em}\hspace*{0.333em}$\operatorname{prox}_{f_i,\rho_i}(w) = \mathcal{P}_{\mathcal{C}_i}(w)$~for~$i=1, 2, \dots, p$~\texttt{//norm/bound/cardinality/...\phantom{\ }projectors}\\\hspace*{0.333em}\hspace*{0.333em}$\operatorname{prox}_{f_i,\rho_{p+1}}(w) = (m+\rho_{p+1} w)/(1+\rho_i)$~\texttt{//prox\phantom{\ }for\phantom{\ }the\phantom{\ }squared\phantom{\ }distance\phantom{\ }from}~$m$\\\hspace*{0.333em}\hspace*{0.333em}select~$\rho_i^0$,~$\gamma_i^0$,~$\text{update-freqency}$~for~$\gamma$~and~$\rho$\\\hspace*{0.333em}\hspace*{0.333em}optional:~initial~guess~for~$x$,~$y_i$~and~$v_i$

$\textbf{initialize:}$~~\\\hspace*{0.333em}\hspace*{0.333em}\hspace*{0.333em}$B_i = A_i^\top A_i$~\texttt{//pre-compute\phantom{\ }for\phantom{\ }all}~$i$\\\hspace*{0.333em}\hspace*{0.333em}\hspace*{0.333em}$C = \sum_{i=1}^{p+1} (\rho_i B_i )$~\texttt{//pre-compute}\\\hspace*{0.333em}\hspace*{0.333em}\hspace*{0.333em}$k=1$\\\textbf{WHILE}~not~converged\\\hspace*{0.333em}\hspace*{0.333em}\hspace*{0.333em}\hspace*{0.333em}\hspace*{0.333em}\hspace*{0.333em}$x^{k+1} = C^{-1} \sum_{i=1}^{p+1} \Big[A_i^\top(\rho_i^k y_i^{k} + v_i^k) \Big]$~\texttt{//CG,\phantom{\ }stop\phantom{\ }when}~\eqref{cg_stop}~\texttt{holds}\\\hspace*{0.333em}\hspace*{0.333em}\hspace*{0.333em}\hspace*{0.333em}\hspace*{0.333em}\hspace*{0.333em}\textbf{FOR}~$i=1, 2, \dots, p+1$~\texttt{//compute\phantom{\ }in\phantom{\ }parallel}\\\hspace*{0.333em}\hspace*{0.333em}\hspace*{0.333em}\hspace*{0.333em}\hspace*{0.333em}\hspace*{0.333em}\hspace*{0.333em}\hspace*{0.333em}\hspace*{0.333em}\hspace*{0.333em}\hspace*{0.333em}$s_i^{k+1} = A_i x^{k+1}$\\\hspace*{0.333em}\hspace*{0.333em}\hspace*{0.333em}\hspace*{0.333em}\hspace*{0.333em}\hspace*{0.333em}\hspace*{0.333em}\hspace*{0.333em}\hspace*{0.333em}\hspace*{0.333em}\hspace*{0.333em}$\bar{x}_i^{k+1} = \gamma_i^k s_i^{k+1} + ( 1-\gamma_i^k ) y_i^{k}$\\\hspace*{0.333em}\hspace*{0.333em}\hspace*{0.333em}\hspace*{0.333em}\hspace*{0.333em}\hspace*{0.333em}\hspace*{0.333em}\hspace*{0.333em}\hspace*{0.333em}\hspace*{0.333em}\hspace*{0.333em}$y_i^{k+1} = \operatorname{prox}_{f_i,\rho_i}(\bar{x}_i^{k+1} - \frac{v_i^k}{\rho_i^k} )$\\\hspace*{0.333em}\hspace*{0.333em}\hspace*{0.333em}\hspace*{0.333em}\hspace*{0.333em}\hspace*{0.333em}\hspace*{0.333em}\hspace*{0.333em}\hspace*{0.333em}\hspace*{0.333em}\hspace*{0.333em}$v_i^{k+1} = v_i^k + \rho_i^k (y_i^{k+1} - \bar{x}_i^{k+1})$\\\hspace*{0.333em}\hspace*{0.333em}\hspace*{0.333em}\hspace*{0.333em}\hspace*{0.333em}\hspace*{0.333em}\hspace*{0.333em}\hspace*{0.333em}\hspace*{0.333em}\hspace*{0.333em}\hspace*{0.333em}stop~if~conditions~\eqref{stopping_conditions}~hold\\\hspace*{0.333em}\hspace*{0.333em}\hspace*{0.333em}\hspace*{0.333em}\hspace*{0.333em}\hspace*{0.333em}\hspace*{0.333em}\hspace*{0.333em}\hspace*{0.333em}\hspace*{0.333em}\hspace*{0.333em}If~$\text{mod}(k,\text{update-freqency}) = 1$\\\hspace*{0.333em}\hspace*{0.333em}\hspace*{0.333em}\hspace*{0.333em}\hspace*{0.333em}\hspace*{0.333em}\hspace*{0.333em}\hspace*{0.333em}\hspace*{0.333em}\hspace*{0.333em}\hspace*{0.333em}\hspace*{0.333em}\hspace*{0.333em}\hspace*{0.333em}$\{\rho_i^{k+1},\gamma_i^{k+1}\}=\text{adapt-rho-gamma}(v_i^k,v_i^{k+1},y_i^{k+1},s_i^{k+1},\rho_i^k)$\\\hspace*{0.333em}\hspace*{0.333em}\hspace*{0.333em}\hspace*{0.333em}\hspace*{0.333em}\hspace*{0.333em}\hspace*{0.333em}\hspace*{0.333em}\hspace*{0.333em}\hspace*{0.333em}End~if\\\hspace*{0.333em}\hspace*{0.333em}\hspace*{0.333em}\hspace*{0.333em}\hspace*{0.333em}\hspace*{0.333em}\textbf{END}\\\hspace*{0.333em}\hspace*{0.333em}\hspace*{0.333em}\hspace*{0.333em}\hspace*{0.333em}\hspace*{0.333em}\textbf{FOR}~$i=1, 2, \dots, p+1$~\texttt{//update}~$C$~\texttt{if\phantom{\ }necessary}\\\hspace*{0.333em}\hspace*{0.333em}\hspace*{0.333em}\hspace*{0.333em}\hspace*{0.333em}\hspace*{0.333em}\hspace*{0.333em}\hspace*{0.333em}\hspace*{0.333em}\hspace*{0.333em}If~$\rho_i^{k+1} \neq \rho_i^{k}$\\\hspace*{0.333em}\hspace*{0.333em}\hspace*{0.333em}\hspace*{0.333em}\hspace*{0.333em}\hspace*{0.333em}\hspace*{0.333em}\hspace*{0.333em}\hspace*{0.333em}\hspace*{0.333em}\hspace*{0.333em}\hspace*{0.333em}\hspace*{0.333em}\hspace*{0.333em}\hspace*{0.333em}$C \leftarrow C + B_i (\rho_i^{k+1} - \rho_i^{k})$\\\hspace*{0.333em}\hspace*{0.333em}\hspace*{0.333em}\hspace*{0.333em}\hspace*{0.333em}\hspace*{0.333em}\hspace*{0.333em}\hspace*{0.333em}\hspace*{0.333em}\hspace*{0.333em}End~if\\\hspace*{0.333em}\hspace*{0.333em}\hspace*{0.333em}\hspace*{0.333em}\hspace*{0.333em}\hspace*{0.333em}\textbf{END}\\\hspace*{0.333em}\hspace*{0.333em}\hspace*{0.333em}\hspace*{0.333em}\hspace*{0.333em}\hspace*{0.333em}$k \leftarrow k+1$\\\textbf{END}\\$\textbf{output:}$~$x$
\caption{Projection Adaptive Relaxed Simultaneous Direction Method of
Multipliers (PARSDMM) to compute the projection onto an intersection,
including automatic selection of the penalty parameters and
relaxation.}\label{alg:PARSDMM}
\end{scholmdAlgorithm}

\begin{scholmdAlgorithm}
Algorithm~\texttt{adapt-rho-gamma}\\$\textbf{input:}$~$v_i^k,v_i^{k+1},y_i^{k+1},s_i^{k+1},\rho_i^k$

$\varepsilon^\text{corr} = 0.3$\\$\hat{v}^{k+1} = v_i^{k} + \rho_i^k (y_i^{k} - s_i^{k+1})$

$\Delta \hat{v} = \hat{v}_i^{k+1} - \hat{v}^{k_0}$\\$\Delta v = v_i^{k+1} - v^{k_0}$\\$\Delta \hat{h} = s_i^{k+1} - s^{k_0} )$\\$\Delta \hat{g} = -(y_i^{k+1} - y^{k_0})$

$\alpha^\text{corr} = \frac{\Delta\hat{h}^\top \Delta \hat{v}}{\| \Delta\hat{h} \| \| \Delta \hat{v} \|}$\\$\beta^\text{corr} = \frac{\Delta\hat{g}^\top \Delta v}{\| \Delta\hat{g} \| \| \Delta v \|}$

~~If~$\alpha^\text{corr} > \varepsilon^\text{corr}$\\\hspace*{0.333em}\hspace*{0.333em}\hspace*{0.333em}\hspace*{0.333em}\hspace*{0.333em}\hspace*{0.333em}\hspace*{0.333em}$\hat{\alpha}^{\text{MG}} = \frac{\Delta\hat{h}^\top \Delta \hat{v}}{ \Delta\hat{h}^\top \Delta\hat{h}}$,~$\hat{\alpha}^{\text{SD}} = \frac{\Delta \hat{v}^\top \Delta \hat{v}}{\Delta\hat{h}^\top \Delta \hat{v}}$,~$\hat{\alpha} = \begin{cases} \hat{\alpha}^{\text{MG}} & \text{if } 2 \hat{\alpha}^{\text{MG}} > \hat{\alpha}^{\text{SD}} \\ \hat{\alpha}^{\text{SD}} - 0.5 \hat{\alpha}^{\text{MG}} & \text{if } \text{else} \end{cases}$~\\\hspace*{0.333em}\hspace*{0.333em}End\\\hspace*{0.333em}\hspace*{0.333em}If~$\beta^\text{corr} > \varepsilon^\text{corr}$~\\\hspace*{0.333em}\hspace*{0.333em}\hspace*{0.333em}\hspace*{0.333em}\hspace*{0.333em}\hspace*{0.333em}$\hat{\beta}^{\text{MG}} = \frac{\Delta\hat{g}^\top \Delta v}{\Delta\hat{g}^\top \Delta\hat{g}}$,~$\hat{\beta}^{\text{SD}} = \frac{\Delta v^\top \Delta v}{\Delta\hat{g}^\top \Delta v}$,~$\hat{\beta} = \begin{cases} \hat{\beta}^{\text{MG}} & \text{if } 2 \hat{\beta}^{\text{MG}} > \hat{\beta}^{\text{SD}} \\ \hat{\beta}^{\text{SD}} - 0.5 \hat{\beta}^{\text{MG}} & \text{if } \text{else} \end{cases}$~\\\hspace*{0.333em}\hspace*{0.333em}End

$\{ \rho^{k+1},\gamma^{k+1} \} = \begin{cases} \{ \sqrt{\hat{\alpha} \hat{\beta}} , 1+\frac{2 \sqrt{\hat{\alpha} \hat{\beta}}}{\hat{\alpha} + \hat{\beta}} \} & \text{if } \alpha^\text{corr} > \varepsilon^\text{corr} \: \& \: \beta^\text{corr} > \varepsilon^\text{corr} \\ \{ \hat{\alpha} , 1.9 \} & \text{if } \alpha^\text{corr} > \varepsilon^\text{corr} \: \& \: \beta^\text{corr} \leq \varepsilon^\text{corr} \\ \{ \hat{\beta} , 1.1 \} & \text{if } \alpha^\text{corr} \leq \varepsilon^\text{corr} \: \& \: \beta^\text{corr} > \varepsilon^\text{corr} \\ \{ \rho^k , 1.5 \} & \text{if } \alpha^\text{corr} \leq \varepsilon^\text{corr} \: \& \: \beta^\text{corr} \leq \varepsilon^\text{corr} \\ \end{cases}$

set~$\hat{v}^{k_0} \leftarrow \hat{v}_i^{k+1}$,~$v^{k_0} \leftarrow v_i^{k+1}$,~$s^{k_0} \leftarrow s_i^{k+1}$,~$y^{k_0} \leftarrow y_i^{k+1}$~and~save~for~next~call~to~\texttt{adapt-rho-gamma}~\\save~$v_i^{k+1},y_i^{k+1}$~for~next~call~to~\texttt{adapt-rho-gamma}\\$\textbf{output:}$~$\rho_i^{k+1}$,~$\gamma_i^{k+1}$
\caption{Adapt $\rho$ and $\gamma$ according to \citep{Xu_2017_CVPR}
with some modifications to save computational work. The constant
$\varepsilon^\text{corr}$ is in the range $[0.1 - 0.4]$ as suggested by
\citep{Xu_2017_CVPR}. Quantities from the previous call to
adapt-rho-gamma have the indication $k_0$. Actual implementation
computes and re-uses some of the inner products and
norms.}\label{alg:adapt-rho-gamma}
\end{scholmdAlgorithm}

\subsection{Multilevel PARSDMM}\label{multilevel-parsdmm}

Inverse problems with PDE forward models typically need a fine grid for
stable physical simulations. At the same time, we often use constraints
to estimate `simple' models---i.e.~models that are smooth, have a
low-rank, are sparse in some transform-domain, and that may not need
many grid points for accurate representations of the image/model. This
suggests we can reduce the total computational time of PARSDMM
(Algorithm~\ref{alg:PARSDMM}) by using a multilevel continuation
strategy \citep[see,
e.g.,][]{doi:10.1080/10556780008805795, Ascher_2001, doi:10.1080/10556788.2012.759571, doi:10.1137/16M1082299}.
The multilevel idea presented in this section applies to the projection
onto the intersection of constraint sets only and not to the grids for
solving PDEs. Our approach proceeds as follows: we start at a coarse
grid and continue towards finer grids without cycling between coarse and
fine grids. By using the solution at the coarse grid as the initial
guess for the solution on the finer grid, the convergence guarantees are
the same as for the single level version of our algorithm. This
multilevel approach is similar to multilevel ADMM by
\citet{Macdonald2018} for non-convex linear equality constrained
problems. Numerical experiments in the following section show reduced
solution times and better performance on non-convex problems.

Different from many applications of multilevel algorithms, is that we
are not interested in approximating the primal variable $x$, because
ADMM-type iterations compute this quantity in the first step, see
Algorithm~\ref{alg:PARSDMM}. Instead, we need to concern ourselves with
the initial guesses of auxilliary variables $y_i$ and Lagrangian
multipliers $v_i$ for all $i \in \{1,\dots,p,\,p+1 \}$. After
initialization of the coarsest grid with all zero vectors, we move to a
finer grid by interpolating $y_i$ and $v_i$. Since the solution estimate
$x \in \mathbb{R}^N$ always refers to an image (2D/3D) for our
applications, we know that $v_i$ and $y_i$ relate to images as well and
their dimensions depend on the corresponding $A_i$. Therefore we
interpolate $y_i$ and $v_i$ as images.

\textbf{Example.} When $A_i$ is a discrete derivative matrix, then the
vectors $v_i$ and $y_i$ live on a grid that we know at every level of
the multilevel scheme. If we have $A_i = D_z \otimes I_x$, where $D_z$
is the first-order finite-difference matrix as in~\eqref{FD_mat}, we
know that $A_i \in \mathbb{R}^{((n_z-1)n_x) \times (n_z \times n_x)}$
and therefore $v_i \in \mathbb{R}^{(n_z-1)n_x}$ and
$y_i \in \mathbb{R}^{(n_z-1)n_x}$. We can thus reshape the associated
vectors $v_i$ and $y_i$ as an image (in 2D) of size $(n_z-1 \times n_x)$
and interpolate it to the finer grid for the next level, working from
coarse to fine. In 3D, we follow the same approach. We also need a
coarse version of $m$ at each level: $m_l$ for
$l=n_\text{levels},n_\text{levels}-1,\dots,1$. We simply obtain the
coarse models by applying anti-alias filtering and subsampling of the
original $m$. In principle, any subsampling and interpolation technique
may be used in this multilevel framework, because it just constructs
initial guesses for the next levels.

We decide the number of levels ($n_\text{levels}$) and the coarsening
factor ahead of time. Together with the original grid, these determine
the grid at all levels so we can set up the linear operators and
proximal mappings at each level. This set-up phase is a one time cost
since its result is reused every time we project a model $m$ onto the
intersection of constraint sets. The additional computational costs of
the multilevel scheme are the interpolation of all $y_i$ and $v_i$ to a
finer grid, but this happens only once per level and not every
ML-PARSDMM (Algorithm~\ref{alg:ML-PARSDMM}) iteration. So the
computational overhead we incur from the interpolations is small
compared to the speedups that Algorithm~\ref{alg:ML-PARSDMM} offers.

\begin{scholmdAlgorithm}
$\textbf{inputs:}$\\\hspace*{0.333em}\hspace*{0.333em}$n_\text{levels}$~\texttt{//number\phantom{\ }of\phantom{\ }levels}\\\hspace*{0.333em}\hspace*{0.333em}$l=\{ n_\text{levels},n_\text{levels}-1, \dots, 1\}$\\\hspace*{0.333em}\hspace*{0.333em}$\text{grid}_l$~~~\texttt{//grid\phantom{\ }info\phantom{\ }at\phantom{\ }each\phantom{\ }level}~$l$\\\hspace*{0.333em}\hspace*{0.333em}$m_{l}$~~~~~~~\texttt{//model\phantom{\ }to\phantom{\ }project\phantom{\ }at\phantom{\ }every\phantom{\ }level}~$l$\\\hspace*{0.333em}\hspace*{0.333em}$A_{1,l},A_{2,l},\dots,A_{p+1,l}$~\texttt{//linear\phantom{\ }operators\phantom{\ }at\phantom{\ }every\phantom{\ }level}\\\hspace*{0.333em}\hspace*{0.333em}\texttt{//\phantom{\ }norm/bound/cardinality/...\phantom{\ }projectors\phantom{\ }at\phantom{\ }each\phantom{\ }level:}\\\hspace*{0.333em}\hspace*{0.333em}$\operatorname{prox}_{f_{i,l},\rho_i}(w) = \mathcal{P}_{\mathcal{C}_{i,l}}(w)$~for~$i=1, 2, \dots, p$\\\hspace*{0.333em}\hspace*{0.333em}\texttt{//\phantom{\ }proximal\phantom{\ }map\phantom{\ }for\phantom{\ }the\phantom{\ }squared\phantom{\ }distance\phantom{\ }from}~$m$~\texttt{at\phantom{\ }each\phantom{\ }level:}\\\hspace*{0.333em}\hspace*{0.333em}$\operatorname{prox}_{f_{p+1,l},\rho_{p+1}}(w) = (m_l+\rho_{p+1} w)/(1+\rho_{p+1})$

\texttt{//start\phantom{\ }at\phantom{\ }coarsest\phantom{\ }grid}\\\textbf{FOR}~$l=n_\text{levels},n_\text{levels}-1,\dots,1$\\\hspace*{0.333em}\hspace*{0.333em}\hspace*{0.333em}\hspace*{0.333em}\hspace*{0.333em}\hspace*{0.333em}\texttt{//solve\phantom{\ }on\phantom{\ }current\phantom{\ }grid}\\\hspace*{0.333em}\hspace*{0.333em}\hspace*{0.333em}\hspace*{0.333em}\hspace*{0.333em}\hspace*{0.333em}$(x_l,\{y_{i,l}\},\{v_{i,l}\})=\text{PARSDMM}(m_l,\{ A_{i,l} \},\{ \operatorname{prox}_{f_{i,l},\rho_i} \},x_l,\{y_{i,l}\},\{v_{i,l}\})$\\\hspace*{0.333em}\hspace*{0.333em}\hspace*{0.333em}\hspace*{0.333em}\hspace*{0.333em}\hspace*{0.333em}$x_l \rightarrow x_{l-1}$~\texttt{//interpolate\phantom{\ }to\phantom{\ }finer\phantom{\ }grid}\\\hspace*{0.333em}\hspace*{0.333em}\hspace*{0.333em}\hspace*{0.333em}\hspace*{0.333em}\hspace*{0.333em}\textbf{FOR}~$i=1, 2, \dots, p+1$\\\hspace*{0.333em}\hspace*{0.333em}\hspace*{0.333em}\hspace*{0.333em}\hspace*{0.333em}\hspace*{0.333em}\hspace*{0.333em}\hspace*{0.333em}\hspace*{0.333em}\hspace*{0.333em}$y_{i,l} \rightarrow y_{i,l-1}$~\texttt{//interpolate\phantom{\ }to\phantom{\ }finer\phantom{\ }grid}\\\hspace*{0.333em}\hspace*{0.333em}\hspace*{0.333em}\hspace*{0.333em}\hspace*{0.333em}\hspace*{0.333em}\hspace*{0.333em}\hspace*{0.333em}\hspace*{0.333em}\hspace*{0.333em}$v_{i,l} \rightarrow v_{i,l-1}$~\texttt{//interpolate\phantom{\ }to\phantom{\ }finer\phantom{\ }grid}\\\hspace*{0.333em}\hspace*{0.333em}\hspace*{0.333em}\hspace*{0.333em}\hspace*{0.333em}\hspace*{0.333em}\textbf{END}~\\\textbf{END}\\$\textbf{output:}$~$x$~at~original~grid~(level~$1$)
\caption{Multilevel PARSDMM to compute the projection onto an
intersection of sets.}\label{alg:ML-PARSDMM}
\end{scholmdAlgorithm}

\section{Software and numerical
examples}\label{software-and-numerical-examples}

The software corresponding to this paper is available at
https://github.com/slimgroup/SetIntersectionProjection.jl. All
algorithms, examples, and utilities to set up the projectors and linear
operators are included. Our software is specialized to the specific and
fixed problem structure~\eqref{proj_intersect_indicator} with the
flexibility to work with multiple linear operators and projectors.
Because of these design choices, the user only needs to provide the
model to project, $m$, and pairs of linear operators and projectors onto
simple sets:
$\{ (A_1, \mathcal{P}_{\mathcal{C}_1} ), ( A_2, \mathcal{P}_{\mathcal{C}_2} ), \dots, ( A_p, \mathcal{P}_{\mathcal{C}_p} ) \}$.
The software adds the identity matrix and the proximal map for the
distance squared from $m$. These are all computational components
required to solve intersection projection problems as formulated
in~\eqref{PARSDMM_2}. No internal reformulation is required by our
software.

To reap benefits from modern programming language design, including
just-in-time compilation, multiple dispatch, and mixing distributed and
multi-threaded computations, we wrote our software package in Julia. Our
code uses parametric typing, which means that the same scripts can run
in \texttt{Float32} (single) and \texttt{Float64} (double) precision. As
expected, most components of our software run faster in \texttt{Float32}
with reduced memory consumption. The timings in the following examples
use \texttt{Float32}.

We provide scripts that the set up the linear operators and projectors
for regular grids in 2D and 3D. It is not necessary to use these scripts
as the solver is agnostic to the specific construction of the projectors
or linear operators. Table~\eqref{set-overview} displays the constraints
we currently support. For example, when the user requests the script to
set up minimum and maximum bounds on the discrete gradient in the
$z$-direction of the model, the script returns the discrete derivative
matrix $A=D_z \otimes I_x$ and a function
$\mathcal{P}_\text{bounds}(\cdot)$ that projects the input onto the
bounds. The software currently supports the identity matrix, matrices
representing the discrete gradient and the operators that we apply
matrix-free: the discrete cosine/Fourier/wavelet/curvelet
\citep{ying20053d} transforms.

\begin{table}
\centering
\begin{tabular}{ll}
\toprule\addlinespace
descriptions & set\tabularnewline
\midrule
bounds & $\{ m \: | \: l[i] \leq (A m)[i] \leq b[i] \}$\tabularnewline
$\ell_1$ & $\{ m \: | \: \| A m \|_1 \leq \sigma \}$\tabularnewline
$\ell_2$ & $\{ m \: | \: \| A m \|_2 \leq \sigma \}$\tabularnewline
annulus &
$\{ m \: | \: \sigma_l \leq \| A m \|_2 \leq \sigma_u \}$\tabularnewline
nuclear norm & $\{ m \: | \: \sum_{j=1}^k \lambda[j] \leq \sigma \}$,
with $Am = \operatorname{vec}( \sum_{j=1}^{k}\lambda[j] u_j v_j^\top )$
is the SVD.\tabularnewline
cardinality & $\{ m \: | \: \operatorname{card}(Am) \leq k \}$, $k$ is a
positive integer\tabularnewline
rank & $\{ m \: | \operatorname{card}(\lambda) \leq r \}$, with
$Am = \operatorname{vec}( \sum_{j=1}^{k}\lambda[j] u_j v_j^\top) \}$ is
the SVD and $r < \text{min}(n_z,n_x)$\tabularnewline
subspace constraints &
$\{ m \: | \: m = A c, \:\: c \in \mathbb{R}^M \}$\tabularnewline
\bottomrule
\end{tabular}
\caption{Overview of constraint sets that the software currently
supports. A new constraint requires the projector onto the set (without
linear operator) and a linear operator or equivalent matrix-vector
product together with its adjoint. Vector entries are indexed as $m[i]$,
$u_j$ and $v_j^\top$ indicate singular vectors.}\label{set-overview}
\end{table}

For the special case of orthogonal linear operators, we leave the linear
operator inside the set definition because we know the projection onto
$\mathcal{V}$ in closed form. For example, if
$\mathcal{V} = \{ x \: | \: \| Ax \|_1 \leq \sigma \}$ with discrete
Fourier transform (DFT) matrix $A \in \mathbb{C}^{N \times N}$, the
projection is known in closed form as
$\mathcal{P}_\mathcal{V}(x) = A^* \mathcal{P}_{\| \cdot \| \leq \sigma}( A x)$,
where $^*$ denotes the complex-conjugate transpose and
$\mathcal{P}_{\| \cdot \| \leq \sigma}$ is the projection onto the
$\ell_1$-ball. We do this to keep all other computations in PARSDMM
(Algorithm~\ref{alg:PARSDMM}) real, because complex-valued vectors
require more storage and will slow down most computations.

Our software also allows to simultaneously use constraints that apply to
the 2D/3D model and constraints that apply to each column/row/fiber
separately. The linear operator remains the same if we define
constraints for all rows, columns, or both. The difference is that the
projection onto a simple set is now applied to each row/column
independently in parallel via a multi-threaded loop.

As an example of our code, we show how to project a 2D model, $m$, onto
the intersection of bound constraints and the set of models that have
monotonically increasing parameter values in the z-direction.

\begin{Shaded}
\begin{Highlighting}[]
\NormalTok{using SetIntersectionProjection}

\CommentTok{#set up stucture with grid information}
\NormalTok{mutable struct compgrid}
  \NormalTok{d :: }\DataTypeTok{Tuple}
  \NormalTok{n :: }\DataTypeTok{Tuple}
\KeywordTok{end}

\CommentTok{#load a model}
\CommentTok{#m = }

\CommentTok{#the following optional lines of }
\CommentTok{#code set up linear operators and projectors}

\CommentTok{#grid information ( (dz,dx),(nz,nx) )}
\NormalTok{comp_grid = compgrid( (}\FloatTok{25.0}\NormalTok{, }\FloatTok{6.0}\NormalTok{), (size(m,}\FloatTok{1}\NormalTok{), size(m,}\FloatTok{2}\NormalTok{)) ) }
\NormalTok{m = vec(m) }\CommentTok{#algorithms take vectorized input}

\CommentTok{#initialize constraint information}
\NormalTok{constraint = }\DataTypeTok{Vector}\NormalTok{\{SetIntersectionProjection.set_definitions\}() }

\CommentTok{#set up bound constraints}
\NormalTok{min          = }\FloatTok{1500.0}           \CommentTok{#minimum velocity}
\NormalTok{max          = }\FloatTok{4500.0}           \CommentTok{#maximum velocity}
\NormalTok{set_type     = }\StringTok{"bounds"}         \CommentTok{#bound constraint set}
\NormalTok{TD_OP        = }\StringTok{"identity"}       \CommentTok{#identity matrix in the set definition}
\NormalTok{app_mode     = (}\StringTok{"matrix"}\NormalTok{,}\StringTok{""}\NormalTok{)    }\CommentTok{#bounds applied to the model as a matrix}
\NormalTok{custom_OP    = ([],false)       }\CommentTok{#no custom linear operators}

\NormalTok{push!(constraint,set_definitions(set_type,TD_OP,min,max,app_mode,custom_OP))}

\CommentTok{#bounds on parameters in a transform-domain (vertical slope constraint)}
\NormalTok{min          = }\FloatTok{0.0}              
\NormalTok{max          = }\FloatTok{1e6}          
\NormalTok{set_type     = }\StringTok{"bounds"}
\NormalTok{TD_OP        = }\StringTok{"D_z"}            \CommentTok{#discrete derivative in z-direction}
\NormalTok{app_mode     = (}\StringTok{"matrix"}\NormalTok{,}\StringTok{""}\NormalTok{)}
\NormalTok{custom_TD_OP = ([],false)}

\NormalTok{push!(constraint,set_definitions(set_type,TD_OP,min,max,app_mode,custom_OP))}

\NormalTok{options = PARSDMM_options() }\CommentTok{#get default options}
\NormalTok{options.FL = typeof(m[}\FloatTok{1}\NormalTok{,}\FloatTok{1}\NormalTok{]) }\CommentTok{# get precision}

\CommentTok{#get projectors onto simple sets, linear operators, set information}
\NormalTok{(P_sub,TD_OP,set_Prop) = setup_constraints(constraint,comp_grid,options.FL) }

\CommentTok{#precompute and distribute quantities once, reuse later}
\NormalTok{(TD_OP,B,~,~) = PARSDMM_precompute_distribute(TD_OP,set_Prop,comp_grid,options)}

\CommentTok{#project onto intersection}
\NormalTok{(x,log_PARSDMM) = PARSDMM(m,B,TD_OP,set_Prop,P_sub,comp_grid,options)}
\end{Highlighting}
\end{Shaded}

\subsection{Parallel Dykstra versus
PARSDMM}\label{parallel-dykstra-versus-parsdmm}

To see if our new algorithm is faster than black-box type projection
algorithms, such as parallel Dykstra's algorithm (see Appendix A), we
use Adaptive Relaxed ADMM (ARADMM) \citep{Xu_2017_CVPR} for the
projection sub-problems of parallel Dykstra's algorithm. Both PARSDMM
(Algorithm~\ref{alg:PARSDMM}) and Parallel Dykstra-ARADMM have the same
computational components. ARADMM also uses the same update scheme for
the augmented Lagrangian penalty and relaxation parameters as we use in
PARSDMM. This similarity allows for a comparison of the convergence as a
function of the basic computational components. We manually tuned ARADMM
stopping conditions to achieve the best performance for parallel
Dykstra's algorithm overall.

The first numerical experiment is the projection of a 2D geological
model ($341 \times 400$ pixels) onto the intersection of three
constraint sets that are of interest to seismic imaging
\citep{Esser2016arch, ournewpreprint, TVWRI2}:

\begin{enumerate}
\def\labelenumi{\arabic{enumi}.}
\itemsep1pt\parskip0pt\parsep0pt
\item
  $\{ m \: | \: \sigma_1 \leq m[i] \leq \sigma_2 \}$ : bound constraints
\item
  $\{ m \: | \: \| A m \|_1 \leq \sigma \}$ with
  $A = \begin{pmatrix} D_z \otimes I_x \\ I_z \otimes D_x \end{pmatrix}$
  : anisotropic total-variation constraints
\item
  $\{ m \: | \: 0 \leq ((D_z \otimes I_x) m)[i] \leq \infty \}$ :
  vertical monotonicity constraints
\end{enumerate}

For these sets, the primary computational components are

\begin{itemize}
\item
  matrix-vector products in the conjugate-gradient algorithm. Parallel
  Dykstra's algorithms uses matrix-vector products with $A^\top A$,
  $(D_z \otimes I_x)^\top (D_z \otimes I_x)$, and $I$ in parallel.
  PARSDMM computes matrix-vector producs with the sparsity pattern of
  $A^\top A$ (this pattern overlaps with the the linear operators in the
  other two sets).
\item
  projections onto the box constraint set and the $\ell_1$-ball. Both
  parallel Dykstra's algorithm and PARSDMM compute these in parallel.
\item
  parallel communication that sends a vector from one to all parallel
  processes ($x^{k+1}$ in Algorithm~\ref{alg:PARSDMM}), and one
  map-reduce parallel sum that gathers the sum of vectors on all workers
  (the right-hand side for the $x^{k+1}$ computation in
  Algorithm~\ref{alg:PARSDMM}). The communication is the same for
  PARSDMM and parallel Dykstra's algorithm so we ignore it in the
  experiments below.
\end{itemize}

Before we show the numerical results, we discuss how we count the
computational operations mentioned above.

\begin{itemize}
\item
  Matrix-vector products in CG: Parallel Dykstra's algorithm
  simultaneously computes three projections by running three instances
  of ARADMM in parallel. For each parallel Dykstra iteration, we add the
  maximum number of CG iterations, corresponding to one of the
  sub-problems.
\item
  $\ell_1$-ball projections: PARSDMM projects onto the $\ell_1$ ball
  once per iteration. Parallel Dykstra projects (number of parallel
  Dykstra iterations) $\times$ (number of ARADMM iterations for set
  number two) times onto the $\ell_1$ ball. We focus on the
  $\ell_1$-ball projections \citep{duchi:icml08}, because these
  projections are computationally more intensive compared to projections
  onto the box (element-wise comparison) and also less suitable for
  multi-threaded parallelization.
\end{itemize}

\begin{figure*}
\centering
\captionsetup[subfigure]{labelformat=empty}
\subfloat[]{\includegraphics[width=0.330\hsize]{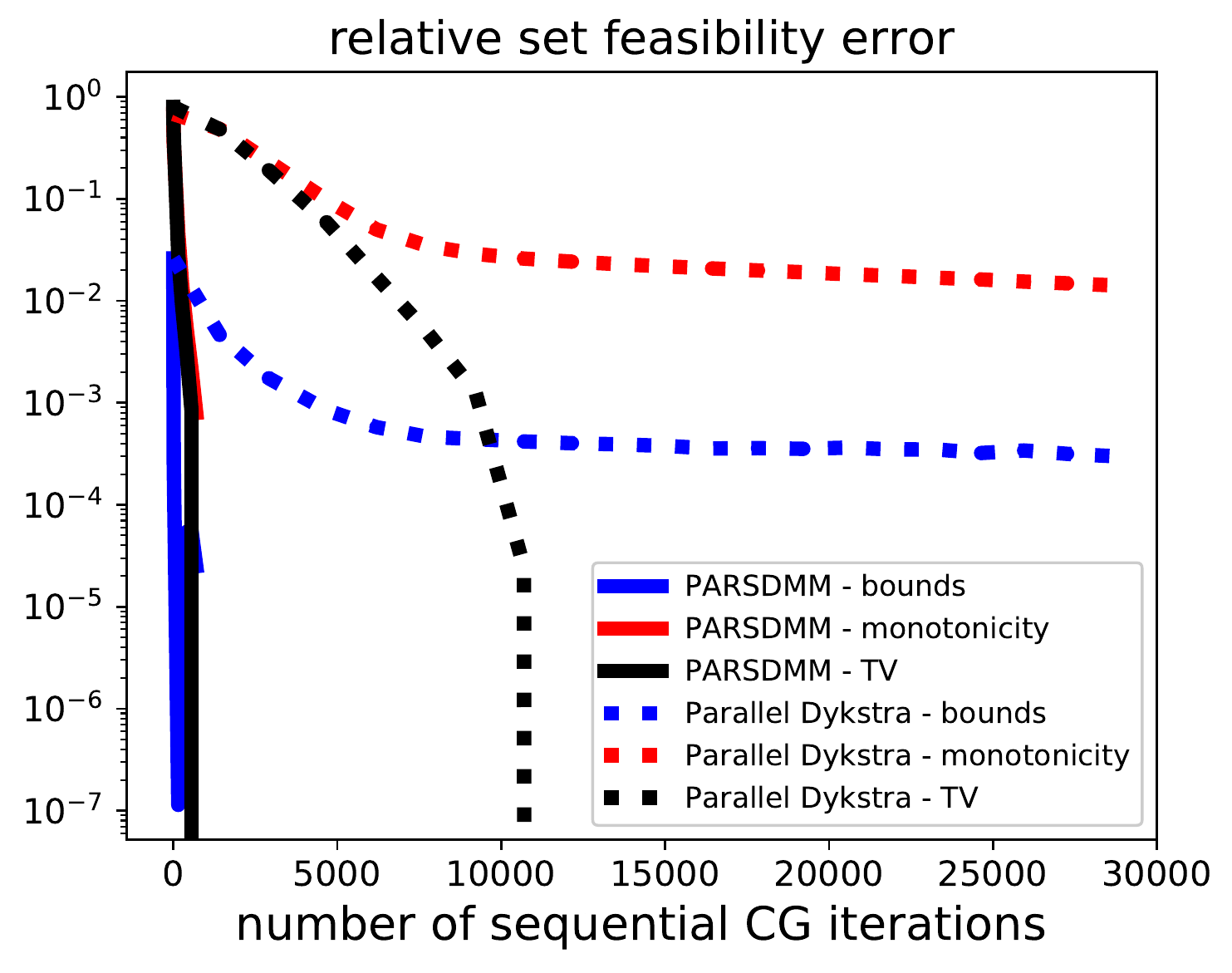}}
\subfloat[]{\includegraphics[width=0.330\hsize]{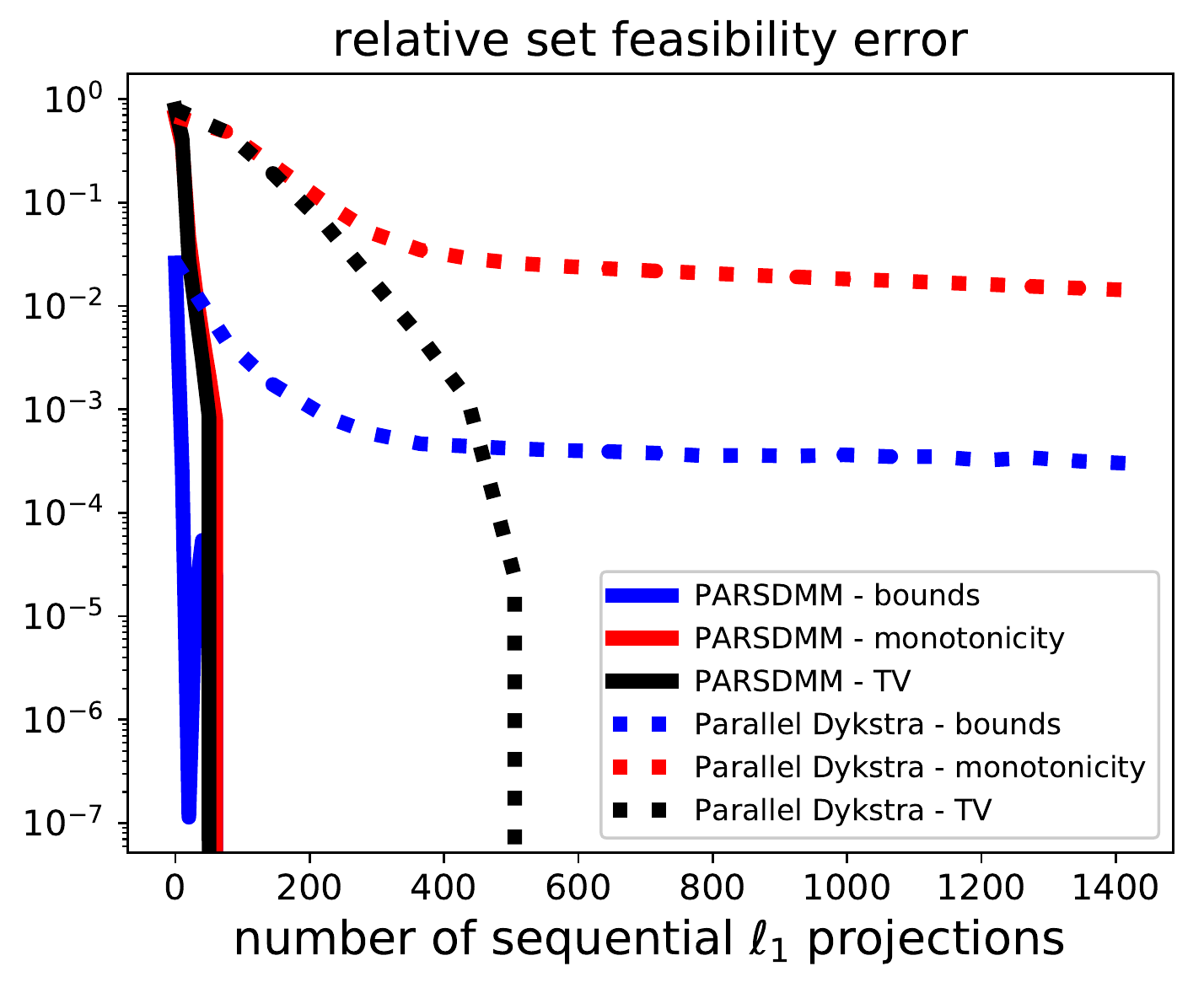}}
\subfloat[]{\includegraphics[width=0.330\hsize]{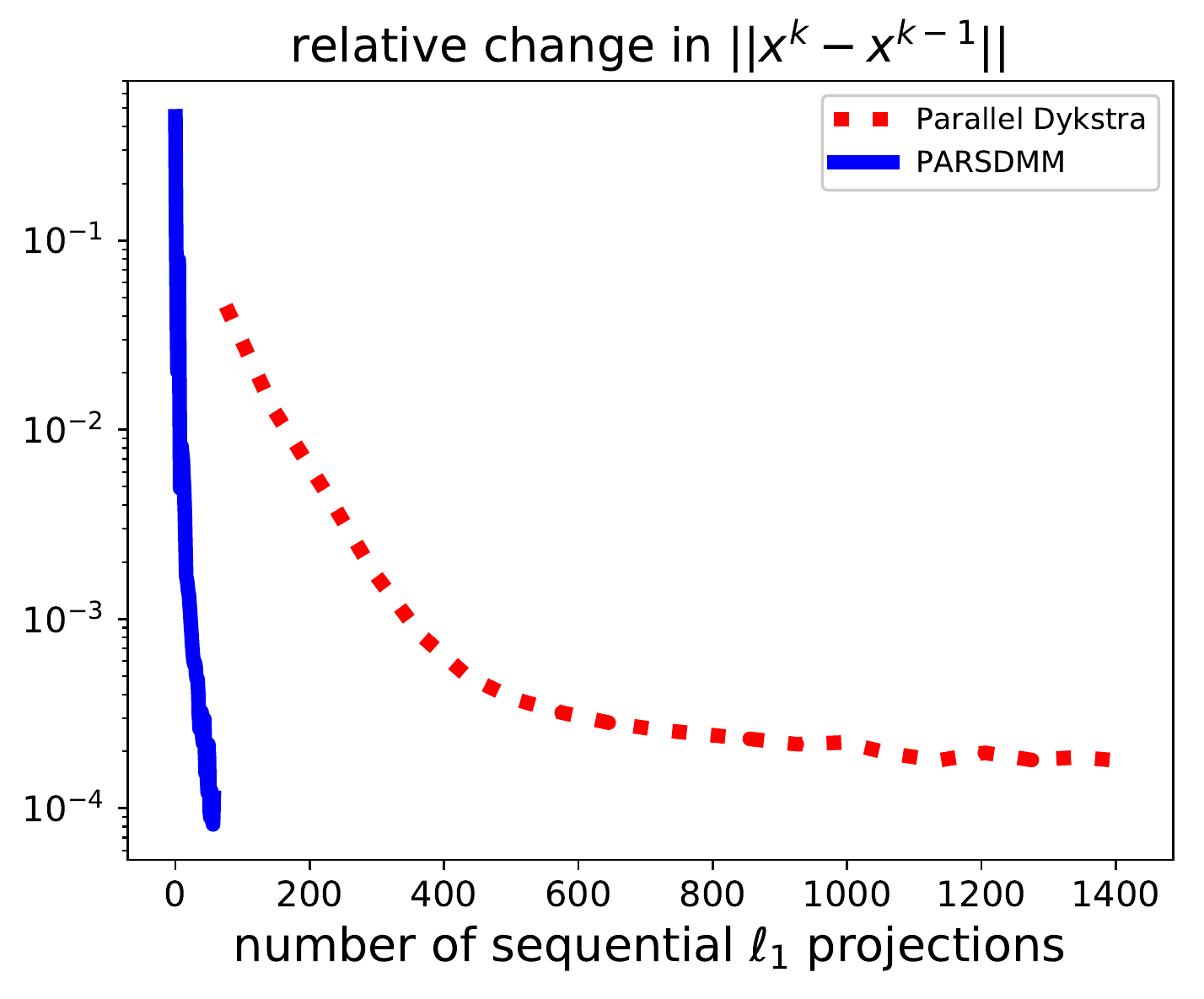}}
\caption{Relative transform-domain set feasibility
(equation~\ref{feas_stop}) as a function of the number of
conjugate-gradient iterations and projections onto the 1-norm ball. This
figure also shows relative change per iteration in the
solution.}\label{Fig:Dyk-vs-PARSDMM}
\end{figure*}

The results in Figure~\ref{Fig:Dyk-vs-PARSDMM} show that PARSDMM
requires much fewer CG iterations and $\ell_1$-ball projections to
achieve the same relative set feasibility error in the transform-domain
as defined in equation~\eqref{feas_stop}. We observe a somewhat
oscillatory convergence of PARSDMM, which is caused by changing the
relaxation and augmented-Lagrangian penalty parameters.

Because non-convex sets are an important application for us, we compare
the performance for a non-convex intersection as well:

\begin{enumerate}
\def\labelenumi{\arabic{enumi}.}
\itemsep1pt\parskip0pt\parsep0pt
\item
  $\{ m \: | \: \sigma_1 \leq m[i] \leq \sigma_2 \}$: bound constraints
\item
  $\{ m \: | \: (D_z \otimes I_x) m = \operatorname{vec}( \sum_{j=1}^{r}\lambda[j] u_j v_j^\top )\}$,
  where $r < \text{min}(n_z,n_x)$, $\lambda[j]$ are the singular values,
  and $u_j$, $v_j$ are singular vectors: rank constraints on the
  vertical gradient of the image
\end{enumerate}

\begin{figure*}
\centering
\captionsetup[subfigure]{labelformat=empty}
\subfloat[]{\includegraphics[width=0.330\hsize]{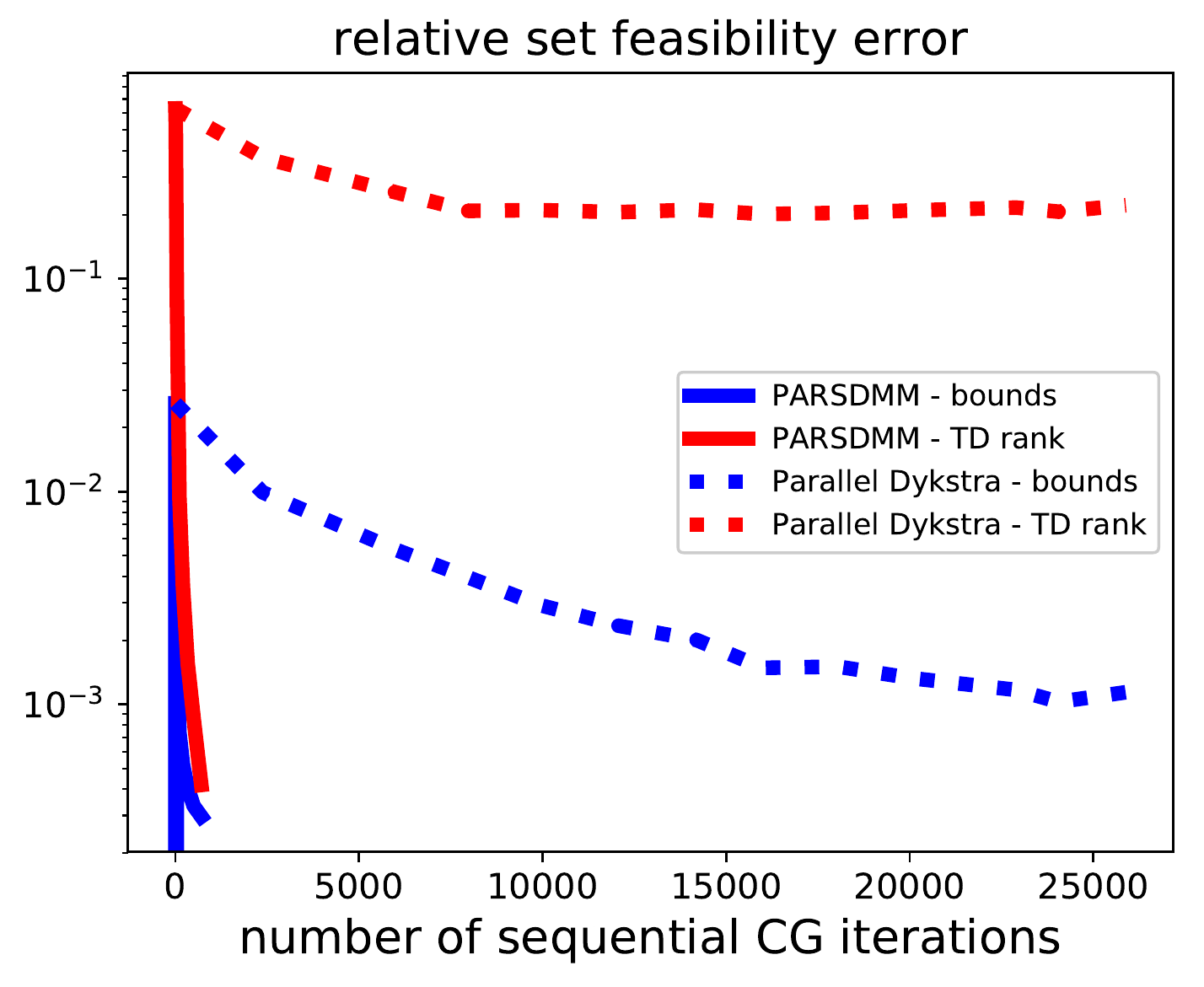}}
\subfloat[]{\includegraphics[width=0.330\hsize]{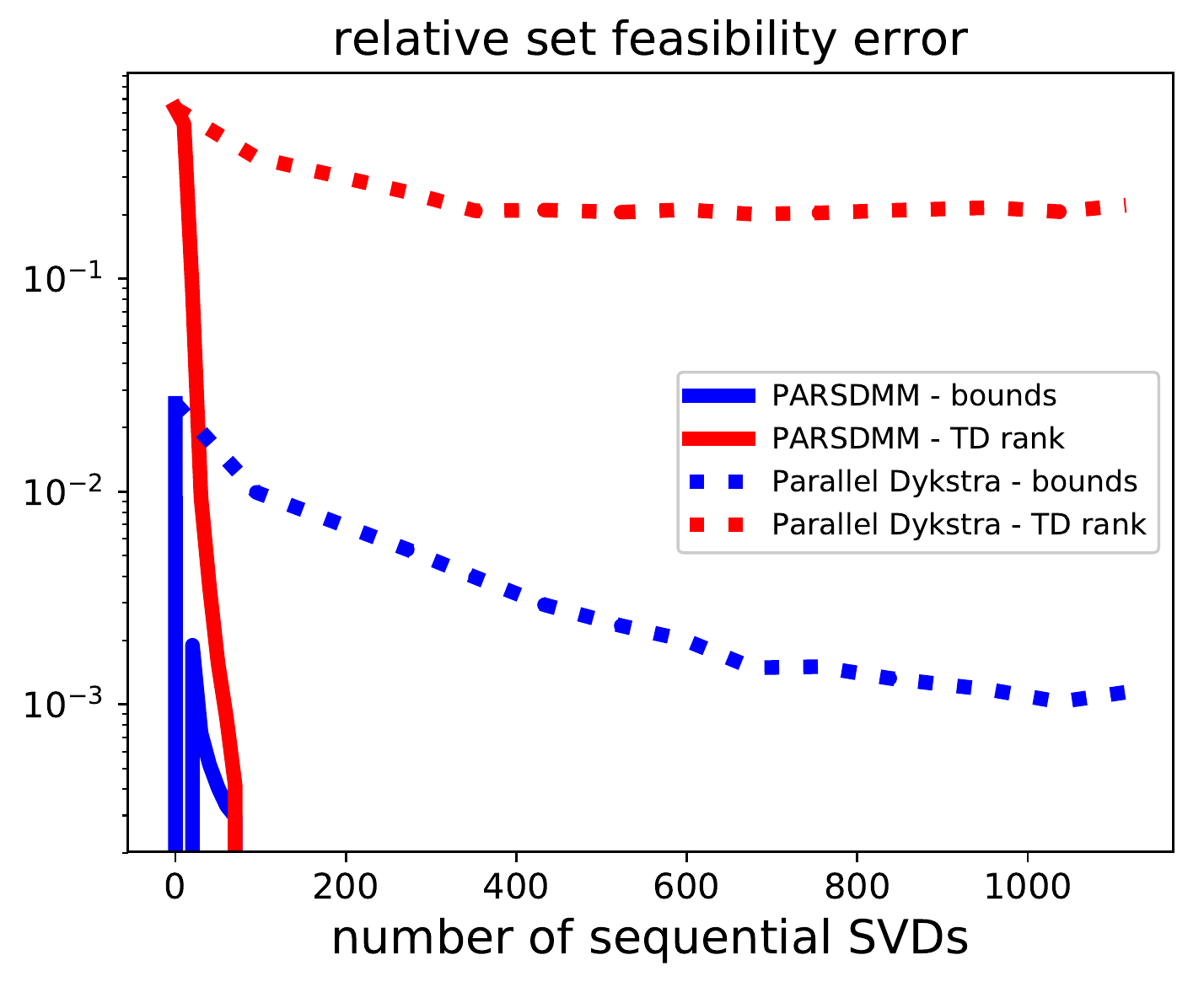}}
\subfloat[]{\includegraphics[width=0.330\hsize]{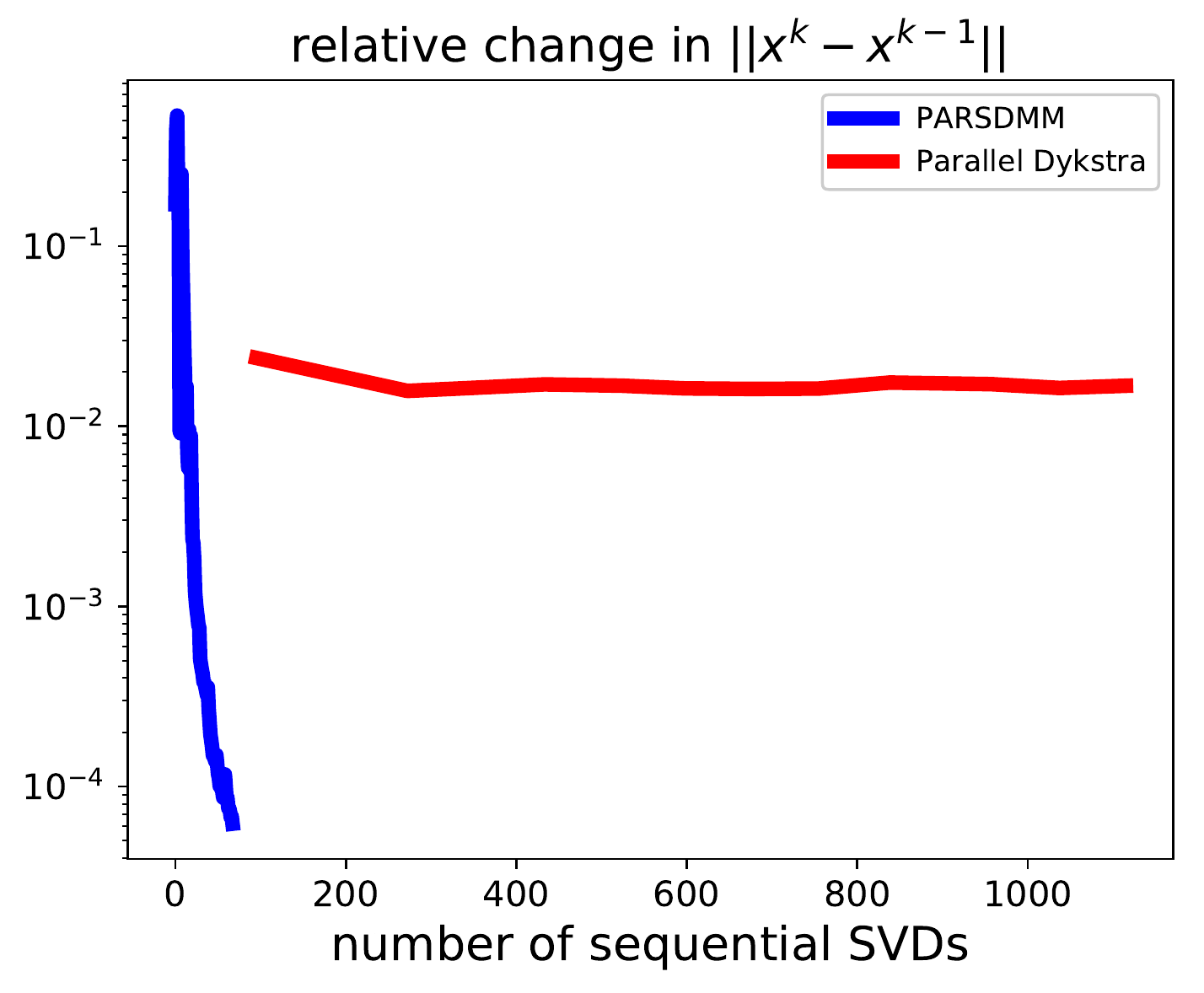}}
\caption{Relative transform-domain set feasibility
(equation~\ref{feas_stop}) as a function of the number of
conjugate-gradient iterations and projections onto the set of matrices
with limited rank via the SVD. This figure also shows relative change
per iteration in the solution.}\label{Fig:Dyk-vs-PARSDMM2}
\end{figure*}

We count the computational operations in the same way as in the previous
example, but this time the computationally most costly projection is
onto the set of matrices with limited rank via the singular value
decomposition. The results in Figure~\ref{Fig:Dyk-vs-PARSDMM2} show that
the convergence of parallel Dykstra's algorithm almost stalls: the
solution estimate gets closer to satisfying the bound constraints, but
there is hardly any progress towards the rank constraint set. PARSDMM
does not seem to struggle with the non-convex set in this particular
example.

We used the single-level version of PARSDMM such that we can compare the
computational cost with Parallel Dykstra. The PARSDMM results in this
section are therefore pessimistic in general, as the multilevel version
can offer additional speedups, which we show next.

\subsection{Timings for 2D and 3D
projections}\label{timings-for-2d-and-3d-projections}

To get an idea about solution times versus model size, as well as how
beneficial the parallelism and multilevel continuation are, we show
timings for projections of geological models onto an intersection for
the four modes of operation: PARSDMM, parallel PARSDMM, multilevel
PARSDMM, and multilevel parallel PARSDMM. As we mentioned, the
multilevel version has a small additional overhead compared to
single-level PARSDMM because of one interpolation procedure per level.
Parallel PARSDMM has communication overhead. Note that serial PARSDMM
still uses multi-threading for projections, the matrix-vector product in
the conjugate-gradient method, and BLAS operations, but the $y_i$ and
$v_i$ computations in Algorithm~\ref{alg:PARSDMM} remain sequential for
every $i=1,2,\cdots,p, p+1$, contrary to parallel PARSDMM. We carry our
computations out on a dedicated cluster node with $2$ CPUs per node with
10 cores per CPU (Intel Ivy Bridge 2.8 GHz E5-2680v2) and 128 GB of
memory per node.

The following sets are used in \citet{ournewpreprint} to regularize a
geophysical inverse problem and form the intersection for our test case:

\begin{enumerate}
\def\labelenumi{\arabic{enumi}.}
\itemsep1pt\parskip0pt\parsep0pt
\item
  $\{ m \: | \: \sigma_1 \leq m[i] \leq \sigma_2 \}$ : bound constraints
\item
  $\{ m \: | \: -\sigma_3 \leq ((I_z \otimes D_x) m)[i] \leq \sigma_3 \}$:
  lateral smoothness constraints. There are two of these constraints in
  the 3D case: for the $x$ and $y$ direction separately.
\item
  $\{ m \: | \: 0 \leq ((D_z \otimes I_x) m)[i] \leq \infty \}$ :
  vertical monotonicity constraints
\end{enumerate}

\begin{figure*}
\centering
\captionsetup[subfigure]{labelformat=empty}
\subfloat[]{\includegraphics[width=0.500\hsize]{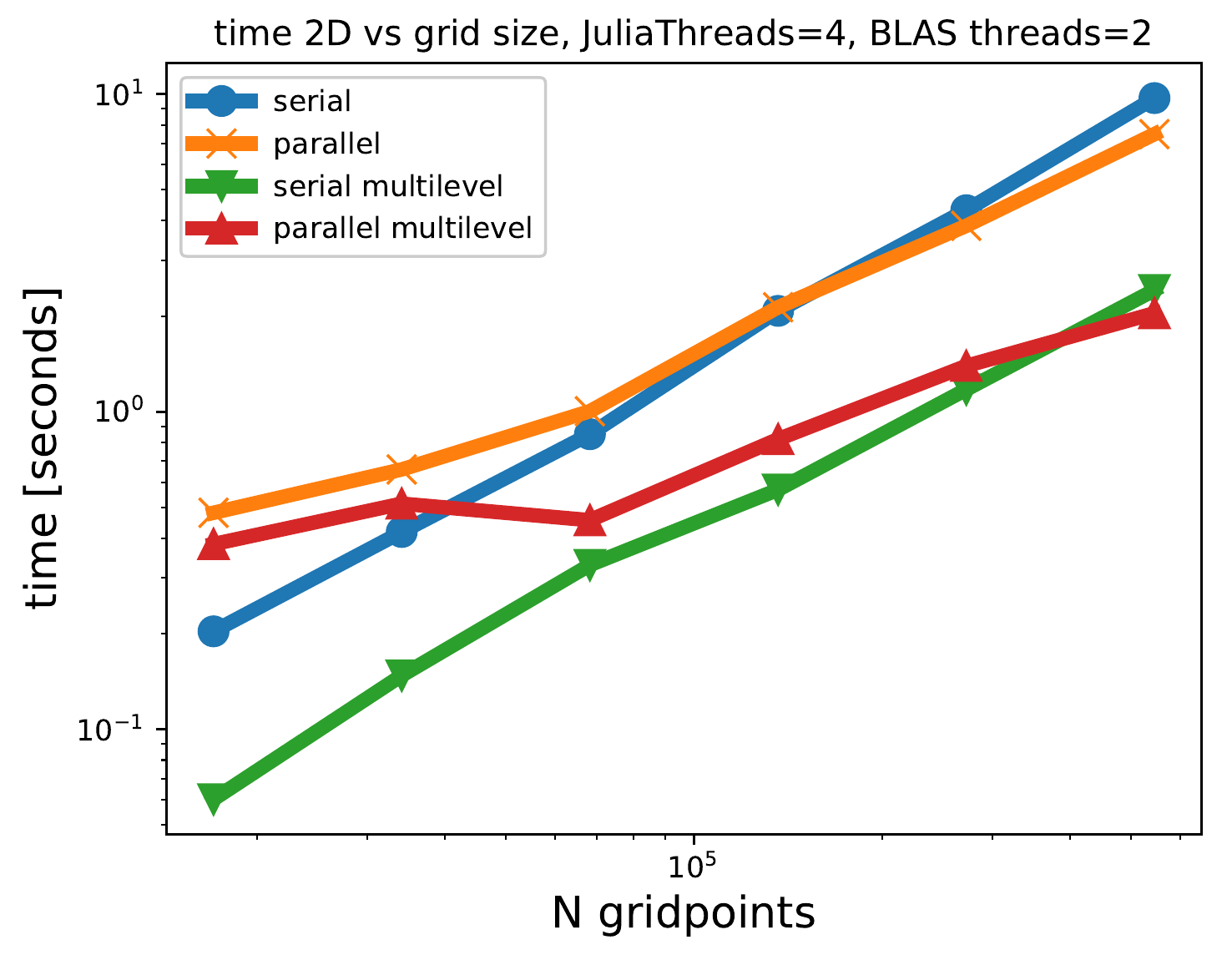}}
\subfloat[]{\includegraphics[width=0.500\hsize]{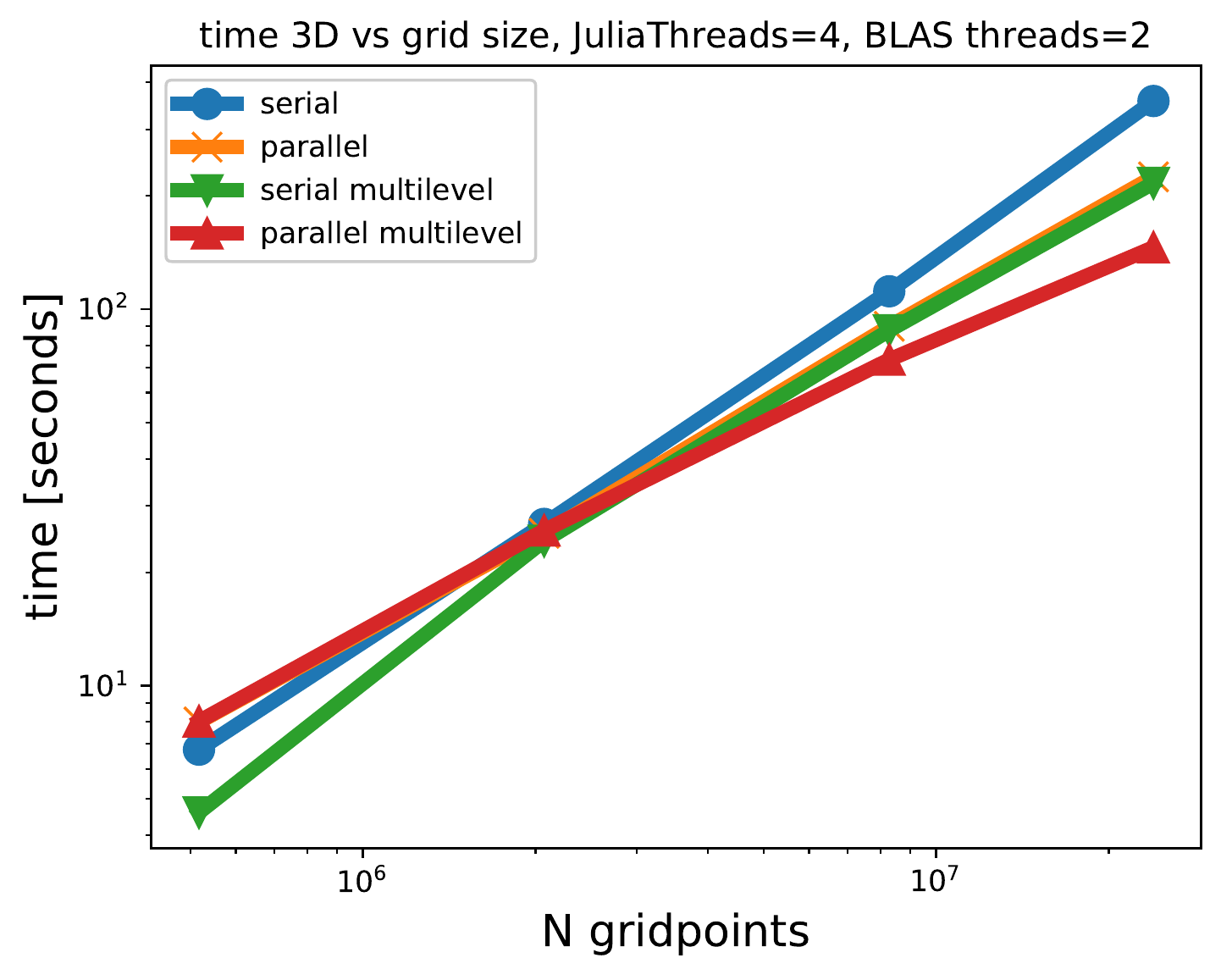}}
\caption{Timings for a 2D and 3D example where we project a geological
model onto the intersection of bounds, lateral smoothness, and vertical
monotonicity constraints.}\label{Fig:timings-1}
\end{figure*}

The results in Figure~\ref{Fig:timings-1} show that the multilevel
strategy is much faster than the single-level version of PARSDMM. The
multilevel overhead costs are thus small compared to the speedup. It
also shows that, as expected, the parallel versions require some
communication time, so the problems need to be large enough for the
parallel version of PARSDMM to offer speedups.

\subsection{Geophysical parameter estimation with
constraints}\label{geophysical-parameter-estimation-with-constraints}

Seismic full-waveform inversion (FWI) estimates rock properties
(acoustic velocity in this example) from seismic (pressure) signals
measured by hydrophones. FWI is a partial-differential-equation (PDE)
constrained optimization problem where after eliminating the PDE
constraint, the simulated data, $d_\text{predicted}(m)$, are connected
nonlinearly to the unknown model parameters, $m \in \mathbb{R}^N$. We
assume that we know the source and receiver locations, as well as the
source function. A classic example of an objective for FWI is the
nonlinear least-squares misfit
$f(m)=1/2 \| d_\text{obs} - d_\text{predicted}(m) \|_2^2$, which we use
for this numerical experiment.

FWI is a problem hampered by local minima. Empirical evidence
\citep{Esser2016arch, doi:10.1190/tle36010094.1, TVWRI2} suggests that
we can mitigate issues with parasitic local minima by insisting that all
model iterates be elements of the intersection of multiple constraint
sets. This means that we add regularization to the objective
$f(m) : \mathbb{R}^N \rightarrow \mathbb{R}$ in the form of multiple
constraints---i.e., we have
\begin{equation}
\min_{m} f(m) \quad \text{s.t.} \quad m \in \mathcal{V} = \bigcap_{i=1}^p \mathcal{V}_i.
\label{FWI_prob}
\end{equation}
 We use the spectral projected gradient (SPG) algorithm with a
non-monotone line search \citep{Birgin:1999:NSP:588891.589081} to solve
the above problem. SPG uses information from the current and previous
gradient of $f(m)$ to approximate the action of the Hessian of $f(m^k)$
with the scalar $\alpha$: the Barzilai-Borwein step length. At iteration
$k$, SPG updates the model parameters as follows:
\begin{equation}
m^{k+1} = (1-\gamma) m^k - \gamma \mathcal{P}_{\mathcal{V}} (m^k - \alpha \nabla_{m}f(m^k)),
\label{SPG_iter}
\end{equation}
 where the non-monotone line search determines $\gamma \in (0,1]$ and
backtracks between two points in a convex set if all $\mathcal{V}_i$ are
convex. We compute the projection onto the intersection using PARSDMM
(Algorithm~\ref{alg:PARSDMM}). The total number of SPG iterations plus
line-search steps is limited to the relatively small number of ten,
because these require the solution of multiple PDEs, which is
computationally intensive, especially in 3D.

\begin{figure*}
\centering
\captionsetup[subfigure]{labelformat=empty}
\subfloat[]{\includegraphics[width=0.500\hsize]{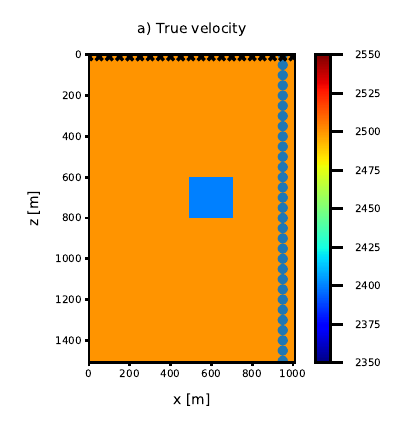}}
\subfloat[]{\includegraphics[width=0.500\hsize]{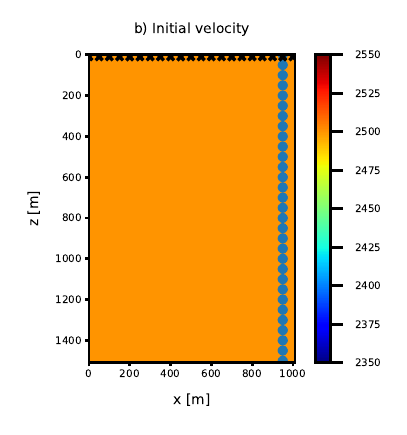}}
\\
\subfloat[]{\includegraphics[width=0.500\hsize]{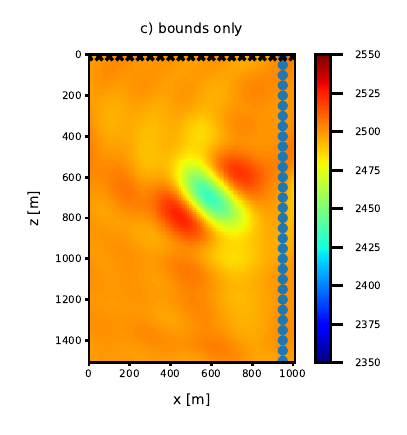}}
\subfloat[]{\includegraphics[width=0.500\hsize]{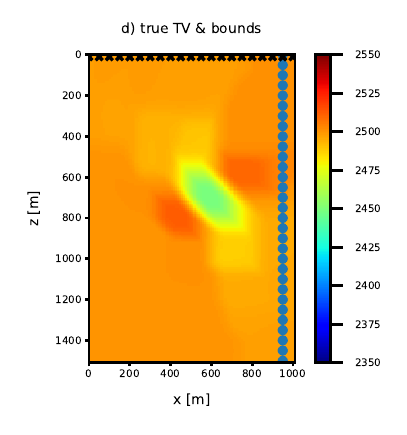}}
\caption{True, initial, and estimated models with convex constraints for
the full-waveform inversion example. Crosses and circles represent
sources and receivers, respectively. All projections inside the spectral
projected gradient algorithm are computed using PARSDMM.}\label{Fig:FWI}
\end{figure*}

The experimental setting is as follows: The Helmholtz equation models
the wave propagation in an acoustic model. The data acquisition system
is a vertical-seismic-profiling experiment with sources at the surface
and receivers in a well, see Figure~\ref{Fig:FWI}. All boundaries are
perfectly-matched-layers (PML) that absorb outgoing waves as if the
model is spatially unbounded. The challenges that we address by
constraining the model parameters are: one-sided `source illumination'
that often leads to spurious artifacts in the source-receiver direction,
a limited frequency range ($3-10$ Hertz), and the non-convexity of the
data-misfit $f(m)$. We use the software by \citet{waveformCurt} to
simulate seismic data and compute $f(m)$ and $\nabla_m f(m)$.

The prior knowledge consists of: \emph{(a)} minimum and maximum
velocities ($2350 - 2650$ m/s); \emph{(b)} The anomaly is rectangular,
but we do not know the size, aspect ratio, or location.

We start simple and look at what happens if we add bounds and
total-variation constraints. Figure~\ref{Fig:FWI} shows the true model,
initial guess, and the estimated models using convex constraints. The
data acquisition geometry causes the model estimate with bound
constraints to be an elongated diagonal anomaly that is incorrect in
terms of size, shape, orientation, and parameter values.
Figure~\ref{Fig:FWI}(d) shows that even in the unusual case that we know
and use an anisotropic total-variation (TV) constraint equal to the TV
of the true model, we obtain a model estimate where the shape of the
anomaly is still far from the truth, although many of the spurious
oscillations are damped.

As we will demonstrate, the inclusion of multiple non-convex cardinality
and rank constraints help the parameter estimation in this example. From
the prior information that the anomaly is rectangular and aligned with
the domain boundaries, we deduce that the rank of the model is equal to
two. We also know that the cardinality of the discrete gradient of each
row and each column is less than or equal to two as well. If we assume
that the anomaly is not larger than half the total domain extent in each
direction, we know that the cardinality of the discrete derivative of
the model (in matrix format) is not larger than the number of grid
points in each direction. To summarize, the following constraint sets
follow from the prior information:

\begin{enumerate}
\def\labelenumi{\arabic{enumi}.}
\itemsep1pt\parskip0pt\parsep0pt
\item
  $\{ x \: | \: \operatorname{card}( (D_z \otimes I_x) x ) \leq n_x \}$
\item
  $\{ x \: | \: \operatorname{card}( (I_z \otimes D_x) x ) \leq n_z \}$
\item
  $\{ x \: | \: \operatorname{rank}(x) \leq 3 \}$
\item
  $\{ x \: | \: 2350 \leq x[i] \leq 2650 \: \forall i\}$
\item
  $\{ x \: | \: \operatorname{card}( D_x X[i,:] ) \leq 2 \:\: \text{for} \:\: i \in \{1,2,\dots,n_z\} \}$,
  $X[i,:]$ is a row of the 2D model
\item
  $\{ x \: | \: \operatorname{card}( D_z X[:,j] ) \leq 2 \:\: \text{for} \:\: j \in \{1,2,\dots,n_x\}\}$,
  $X[:,j]$ is a column of the 2D model
\end{enumerate}

\begin{figure*}
\centering
\captionsetup[subfigure]{labelformat=empty}
\subfloat[]{\includegraphics[width=0.330\hsize]{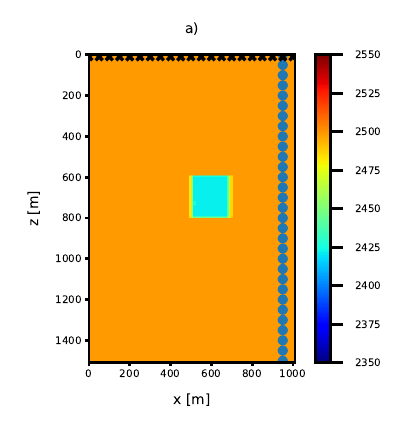}}
\subfloat[]{\includegraphics[width=0.330\hsize]{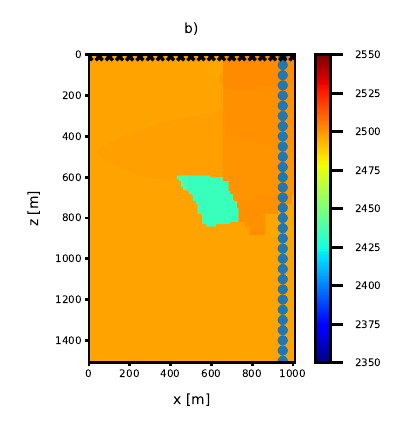}}
\subfloat[]{\includegraphics[width=0.330\hsize]{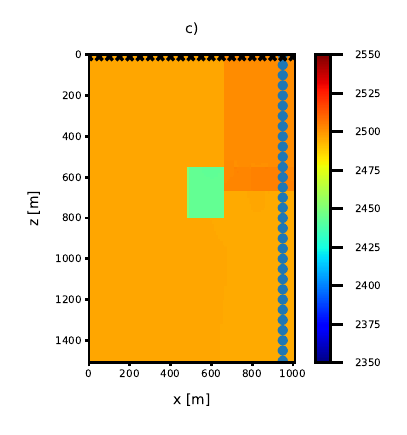}}
\caption{FWI results using various combinations of convex and non-convex
constraints using PARSDMM.}\label{Fig:FWI-1}
\end{figure*}

We use slightly overestimated rank and matrix cardinality constraints
compared to the true model to mimic the more realistic situation that we
typically have a priori access to over-estimated model properties. The
results in Figure~\ref{Fig:FWI-1} use PARSDMM to compute projections
onto the intersection of constraints, and show that non-convex
constraints can lead to improved model estimates.
Figure~\ref{Fig:FWI-1}(a) is the result of working with constraints
$[1,2,3,4]$, Figure~\ref{Fig:FWI-1}(b) uses constraints $[1,2,4,5,6]$,
and Figure~\ref{Fig:FWI-1}(c) uses all constraints $[1,2,3,4,5,6]$. The
result with rank constraints and both matrix and row/column-based
cardinality constraints on the discrete gradient of the model is the
most accurate in terms of the recovered anomaly shape. All results in
Figure~\ref{Fig:FWI} that work with non-convex sets are at least as
accurate as the result obtained with the true TV in terms of anomaly
shape. Another important observation is that all non-convex results
estimate a lower-than-background velocity anomaly, although not as low
as the true anomaly. Contrary, the models obtained using convex sets
show incorrect higher-than-background velocity artifacts in the vicinity
of the true anomaly location.

\subsection{Learning a parametrized intersection from a few training
examples}\label{learning-a-parametrized-intersection-from-a-few-training-examples}

In the introduction, we discussed projection or feasibility
problem~\eqref{proj_intersect_lininvprob} formulations of inverse
problems. The PARSDMM algorithm~\eqref{PARSDMM_1} is a good candidate to
solve these types of problems because we mitigate rapidly increasing
computation times for problems with many sets, by taking the similarity
between linear operators in set definitions into account. Of course, we
can only use multiple constraint sets if we have multiple pieces of
prior information. \citet{TVLearn} present a simple solution and note
that for $15$ out of $20$ investigated data-sets, $99\%$ of the images
have a total-variation within $20\%$ of the average total variation of
the data-set. The average total-variation serves as a constraint set
that typically leads to good results. Here we follow the same reasoning,
but we will work with many constraint sets that we learn from a few
example images. To summarize, our strategy is as follows:

\begin{enumerate}
\def\labelenumi{\arabic{enumi}.}
\itemsep1pt\parskip0pt\parsep0pt
\item
  Observe the constraint parameters in various transform-domains for all
  training examples (independently in parallel for each example and each
  constraint).
\item
  Add a data-fit constraint.
\item
  The solution of the inverse problem is the projection of an initial
  guess $m$ onto the intersection of sets that describe model properties
  and data-fit
  \begin{equation}
  \min_{x,\{y_i\}} \frac{1}{2}\| x - m \|_2^2 + \sum_{i=1}^{p-1}  \iota_{\mathcal{C}_i}(y_i) + \iota_{\mathcal{C}_p^\text{data}}(y_p)\quad \text{s.t.}    \quad \begin{cases}
  A_i x = y_i \\ Fx=y_p
  \end{cases},
  \label{proj_intersect_lininvprob2}
  \end{equation}
   where $F$ is a linear forward modeling operator.
\end{enumerate}

Before we proceed to the examples, it is worth mentioning the main
advantages of this strategy. Because all set definitions are independent
of all other sets, there are no penalty/weight parameters, and we avoid
hand-tuning the constraint definitions. We can observe `good'
constraints from just one or a few example images. Methods that do not
require training, such as basis-pursuit type formulations
\citep[e.g.,][]{MRM:MRM21391, Candes2009, doi:10.1137/080714488, Becker2011, doi:10.1137/130919210},
often minimize the $\ell_1$ norm or nuclear norm of transform-domain
coefficients (e.g., Fourier, wavelet) of an image subject to a data-fit
constraint. However, without learning, these methods require hand
picking a suitable transform for each class of images. We will work with
many transform-domain operators simultaneously, so that at least some of
the constraint/linear operator combinations will describe uncorrupted
images with small norms/bounds/cardinality, but not noisy/blurred/masked
images. Note that we are not learning any dictionaries, but work with
pre-defined transforms such as the Fourier basis, wavelets, and linear
operators based on discrete gradients.

For both of the next two examples we observe the following constraint
parameters from exemplar images:

\begin{enumerate}
\def\labelenumi{\arabic{enumi}.}
\itemsep1pt\parskip0pt\parsep0pt
\item
  $\{ m \: | \: \sigma_1 \leq m[i] \leq \sigma_2 \}$ (upper and lower
  bounds)
\item
  $\{ m \: | \: \sum_{j=1}^k \lambda[j] \leq \sigma_3 \}$ with
  $m = \operatorname{vec}( \sum_{j=1}^{k}\lambda[j] u_j v_j^\top )$ is
  the SVD of the image (nuclear norm)
\item
  $\{ m \: | \: \sum_{j=1}^k \lambda[j] \leq \sigma_4 \}$, with
  $(D_z \otimes I_x)m = \operatorname{vec}( \sum_{j=1}^{k}\lambda[j] u_j v_j^* )$
  is the SVD of the vertical derivative of the image (nuclear norm of
  discrete gradients of the image, total-nuclear-variation). Use similar
  constraint for the x-direction.
\item
  $\{ m \: | \: \| A m \|_1 \leq \sigma_5 \}$ with
  $A = \begin{pmatrix} D_z \otimes I_x \\ I_z \otimes D_x \end{pmatrix}$
  (anisotropic total-variation)
\item
  $\{ m \: | \: \sigma_6 \leq \| m \|_2 \leq \sigma_7 \}$ (annulus)
\item
  $\{ m \: | \: \sigma_8 \leq \| A m \|_2 \leq \sigma_9 \}$ with
  $A = \begin{pmatrix} D_z \otimes I_x \\ I_z \otimes D_x \end{pmatrix}$
  (annulus of the discrete gradients of the training images)
\item
  $\{ m \: | \: \| A m \|_1 \leq \sigma_{10} \}$ with $A = $ discrete
  Fourier transform ($\ell_1$-norm of DFT coefficients)
\item
  $\{ m \: | \: - \sigma_{11} \leq ((D_z \otimes I_x) m)[i] \leq \sigma_{12} \}$
  (slope-constraints in the $z$-direction, bounds on the discrete
  gradients of the image). Use similar constraint for the $x$-direction.
\item
  $\{ m \: | \: \| A m \|_1 \leq \sigma_{11} \}$ with $A = $ discrete
  wavelet transform
\end{enumerate}

These are nine types of convex and non-convex constraints on the model
properties ($11$ sets passed to PARSDMM because sets three and eight are
applied to the two directions separately). For data-fitting, we add a
point-wise constraint,
$\{ x \: | \: l \leq (F x - d_\text{obs}) \leq u \}$ with a linear
forward model $F \in \mathbb{R}^{M \times N}$.

\subsubsection{Joint
deblurring-denoising-inpainting}\label{joint-deblurring-denoising-inpainting}

\begin{figure*}
\centering
\includegraphics[width=1.000\hsize]{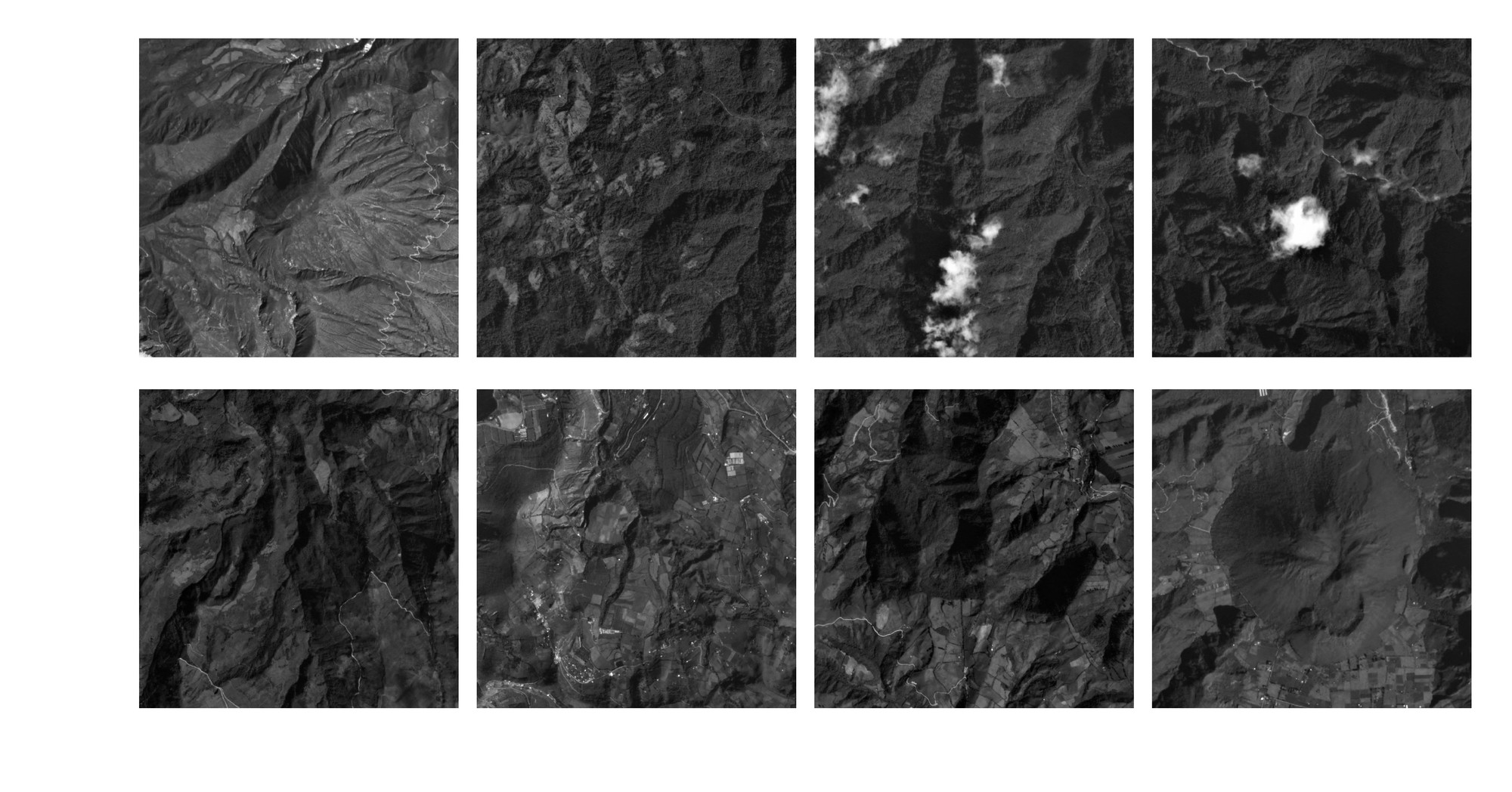}
\caption{A sample of $8$ out of $35$ training
images.}\label{Fig:inpainting-deblurring-training}
\end{figure*}

The goal of the first example is to recover a $[0 - 255]$ grayscale
image from $20\%$ observed pixels of a blurred image ($25$ pixels known
motion blur), where each observed data point also contains zero-mean
random noise in the interval $[-10 - 10]$. The forward operator $F$ is
thus a restriction of an averaging matrix. As an additional challenge,
we do not assume exact knowledge of the noise level and work with the
over-estimation $[-15 - 15]$. The data set contains a series of images
from `Planet Labs PlanetScope Ecuador' with a resolution of three
meters, available at openaerialmap.org. There are $35$ patches of
$1100 \times 1100$ pixels for training, some of which are displayed in
Figure~\ref{Fig:inpainting-deblurring-training}.

We compare the results of the projection onto the intersection of the
$11$ learned constraints using the proposed PARSDMM algorithm, to a
basis pursuit denoise (BPDN) formulation that recovers a vector of
wavelet coefficients, $c$, by solving
$\min_c \| c \|_1 \:\: \text{s.t.} \:\: \|F W^* c - d_\text{obs} \|_2 \leq \sigma$
(BPDN-wavelet). The matrix $W$ represents the wavelet transform:
Daubechies Wavelets as implemented by the SPOT linear operator toolbox
(http://www.cs.ubc.ca/labs/scl/spot/index.html) and computed with the
Rice Wavelet Toolbox (RWT, github.com/ricedsp/rwt). We solve
BPDN-wavelet using the SPGL1 algorithm \citep{doi:10.1137/080714488}.

\begin{figure*}
\centering
\includegraphics[width=1.000\hsize]{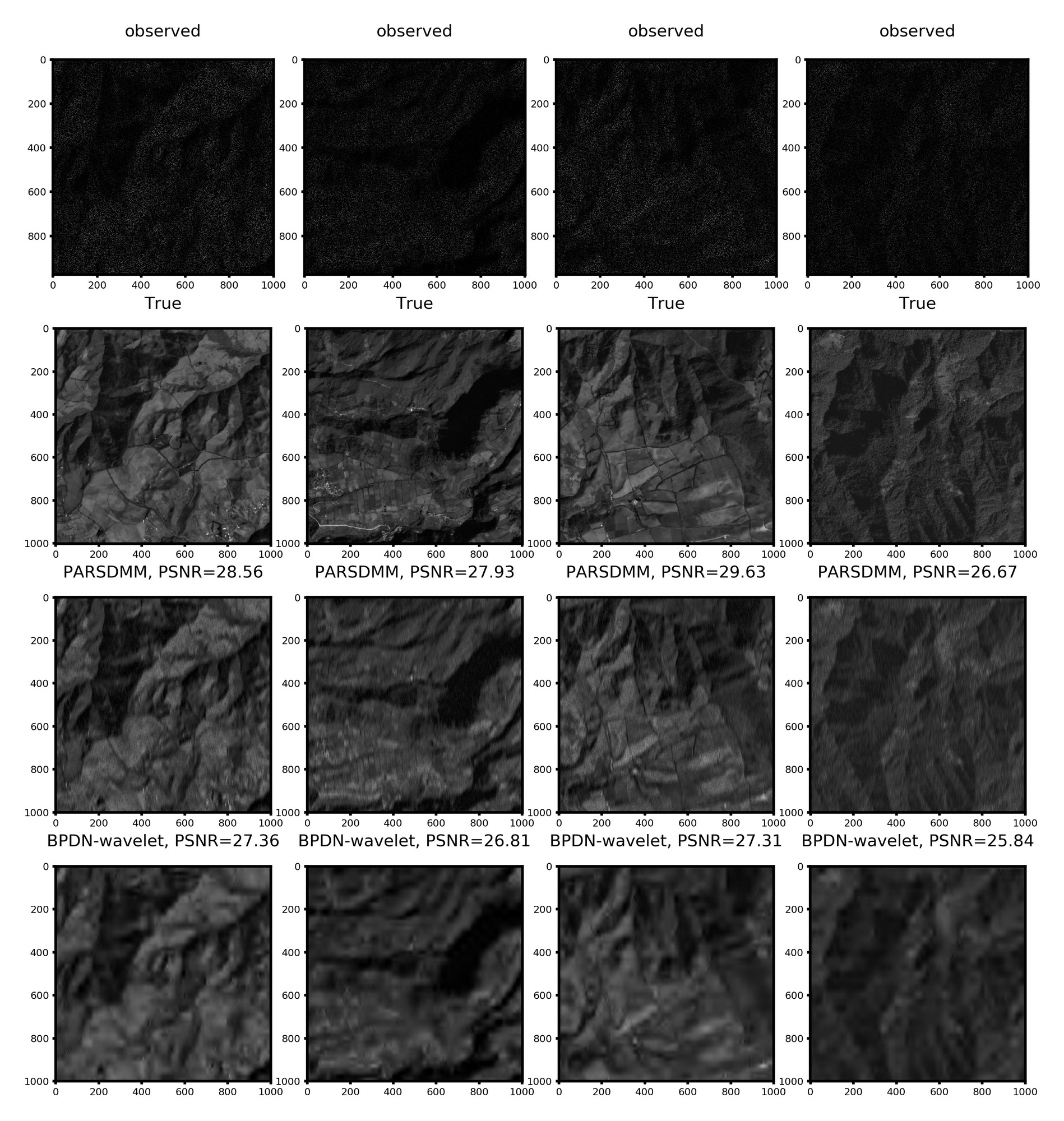}
\caption{Reconstruction results from 80\% missing pixels of an image
with motion blur (25 pixels) and zero-mean random noise in the interval
$[-10, 10]$. Results that are the projection onto an intersection of
multiple learned constraints sets with PARSDMM are visually better than
the BPDN-wavelet results.}\label{Fig:inpainting-deblurring-evaluation}
\end{figure*}

In Figure~\ref{Fig:inpainting-deblurring-evaluation} we see that an
overestimation of $\sigma$ in the BPDN formulation results in
oversimplified images, because the $\ell_2$-ball constraint is too large
which leads to a coefficient vector $c$ that has an $\ell_1$ norm that
is small compared to the true image. The values for $l$ and $u$ in the
data-fit constraint $\{ x \: | \: l \leq (F x - d_\text{obs}) \leq u \}$
are also too large. However, the results from the projection onto the
intersection of multiple constraints suffer much less from overestimated
noise levels, because there are many other constraints that control the
model properties. The results in
Figure~\ref{Fig:inpainting-deblurring-evaluation} show that the learned
set-intersection approach achieves a higher PSNR for all evaluation
images compared to the BPDN formulation.

\subsubsection{Image desaturation}\label{image-desaturation}

\begin{figure*}
\centering
\includegraphics[width=1.000\hsize]{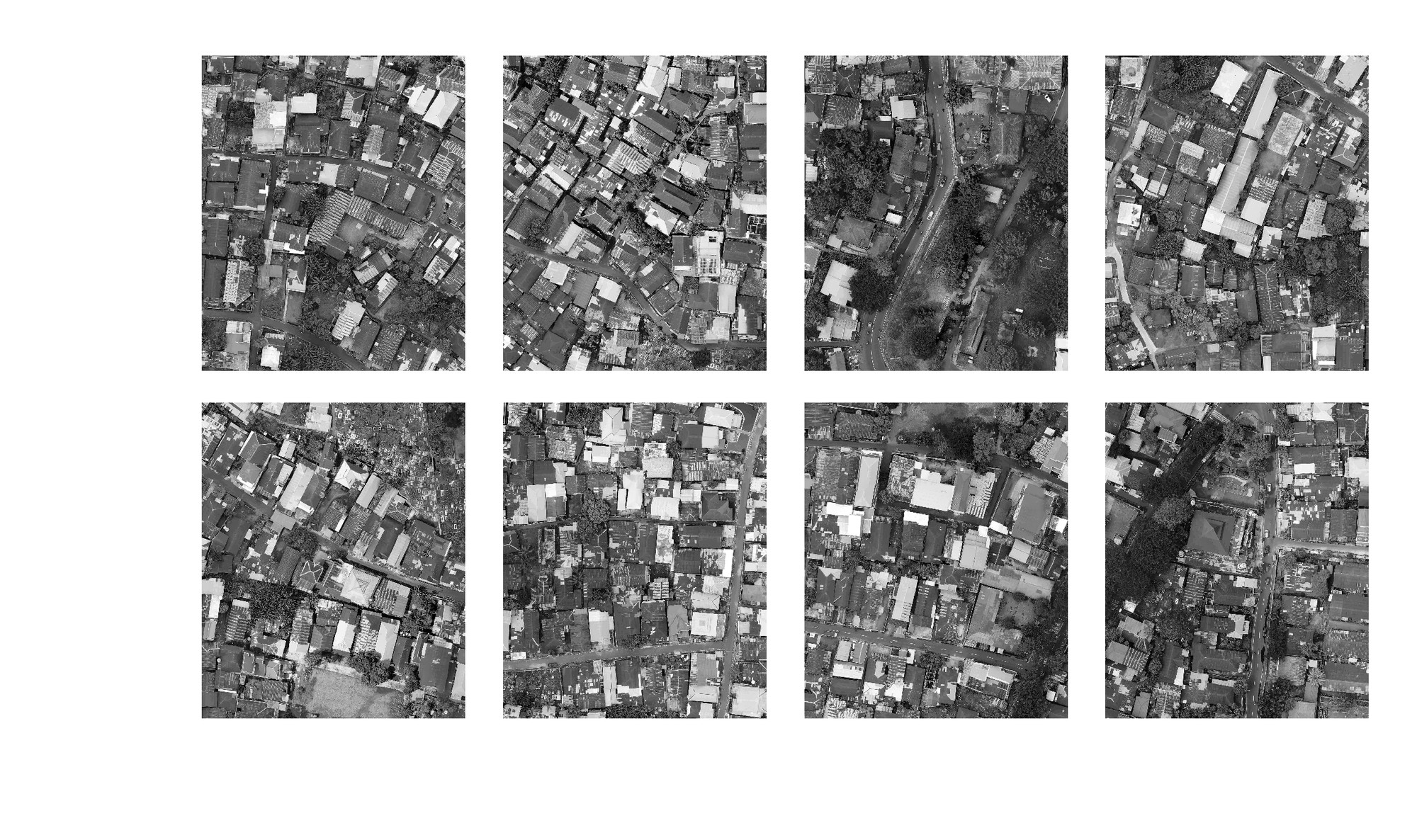}
\caption{A sample of $8$ out of $16$ training
images.}\label{Fig:desaturation-training}
\end{figure*}

To illustrate the versatility of the learning strategy, algorithm, and
constraint sets from the previous example, we now solve an image
desaturation problem for a different data set. The only two things that
we change are the constraint set parameters, which we observe from new
training images (Figure~\ref{Fig:desaturation-training}), and a
different linear forward operator $F$. The data set contains image
patches ($1500 \times 1250$ pixels) from the `Desa Sangaji Kota Ternate'
image with a resolution of $11$ centimeters, available at
openaerialmap.org. The corrupted observed images are saturated grayscale
and generated by clipping the pixel values from $0 - 60$ to $60$ and
from $125 - 255$ to $125$, so there is saturation on both the dark and
bright pixels. If we have no other information about the pixels at the
clipped value, the desaturation problem implies the point-wise bound
constraints \citep[e.g.,][]{BPDNdeSaturation}
\begin{equation}
\begin{cases}
0 \leq x[i] \leq 60 & \text{if } d^{\text{obs}}[i] =60\\
x[i] = d^{\text{obs}}[i] & \text{if } 60 < d^{\text{obs}}[i] < 125\\
125 \leq x[i] \leq 255 & \text{if } d^{\text{obs}}[i] = 125\\
\end{cases}.
\label{saturation_constraint}
\end{equation}
 The forward operator is thus the identity matrix. We solve
problem~\eqref{proj_intersect_lininvprob2} with these point-wise
data-fit constraints and the $11$ model-property constraints listed
earlier.

\begin{figure*}
\centering
\includegraphics[width=1.000\hsize]{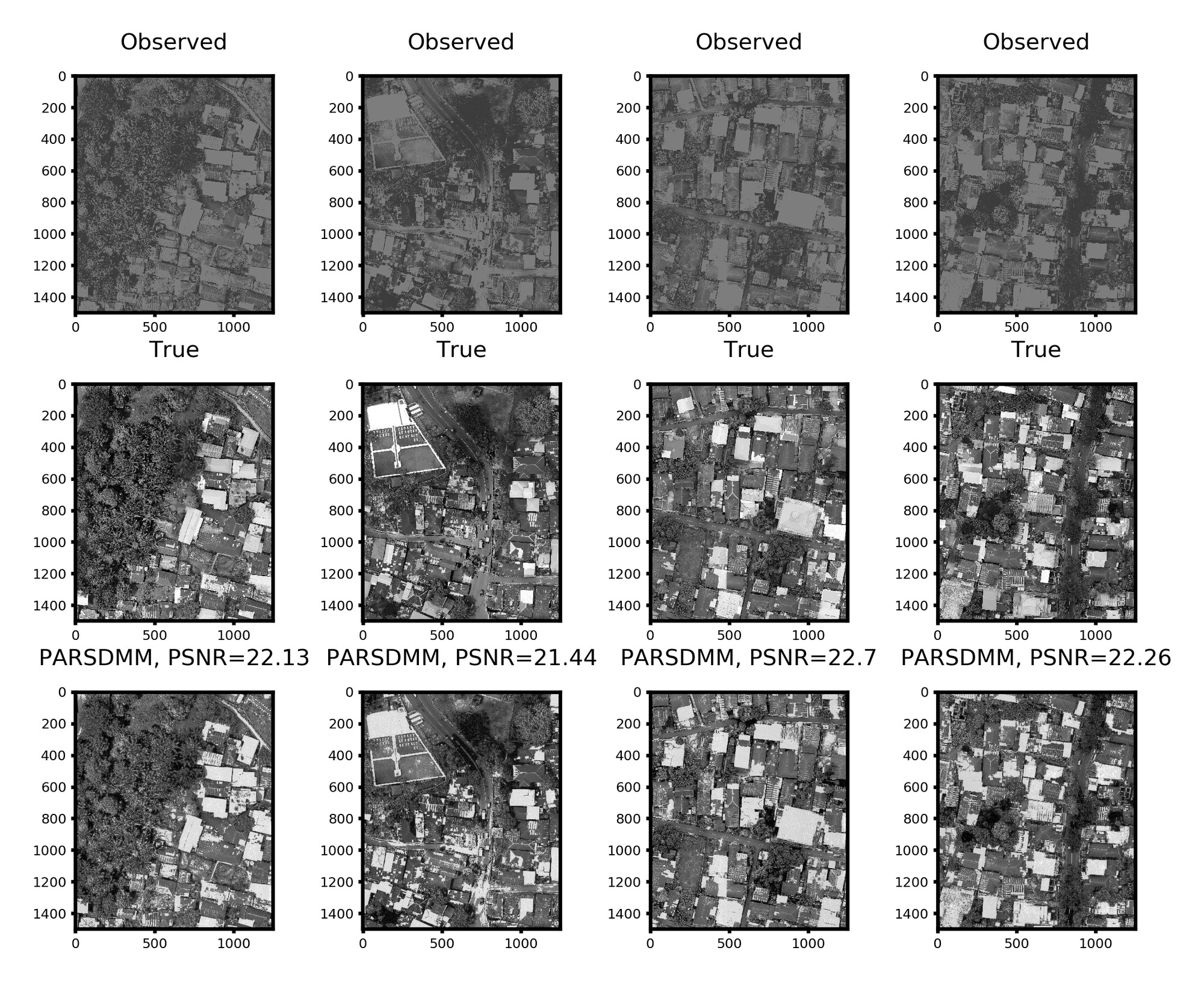}
\caption{Reconstruction results from recovery from saturated images as
the projection onto the intersection of $12$ constraint
sets.}\label{Fig:desaturation-evaluation}
\end{figure*}

Figure~\ref{Fig:desaturation-evaluation} shows the results, true and
observed data for four evaluation images. Large saturated patches are
not desaturated accurately everywhere, because they contain no
non-saturated observed pixels that serve as `anchor' points.

Both the desaturation and the joint deblurring-denoising-inpainting
example show that PARSDMM with multiple convex and non-convex sets
converges to good results, while only a few training examples were
sufficient to estimate the constraint set parameters. Because of the
problem formulation, algorithms, and simple learning strategy, there
were no parameters to hand-pick.

\section{Discussion and future research
directions}\label{discussion-and-future-research-directions}

We developed algorithms to compute projections onto intersections of
multiple sets that help setting up and solving constrained inverse
problems. Our design choices, together with the constrained formulation,
minimize the number of parameters that we need to hand-pick for the
problem formulation, algorithms, and regularization. Our software
package \texttt{SetIntersectionProjection} helps inverse problem
practitioners to test various combinations of constraints for faster
evaluation of their strategies to solve inverse problems. Besides
practicality, we want our work to apply not just toy problems, but also
to models on larger 3D grids. We achieved this via automatic adjustment
of scalar algorithm parameters, parallel implementation, and multilevel
acceleration. There are some limitations, but also opportunities to
increase computational performance that we will now discuss.

Regarding the scope of the \texttt{SetIntersectionProjection} software,
we emphasize that satisfying a constraint for our applications in
imaging inverse problems is different from solving general (non-convex)
optimization problems. When we refer to `reliably' solving a non-convex
problem, we are satisfied with an algorithm that usually approximates
the solution well. For example, if we seek to image a model $m$ that has
$k$ discontinuities, we add the constraint
$\{ m \: | \: \operatorname{card}(Dm) \leq k \}$ where $D$ is a
derivative operator. A satisfying solution for our applications has $k$
large vector entries, whereas all others are small. We do not need to
find a vector that has a cardinality of exactly $k$, because the
estimated model is the same for practical purposes if the results are
assessed qualitatively, or where the expected resolution is much lower
than the fine details we could potentially improve. Moreover, the
forward operator for the inverse problem often is not sensitive to small
changes in the model, and we do not benefit from spending much more
computational time trying to find a more accurate solution to the
non-convex problem. Besides the multilevel projection and automatic
adjustment of augmented-Lagrangian parameters that we already use,
\citep{doi:10.1080/10556788.2017.1304548} present several other
heuristics that can improve the solution of non-convex problems in the
context of ADMM-based algorithms. Future work could test if these
heuristics are computationally feasible for our often large-scale
problems and if they cooperate with our other heuristics.

Besides the limitations and scope of this work, we highlight two ways
how we can reduce computation times for Algorithm~\ref{alg:PARSDMM} and
its multilevel version. First, we recognize that our algorithms use ADMM
as its foundation, which is a synchronous algorithm. This means that the
computations of the projections ($y$-update) in parallel are as slow as
the most time-consuming projection. Without fundamentally changing the
algorithms to asynchronous or randomized projection methods, we can take
a purely software-based approach. Because we compute projections in
parallel, where each projection uses several threads, we are free to
reallocate threads from the fastest projection to the slowest and reduce
the total computational time.

A second computational component that may be improved is the inexact
linear system solve with the conjugate-gradient (CG) method. We do not
use a preconditioner at the moment. Preliminary tests with a simple
diagonal (Jacobi) preconditioner or multigrid V-cycle did reduce the
number of CG iterations, but not the running time for CG in general.
There are a few challenges we face when we design a preconditioner:
\emph{(i)} users may select a variety of linear operators \emph{(ii)}
the system matrix is the weighted sum of multiple linear systems in
normal equation form \emph{(iii)} the weights may change every two
PARSDMM iterations \emph{(iv)} the number of CG iterations varies per
PARSDMM iteration and is often less than ten, which makes it difficult
for preconditioners to reduce the time consumption if they require some
computational overhead or setup time.

Finally we mention how our software package can work cooperatively with
recent developments in plug-and-play regularizers \citep{PandP_ADMM}.
The general idea is to use black-box image processing techniques such as
pre-trained neural networks
\citep{CNNdenoiser, bigdeli2017image, NIPS2017_6831, DPM, MODL8363663, doi:10.1137/17M1122451},
as a map $g(x) : \mathbb{R}^N \rightarrow \mathbb{R}^N$ that behaves
like a proximal operator or projector. Despite the fact that these
plug-and-play algorithms do not, in general, share non-expansiveness
properties with projectors \citep{PPADMM_FPC}, they are successfully
employed in optimization algorithms based on operator-splitting. In our
case, we can use a neural network as the projection operator with the
identity matrix as associated linear operator. In this way we combine
data-constraints and other prior information with a network. A potential
challenge with the plug-and-play concept for constrained optimization is
the difficulty to verify that the intersection of constraints is
effectively non-empty, i.e., can $g(x)$ map to points in the
intersection of the other constraint sets? Some preliminary tests showed
encouraging results and we intent to explore this line of research
further.

\section{Conclusions}\label{conclusions}

We developed novel algorithms and corresponding software, the
\texttt{SetIntersectionProjection} package, for the computation of
projections onto the intersection of multiple constraint sets. These
intersection projections are an important tool for the regularization of
inverse problems. They may be used as the projection part of projected
gradient/(quasi-)Newton algorithms. Projections onto an intersection
also solve set-based formulations for linear image processing problems,
possibly combined with simple learning techniques to extract set
definitions from example images. The algorithms focus on problems with
multiple sets for which we may not have closed-form projections. We
enhance the computational performance for small 2D up to larger 3D
models by specializing the software for projection problems, exploiting
different levels of parallelism on multi-core computing platforms,
automatic selection of scalar (acceleration) parameters, and a
coarse-to-fine grid multilevel implementation. The software is
practical, also for non-expert users, because we do not need manual
step-size selection or related operator norm computations and the
algorithm inputs are pairs of linear operators and projectors, which the
software also generates. Another practical feature is the support for
simultaneous set definitions based on the entire image/tensor and each
slice/row/column. Because we focus on multiple constraints, there is
less of a need to choose the `best' constraint with the `best' linear
operator/transform for a given inverse problem. More constraints are not
much more difficult to deal with than a one or two constraints, also in
terms of computational cost per iteration. We demonstrated the
versatility of the presented algorithms and software using examples from
partial-differential-equation based parameter estimation and image
processing. These examples also show that the algorithms perform well on
problems that include non-convex sets.

\section{Appendix A: Black-box alternating projection
methods}\label{appendix-a-black-box-alternating-projection-methods}

We briefly show that the proposed PARSDMM algorithm
(Algorithm~\ref{alg:PARSDMM}) is different, but closely related to
black-box alternating projection algorithms for computing the projection
onto an intersection of sets. We base this appendix on the alternating
direction method of multipliers (ADMM). The ADMM algorithm is closely
related to Dykstra's algorithm
\citep{doi:10.1080/01621459.1983.10477029, Dyk} for projection problems,
as described by \citep{bauschke2015projection, NIPS2017_6655}, including
the conditions that lead to equivalency.

The parallel version of Dykstra's algorithm
(Algorithm~\ref{alg:ParDykstra}) projects the vector
$m \in \mathbb{R}^N$ onto an intersection of $p$ sets using projections
onto each set separately using projectors
$\mathcal{P}_{\mathcal{V}_1},\mathcal{P}_{\mathcal{V}_2},\dots,\mathcal{P}_{\mathcal{V}_p}$.
If the definitions of the sets $\mathcal{V}_i$ include non-orthogonal
linear operators, these projections are often non-trivial and their
computation requires another iterative algorithm.

\begin{scholmdAlgorithm}
Algorithm~$\text{Parallel-DYKSTRA}(m,\mathcal{P}_{\mathcal{V}_1},\mathcal{P}_{\mathcal{V}_2},\dots,\mathcal{P}_{\mathcal{V}_p})$\\input:~\\\hspace*{0.333em}\hspace*{0.333em}model~to~project:~$m$\\\hspace*{0.333em}\hspace*{0.333em}projectors~onto~sets~$\mathcal{P}_{\mathcal{V}_1},\mathcal{P}_{\mathcal{V}_2},\dots,\mathcal{P}_{\mathcal{V}_p}$\\\texttt{//initialize}\\0a.~~~$x^0=m$,~$k=1$~\\0b.~~~$v_i^0 = x^0$~for~$i=1,2,\dots,p$\\0c.~~~select~weights~$\rho_i$~such~that~$\sum_{i=1}^p \rho_i =1$\\\hspace*{0.333em}\hspace*{0.333em}\textbf{while}~stopping~conditions~not~satisfied~\textbf{do}\\\hspace*{0.333em}\hspace*{0.333em}\hspace*{0.333em}\hspace*{0.333em}\hspace*{0.333em}\hspace*{0.333em}\hspace*{0.333em}\textbf{FOR}~$i=1,2,\dots,p$\\1.~~~~~~~~$y^{k+1}_i = \mathcal{P}_{\mathcal{V}_i}(v^k_{i})$\\\hspace*{0.333em}\hspace*{0.333em}\hspace*{0.333em}\hspace*{0.333em}\hspace*{0.333em}\hspace*{0.333em}\hspace*{0.333em}\textbf{END}\\2.~~~~$x^{k+1} = \sum_{i=1}^p \rho_i y^{k+1}_i$\\\hspace*{0.333em}\hspace*{0.333em}\hspace*{0.333em}\hspace*{0.333em}\hspace*{0.333em}\hspace*{0.333em}\hspace*{0.333em}\textbf{FOR}~$i=1,2,\dots,p$\\3.~~~~~~~~$v^{k+1}_i = x^{k+1} + v^k_{i} - y^{k+1}_i$\\\hspace*{0.333em}\hspace*{0.333em}\hspace*{0.333em}\hspace*{0.333em}\hspace*{0.333em}\hspace*{0.333em}\hspace*{0.333em}\textbf{END}\\4.~~~~$k \leftarrow k+1$\\\hspace*{0.333em}\hspace*{0.333em}\textbf{END}\\output:~$x$
\caption{Parallel Dykstra's algorithm to compute
$\argmin_{x} \frac{1}{2} \| x - m \|_2^2 \:\: \text{s.t.} \:\: x \in \bigcap_{i=1}^p \mathcal{V}_i$.}\label{alg:ParDykstra}
\end{scholmdAlgorithm}

To show the similarity and difference between PARSDMM and parallel
Dykstra's algorithm, we proceed with a derivation similar to
Algorithm~\ref{alg:PARSDMM}, but different in such a way that the final
algorithm is black-box, i.e., it uses projections onto the sets
$\mathcal{V}_i$ and the linear operators are `hidden'.

First we rewrite the projection problem of $m$ onto the intersection of
sets $\mathcal{V}_i$,
\begin{equation}
\min_x \frac{1}{2} \| x - m \|_2^2 + \sum_{i=1}^{p-1} \iota_{\mathcal{V}_i}(x)
\label{A1_proj1}
\end{equation}
 as
\begin{equation}
\min_x \frac{1}{2} \| x - m \|_2^2 + \sum_{i=1}^{p-1} \iota_{\mathcal{C}_i}(A_i x).
\label{A1_proj2}
\end{equation}
 Where we exposed linear operators $A_i$ by rewriting the indicator
functions
$\iota_{\mathcal{V}_i}(x) \rightarrow \iota_{\mathcal{C}_i}(A_i x)$. Now
we introduce additional variables and equality constraints to set up a
parallel algorithm as
\begin{equation}
\min_{x,\{y_i\}} \frac{1}{2} \| y_p - m \|_2^2 + \sum_{i=1}^{p-1} \iota_{\mathcal{C}_i}(A_i y_i) \quad \text{s.t.} \quad x = y_i \: \forall i.
\label{A1_proj3}
\end{equation}
 This problem is suitable for solving with ADMM if we recast it as
\begin{equation}
\min_{x,\tilde{y} } \tilde{f}(\tilde{A} \tilde{y}) \quad \text{s.t.} \quad \tilde{D} x = \tilde{y},
\label{A1_proj4}
\end{equation}
 with
\begin{equation}
\tilde{f}(\tilde{y}) \equiv \frac{1}{2} \| y_p - m \|_2^2 + \sum_{i=1}^{p-1} \iota_{\mathcal{C}_i}(A_i y_i)
\label{A1_f1}
\end{equation}
 and
\begin{equation}
\begin{split}
\tilde{D} \equiv \begin{pmatrix} I_1 \\ \vdots \\ I_{p} \end{pmatrix}, \quad \tilde{y} \equiv \begin{pmatrix} y_1 \\ \vdots \\ y_{p} \end{pmatrix}, \quad \tilde{A} \equiv \begin{pmatrix} A_1 \\ \vdots \\ A_p \end{pmatrix}.
\end{split}
\label{alt_proj_mat_def}
\end{equation}
 The linear equality constraints enforce that all $y_i$ are copies of
$x$ at the solution of problem~\eqref{A1_proj3}. The difference with
PARSDMM is that we leave the $A_i$ inside the indicator functions
instead of moving them to the linear equality constraints. The
corresponding augmented Lagrangian with penalty parameters $\rho_i > 0$
is
\begin{equation}
L_{\rho_1, \dots, \rho_p} (x, y_1, \dots , y_p, v_1, \dots, v_p) = \sum_{i=1}^p \bigg[ \tilde{f}_i(A_i y_i) + v_i^\top (y_i - x ) + \frac{\rho_i}{2} \| y_i - x \|^2_2 \bigg].
\label{A1_auglag}
\end{equation}
 The ADMM iterations with a relaxation parameters $\gamma_i$ are then
given by
\begin{equation}
\begin{aligned}
 x^{k+1} &= \arg\min_{x} \sum_{i=1}^p \Big[ \frac{\rho_i^k}{2} \| y_i^{k} - x + \frac{v_i^k}{\rho_i^k} \|^2_2 \Big] \\
 &= \frac{\sum_{i=1}^p \Big[ \rho_i^k y_i^k + v_i^k \Big]}{\sum_{i=1}^p \rho_i^k} \\ \nonumber
 \bar{x}_i^{k+1} &= \gamma_i^k x_i^{k+1} + ( 1-\gamma_i^k ) y_i^{k} \\ \nonumber
 y_i^{k+1} &= \arg\min_{y_i} \Big[ f_i(A_i y_i) + \frac{\rho_i}{2} \| y_i - \bar{x}_i^{k+1} + \frac{v_i^k}{\rho_i^k} \|^2_2 \Big] \\
 &= \operatorname{prox}_{f_i \circ A_i ,\rho_i^k}(\bar{x}_i^{k+1}-\frac{v_i^k}{\rho_i^k}) \\ \nonumber
 v_i^{k+1} &= v_i^k + \rho_i^k (y_i^{k+1} - \bar{x}_i^{k+1}). \nonumber
\end{aligned}
\label{ADMM_alt_proj_1}
\end{equation}
 The difference with Algorithm~\ref{alg:PARSDMM} is that the linear
operators $A_i$ move from the $x^{k+1}$ computation to the $y_i^{k+1}$
computation. This means the $x^{k+1}$ computation is now a simple
averaging step instead of a linear system solution. The $y_i^{k+1}$
changed from evaluating proximal maps (almost always in closed-form),
into evaluations of proximal maps involving linear operators (usually
not known in closed-form). The proximal maps
$\operatorname{prox}_{f_i \circ A_i ,\rho_i^k}$ for $i=1,\dots,p-1$ are
projections onto $\mathcal{V}_i$, except for $i=p$, which is the
proximal map for $\frac{1}{2} \| y_p - m \|_2^2$. We need another
iterative algorithm to compute the $y_i^{k+1}$ at relatively high
computation cost. The algorithm as a whole becomes more complicated,
because we need additional stopping criteria for the algorithm that
computes the $y_i$ updates.

This iterations from~\eqref{ADMM_alt_proj_1} are similar to parallel
Dykstra (Algorithm~\ref{alg:ParDykstra}) and are, in essence, ADMM
applied to a standard consensus form optimization problem \citep[problem
7.1]{Boyd:2011:DOS:2185815.2185816}.

\bibliographystyle{abbrvnat}
\bibliography{bib_bas}

\begin{thebibliography}{98}
\providecommand{\natexlab}[1]{#1}
\providecommand{\url}[1]{\texttt{#1}}
\expandafter\ifx\csname urlstyle\endcsname\relax
  \providecommand{\doi}[1]{doi: #1}\else
  \providecommand{\doi}{doi: \begingroup \urlstyle{rm}\Url}\fi

\bibitem[Afonso et~al.(2011)Afonso, Bioucas-Dias, and Figueiredo]{SalsaPaper}
M.~V. Afonso, J.~M. Bioucas-Dias, and M.~A.~T. Figueiredo.
\newblock An augmented lagrangian approach to the constrained optimization
  formulation of imaging inverse problems.
\newblock \emph{IEEE Transactions on Image Processing}, 20\penalty0
  (3):\penalty0 681--695, March 2011.
\newblock ISSN 1057-7149.
\newblock \doi{10.1109/TIP.2010.2076294}.

\bibitem[Aggarwal et~al.(2018)Aggarwal, Mani, and Jacob]{MODL8363663}
H.~K. Aggarwal, M.~P. Mani, and M.~Jacob.
\newblock Model based image reconstruction using deep learned priors (modl).
\newblock In \emph{2018 IEEE 15th International Symposium on Biomedical Imaging
  (ISBI 2018)}, pages 671--674, April 2018.
\newblock \doi{10.1109/ISBI.2018.8363663}.

\bibitem[{Antonello} et~al.(2018){Antonello}, {Stella}, {Patrinos}, and {van
  Waterschoot}]{arXiv180301621A}
N.~{Antonello}, L.~{Stella}, P.~{Patrinos}, and T.~{van Waterschoot}.
\newblock {Proximal Gradient Algorithms: Applications in Signal Processing}.
\newblock \emph{ArXiv e-prints}, Mar. 2018.

\bibitem[Arag{\'o}n~Artacho and Campoy(2018)]{AragonArtacho2018}
F.~J. Arag{\'o}n~Artacho and R.~Campoy.
\newblock A new projection method for finding the closest point in the
  intersection of convex sets.
\newblock \emph{Computational Optimization and Applications}, 69\penalty0
  (1):\penalty0 99--132, Jan 2018.
\newblock ISSN 1573-2894.
\newblock \doi{10.1007/s10589-017-9942-5}.
\newblock URL \url{https://doi.org/10.1007/s10589-017-9942-5}.

\bibitem[Aravkin et~al.(2014)Aravkin, Kumar, Mansour, Recht, and
  Herrmann]{doi:10.1137/130919210}
A.~Aravkin, R.~Kumar, H.~Mansour, B.~Recht, and F.~J. Herrmann.
\newblock Fast methods for denoising matrix completion formulations, with
  applications to robust seismic data interpolation.
\newblock \emph{SIAM Journal on Scientific Computing}, 36\penalty0
  (5):\penalty0 S237--S266, 2014.
\newblock \doi{10.1137/130919210}.
\newblock URL \url{https://doi.org/10.1137/130919210}.

\bibitem[Aravkin et~al.(2016)Aravkin, Burke, Drusvyatskiy, Friedlander, and
  Roy]{aravkin2016level}
A.~Y. Aravkin, J.~V. Burke, D.~Drusvyatskiy, M.~P. Friedlander, and S.~Roy.
\newblock Level-set methods for convex optimization.
\newblock \emph{arXiv preprint arXiv:1602.01506}, 2016.

\bibitem[Ascher and Haber(2001)]{Ascher_2001}
U.~M. Ascher and E.~Haber.
\newblock Grid refinement and scaling for distributed parameter estimation
  problems.
\newblock \emph{Inverse Problems}, 17\penalty0 (3):\penalty0 571--590, may
  2001.
\newblock \doi{10.1088/0266-5611/17/3/314}.
\newblock URL \url{https://doi.org/10.1088%2F0266-5611%2F17%2F3%2F314}.

\bibitem[Barzilai and Borwein(1988)]{BARZILAI01011988}
J.~Barzilai and J.~M. Borwein.
\newblock Two-point step size gradient methods.
\newblock \emph{IMA Journal of Numerical Analysis}, 8\penalty0 (1):\penalty0
  141--148, 1988.
\newblock \doi{10.1093/imanum/8.1.141}.
\newblock URL \url{http://imajna.oxfordjournals.org/content/8/1/141.abstract}.

\bibitem[Bauschke and Koch(2015)]{bauschke2015projection}
H.~H. Bauschke and V.~R. Koch.
\newblock Projection methods: Swiss army knives for solving feasibility and
  best approximation problems with halfspaces.
\newblock \emph{Contemporary Mathematics}, 636:\penalty0 1--40, 2015.

\bibitem[Beck(2017)]{doi:10.1137/1.9781611974997}
A.~Beck.
\newblock \emph{First-Order Methods in Optimization}.
\newblock Society for Industrial and Applied Mathematics, Philadelphia, PA,
  2017.
\newblock \doi{10.1137/1.9781611974997}.
\newblock URL \url{http://epubs.siam.org/doi/abs/10.1137/1.9781611974997}.

\bibitem[Becker et~al.(2011)Becker, Cand{\`e}s, and Grant]{Becker2011}
S.~R. Becker, E.~J. Cand{\`e}s, and M.~C. Grant.
\newblock Templates for convex cone problems with applications to sparse signal
  recovery.
\newblock \emph{Mathematical Programming Computation}, 3\penalty0 (3):\penalty0
  165, Jul 2011.
\newblock ISSN 1867-2957.
\newblock \doi{10.1007/s12532-011-0029-5}.
\newblock URL \url{https://doi.org/10.1007/s12532-011-0029-5}.

\bibitem[Bertsekas(1982)]{doi:10.1137/0320018}
D.~P. Bertsekas.
\newblock Projected newton methods for optimization problems with simple
  constraints.
\newblock \emph{SIAM Journal on Control and Optimization}, 20\penalty0
  (2):\penalty0 221--246, 1982.
\newblock \doi{10.1137/0320018}.
\newblock URL \url{https://doi.org/10.1137/0320018}.

\bibitem[Bezanson et~al.(2017)Bezanson, Edelman, Karpinski, and
  Shah]{doi:10.1137/141000671}
J.~Bezanson, A.~Edelman, S.~Karpinski, and V.~B. Shah.
\newblock Julia: A fresh approach to numerical computing.
\newblock \emph{SIAM Review}, 59\penalty0 (1):\penalty0 65--98, 2017.
\newblock \doi{10.1137/141000671}.
\newblock URL \url{https://doi.org/10.1137/141000671}.

\bibitem[Bigdeli and Zwicker(2017)]{bigdeli2017image}
S.~A. Bigdeli and M.~Zwicker.
\newblock Image restoration using autoencoding priors.
\newblock \emph{arXiv preprint arXiv:1703.09964}, 2017.

\bibitem[Birgin et~al.(1999)Birgin, Mart\'{\i}nez, and
  Raydan]{Birgin:1999:NSP:588891.589081}
E.~G. Birgin, J.~M. Mart\'{\i}nez, and M.~Raydan.
\newblock Nonmonotone spectral projected gradient methods on convex sets.
\newblock \emph{SIAM J. on Optimization}, 10\penalty0 (4):\penalty0 1196--1211,
  Aug. 1999.
\newblock ISSN 1052-6234.
\newblock \doi{10.1137/S1052623497330963}.
\newblock URL \url{http://dx.doi.org/10.1137/S1052623497330963}.

\bibitem[Boyd et~al.(2011)Boyd, Parikh, Chu, Peleato, and
  Eckstein]{Boyd:2011:DOS:2185815.2185816}
S.~Boyd, N.~Parikh, E.~Chu, B.~Peleato, and J.~Eckstein.
\newblock Distributed optimization and statistical learning via the alternating
  direction method of multipliers.
\newblock \emph{Found. Trends Mach. Learn.}, 3\penalty0 (1):\penalty0 1--122,
  Jan. 2011.
\newblock ISSN 1935-8237.
\newblock \doi{10.1561/2200000016}.
\newblock URL \url{http://dx.doi.org/10.1561/2200000016}.

\bibitem[Boyle and Dykstra(1986)]{Dyk}
J.~P. Boyle and R.~L. Dykstra.
\newblock \emph{A Method for Finding Projections onto the Intersection of
  Convex Sets in Hilbert Spaces}, pages 28--47.
\newblock Springer New York, New York, NY, 1986.
\newblock ISBN 978-1-4613-9940-7.
\newblock \doi{10.1007/978-1-4613-9940-7_3}.
\newblock URL \url{http://dx.doi.org/10.1007/978-1-4613-9940-7_3}.

\bibitem[Buzzard et~al.(2018)Buzzard, Chan, Sreehari, and
  Bouman]{doi:10.1137/17M1122451}
G.~Buzzard, S.~Chan, S.~Sreehari, and C.~Bouman.
\newblock Plug-and-play unplugged: Optimization-free reconstruction using
  consensus equilibrium.
\newblock \emph{SIAM Journal on Imaging Sciences}, 11\penalty0 (3):\penalty0
  2001--2020, 2018.
\newblock \doi{10.1137/17M1122451}.
\newblock URL \url{https://doi.org/10.1137/17M1122451}.

\bibitem[Cand{\`e}s and Recht(2009)]{Candes2009}
E.~J. Cand{\`e}s and B.~Recht.
\newblock Exact matrix completion via convex optimization.
\newblock \emph{Foundations of Computational Mathematics}, 9\penalty0
  (6):\penalty0 717, Apr 2009.
\newblock ISSN 1615-3383.
\newblock \doi{10.1007/s10208-009-9045-5}.
\newblock URL \url{https://doi.org/10.1007/s10208-009-9045-5}.

\bibitem[Censor(2006)]{CENSOR2006111}
Y.~Censor.
\newblock Computational acceleration of projection algorithms for the linear
  best approximation problem.
\newblock \emph{Linear Algebra and its Applications}, 416\penalty0
  (1):\penalty0 111 -- 123, 2006.
\newblock ISSN 0024-3795.
\newblock \doi{http://dx.doi.org/10.1016/j.laa.2005.10.006}.
\newblock URL
  \url{http://www.sciencedirect.com/science/article/pii/S0024379505004891}.

\bibitem[Censor et~al.(2005)Censor, Elfving, Kopf, and
  Bortfeld]{censor2005multiple}
Y.~Censor, T.~Elfving, N.~Kopf, and T.~Bortfeld.
\newblock The multiple-sets split feasibility problem and its applications for
  inverse problems.
\newblock \emph{Inverse Problems}, 21\penalty0 (6):\penalty0 2071, 2005.

\bibitem[Chan et~al.(2017)Chan, Wang, and Elgendy]{PPADMM_FPC}
S.~H. Chan, X.~Wang, and O.~A. Elgendy.
\newblock Plug-and-play admm for image restoration: Fixed-point convergence and
  applications.
\newblock \emph{IEEE Transactions on Computational Imaging}, 3\penalty0
  (1):\penalty0 84--98, March 2017.
\newblock ISSN 2333-9403.
\newblock \doi{10.1109/TCI.2016.2629286}.

\bibitem[Chang et~al.(2017)Chang, Li, Póczos, and Kumar]{DPM}
J.~H.~R. Chang, C.~Li, B.~Póczos, and B.~V. K.~V. Kumar.
\newblock One network to solve them all — solving linear inverse problems
  using deep projection models.
\newblock In \emph{2017 IEEE International Conference on Computer Vision
  (ICCV)}, pages 5889--5898, Oct 2017.
\newblock \doi{10.1109/ICCV.2017.627}.

\bibitem[Chen et~al.(2001)Chen, Donoho, and
  Saunders]{doi:10.1137/S003614450037906X}
S.~Chen, D.~Donoho, and M.~Saunders.
\newblock Atomic decomposition by basis pursuit.
\newblock \emph{SIAM Review}, 43\penalty0 (1):\penalty0 129--159, 2001.
\newblock \doi{10.1137/S003614450037906X}.
\newblock URL \url{https://doi.org/10.1137/S003614450037906X}.

\bibitem[Combettes(1996)]{COMBETTES1996155}
P.~Combettes.
\newblock The convex feasibility problem in image recovery.
\newblock volume~95 of \emph{Advances in Imaging and Electron Physics}, pages
  155 -- 270. Elsevier, 1996.
\newblock \doi{https://doi.org/10.1016/S1076-5670(08)70157-5}.
\newblock URL
  \url{http://www.sciencedirect.com/science/article/pii/S1076567008701575}.

\bibitem[Combettes(1993)]{STEstimation}
P.~L. Combettes.
\newblock The foundations of set theoretic estimation.
\newblock \emph{Proceedings of the IEEE}, 81\penalty0 (2):\penalty0 182--208,
  Feb 1993.
\newblock ISSN 0018-9219.
\newblock \doi{10.1109/5.214546}.

\bibitem[Combettes and Pesquet(2004)]{TVLearn}
P.~L. Combettes and J.~C. Pesquet.
\newblock Image restoration subject to a total variation constraint.
\newblock \emph{IEEE Transactions on Image Processing}, 13\penalty0
  (9):\penalty0 1213--1222, Sept 2004.
\newblock ISSN 1057-7149.
\newblock \doi{10.1109/TIP.2004.832922}.

\bibitem[Combettes and Pesquet(2011)]{prox_split}
P.~L. Combettes and J.-C. Pesquet.
\newblock Proximal splitting methods in signal processing.
\newblock In H.~H. Bauschke, R.~S. Burachik, P.~L. Combettes, V.~Elser, D.~R.
  Luke, and H.~Wolkowicz, editors, \emph{Fixed-Point Algorithms for Inverse
  Problems in Science and Engineering}, volume~49 of \emph{Springer
  Optimization and Its Applications}, pages 185--212. Springer New York, 2011.
\newblock ISBN 978-1-4419-9568-1.
\newblock \doi{10.1007/978-1-4419-9569-8_10}.
\newblock URL \url{http://dx.doi.org/10.1007/978-1-4419-9569-8_10}.

\bibitem[Constable et~al.(1987)Constable, Parker, and
  Constable]{doi:10.1190/1.1442303}
S.~C. Constable, R.~L. Parker, and C.~G. Constable.
\newblock Occam’s inversion: A practical algorithm for generating smooth
  models from electromagnetic sounding data.
\newblock \emph{GEOPHYSICS}, 52\penalty0 (3):\penalty0 289--300, 1987.
\newblock \doi{10.1190/1.1442303}.
\newblock URL \url{http://dx.doi.org/10.1190/1.1442303}.

\bibitem[Da~Silva and Herrmann(2017)]{waveformCurt}
C.~Da~Silva and F.~J. Herrmann.
\newblock {A Unified 2D/3D Large Scale Software Environment for Nonlinear
  Inverse Problems}.
\newblock \emph{ArXiv e-prints}, Mar. 2017.

\bibitem[Diamond et~al.(2018)Diamond, Takapoui, and
  Boyd]{doi:10.1080/10556788.2017.1304548}
S.~Diamond, R.~Takapoui, and S.~Boyd.
\newblock A general system for heuristic minimization of convex functions over
  non-convex sets.
\newblock \emph{Optimization Methods and Software}, 33\penalty0 (1):\penalty0
  165--193, 2018.
\newblock \doi{10.1080/10556788.2017.1304548}.
\newblock URL \url{https://doi.org/10.1080/10556788.2017.1304548}.

\bibitem[Domahidi et~al.(2013)Domahidi, Chu, and Boyd]{domahidi2013ecos}
A.~Domahidi, E.~Chu, and S.~Boyd.
\newblock Ecos: An socp solver for embedded systems.
\newblock In \emph{Control Conference (ECC), 2013 European}, pages 3071--3076.
  IEEE, 2013.

\bibitem[Duchi et~al.(2008)Duchi, Shalev-Shwartz, Singer, and
  Chandra]{duchi:icml08}
J.~Duchi, S.~Shalev-Shwartz, Y.~Singer, and T.~Chandra.
\newblock Efficient projections onto the l1-ball for learning in high
  dimensions.
\newblock pages 272--279, 2008.

\bibitem[Dykstra(1983)]{doi:10.1080/01621459.1983.10477029}
R.~L. Dykstra.
\newblock An algorithm for restricted least squares regression.
\newblock \emph{Journal of the American Statistical Association}, 78\penalty0
  (384):\penalty0 837--842, 1983.
\newblock \doi{10.1080/01621459.1983.10477029}.
\newblock URL
  \url{http://www.tandfonline.com/doi/abs/10.1080/01621459.1983.10477029}.

\bibitem[Eckstein and Bertsekas(1992)]{Eckstein1992}
J.~Eckstein and D.~P. Bertsekas.
\newblock On the douglas---rachford splitting method and the proximal point
  algorithm for maximal monotone operators.
\newblock \emph{Mathematical Programming}, 55\penalty0 (1):\penalty0 293--318,
  Apr 1992.
\newblock ISSN 1436-4646.
\newblock \doi{10.1007/BF01581204}.
\newblock URL \url{https://doi.org/10.1007/BF01581204}.

\bibitem[Eckstein and Yao(2015)]{eckstein2015understanding}
J.~Eckstein and W.~Yao.
\newblock Understanding the convergence of the alternating direction method of
  multipliers: Theoretical and computational perspectives.
\newblock \emph{Pac. J. Optim. To appear}, 2015.

\bibitem[Esser(2009)]{esser2009applications}
E.~Esser.
\newblock Applications of lagrangian-based alternating direction methods and
  connections to split bregman.
\newblock \emph{CAM report}, 9:\penalty0 31, 2009.

\bibitem[Esser et~al.(2016)Esser, Guasch, Herrmann, and
  Warner]{doi:10.1190/tle35030235.1}
E.~Esser, L.~Guasch, F.~J. Herrmann, and M.~Warner.
\newblock Constrained waveform inversion for automatic salt flooding.
\newblock \emph{The Leading Edge}, 35\penalty0 (3):\penalty0 235--239, 2016.
\newblock \doi{10.1190/tle35030235.1}.
\newblock URL \url{http://dx.doi.org/10.1190/tle35030235.1}.

\bibitem[Esser et~al.(2018)Esser, Guasch, van Leeuwen, Aravkin, and
  Herrmann]{Esser2016arch}
E.~Esser, L.~Guasch, T.~van Leeuwen, A.~Y. Aravkin, and F.~J. Herrmann.
\newblock Total variation regularization strategies in full-waveform inversion.
\newblock \emph{SIAM Journal on Imaging Sciences}, 11\penalty0 (1):\penalty0
  376--406, 2018.
\newblock \doi{10.1137/17M111328X}.
\newblock URL \url{https://doi.org/10.1137/17M111328X}.

\bibitem[Fan et~al.(2017)Fan, Wei, Carin, and Heller]{NIPS2017_6831}
K.~Fan, Q.~Wei, L.~Carin, and K.~A. Heller.
\newblock An inner-loop free solution to inverse problems using deep neural
  networks.
\newblock In I.~Guyon, U.~V. Luxburg, S.~Bengio, H.~Wallach, R.~Fergus,
  S.~Vishwanathan, and R.~Garnett, editors, \emph{Advances in Neural
  Information Processing Systems 30}, pages 2370--2380. Curran Associates,
  Inc., 2017.
\newblock URL
  \url{http://papers.nips.cc/paper/6831-an-inner-loop-free-solution-to-inverse-problems-using-deep-neural-networks.pdf}.

\bibitem[Farrell et~al.(2013)Farrell, Ham, Funke, and
  Rognes]{doi:10.1137/120873558}
P.~Farrell, D.~Ham, S.~Funke, and M.~Rognes.
\newblock Automated derivation of the adjoint of high-level transient finite
  element programs.
\newblock \emph{SIAM Journal on Scientific Computing}, 35\penalty0
  (4):\penalty0 C369--C393, 2013.
\newblock \doi{10.1137/120873558}.
\newblock URL \url{https://doi.org/10.1137/120873558}.

\bibitem[Frigo and Johnson(2005)]{FFTW_paper}
M.~Frigo and S.~G. Johnson.
\newblock The design and implementation of fftw3.
\newblock \emph{Proceedings of the IEEE}, 93\penalty0 (2):\penalty0 216--231,
  Feb 2005.
\newblock ISSN 0018-9219.
\newblock \doi{10.1109/JPROC.2004.840301}.

\bibitem[Gander(1980)]{Gander1980}
W.~Gander.
\newblock Least squares with a quadratic constraint.
\newblock \emph{Numerische Mathematik}, 36\penalty0 (3):\penalty0 291--307, Sep
  1980.
\newblock ISSN 0945-3245.
\newblock \doi{10.1007/BF01396656}.
\newblock URL \url{https://doi.org/10.1007/BF01396656}.

\bibitem[Golub and von Matt(1991)]{Golub1991}
G.~H. Golub and U.~von Matt.
\newblock Quadratically constrained least squares and quadratic problems.
\newblock \emph{Numerische Mathematik}, 59\penalty0 (1):\penalty0 561--580, Dec
  1991.
\newblock ISSN 0945-3245.
\newblock \doi{10.1007/BF01385796}.
\newblock URL \url{https://doi.org/10.1007/BF01385796}.

\bibitem[Haber(2014)]{haber2014computational}
E.~Haber.
\newblock \emph{Computational methods in geophysical electromagnetics}.
\newblock SIAM, 2014.

\bibitem[Haber et~al.(2000)Haber, Ascher, and Oldenburg]{Haber2000}
E.~Haber, U.~M. Ascher, and D.~Oldenburg.
\newblock {On optimization techniques for solving nonlinear inverse problems}.
\newblock \emph{Inverse Problems}, 16\penalty0 (5):\penalty0 1263--1280, Oct.
  2000.
\newblock ISSN 0266-5611.
\newblock \doi{10.1088/0266-5611/16/5/309}.
\newblock URL
  \url{http://stacks.iop.org/0266-5611/16/i=5/a=309?key=crossref.98f435f9ee66231b63da02b10f82a60b}.

\bibitem[Heide et~al.(2016)Heide, Diamond, Nie{\ss}ner, Ragan-Kelley, Heidrich,
  and Wetzstein]{Heide2016PEI28978242925875}
F.~Heide, S.~Diamond, M.~Nie{\ss}ner, J.~Ragan-Kelley, W.~Heidrich, and
  G.~Wetzstein.
\newblock Proximal: Efficient image optimization using proximal algorithms.
\newblock \emph{ACM Trans. Graph.}, 35\penalty0 (4):\penalty0 84:1--84:15, July
  2016.
\newblock ISSN 0730-0301.
\newblock \doi{10.1145/2897824.2925875}.
\newblock URL \url{http://doi.acm.org/10.1145/2897824.2925875}.

\bibitem[Iutzeler and Hendrickx(2017)]{doi:10.1080/10556788.2017.1396601}
F.~Iutzeler and J.~M. Hendrickx.
\newblock A generic online acceleration scheme for optimization algorithms via
  relaxation and inertia.
\newblock \emph{Optimization Methods and Software}, 0\penalty0 (0):\penalty0
  1--23, 2017.
\newblock \doi{10.1080/10556788.2017.1396601}.
\newblock URL \url{https://doi.org/10.1080/10556788.2017.1396601}.

\bibitem[Ivanov et~al.(2013)Ivanov, Vasin, and Tanana]{ivanov2013theory}
V.~K. Ivanov, V.~V. Vasin, and V.~P. Tanana.
\newblock \emph{Theory of linear ill-posed problems and its applications},
  volume~36.
\newblock Walter de Gruyter, 2013.

\bibitem[Jia et~al.(2017)Jia, Cai, and Han]{JIA2017320}
Z.~Jia, X.~Cai, and D.~Han.
\newblock Comparison of several fast algorithms for projection onto an
  ellipsoid.
\newblock \emph{Journal of Computational and Applied Mathematics},
  319:\penalty0 320 -- 337, 2017.
\newblock ISSN 0377-0427.
\newblock \doi{https://doi.org/10.1016/j.cam.2017.01.008}.
\newblock URL
  \url{http://www.sciencedirect.com/science/article/pii/S0377042717300122}.

\bibitem[Kitic et~al.(2016)Kitic, Albera, Bertin, and Gribonval]{CoSpSDMM}
S.~Kitic, L.~Albera, N.~Bertin, and R.~Gribonval.
\newblock Physics-driven inverse problems made tractable with cosparse
  regularization.
\newblock \emph{IEEE Transactions on Signal Processing}, 64\penalty0
  (2):\penalty0 335--348, Jan 2016.
\newblock ISSN 1053-587X.
\newblock \doi{10.1109/TSP.2015.2480045}.

\bibitem[Kotakemori et~al.(2008)Kotakemori, Hasegawa, Kajiyama, Nukada, Suda,
  and Nishida]{kotakemori2008performance}
H.~Kotakemori, H.~Hasegawa, T.~Kajiyama, A.~Nukada, R.~Suda, and A.~Nishida.
\newblock Performance evaluation of parallel sparse matrix-vector products on
  sgi altix3700.
\newblock \emph{Lecture Notes in Computer Science}, 4315:\penalty0 153--166,
  2008.

\bibitem[Kukreja et~al.(2016)Kukreja, Louboutin, Vieira, Luporini, Lange, and
  Gorman]{DevitoDSL}
N.~Kukreja, M.~Louboutin, F.~Vieira, F.~Luporini, M.~Lange, and G.~Gorman.
\newblock Devito: Automated fast finite difference computation.
\newblock In \emph{2016 Sixth International Workshop on Domain-Specific
  Languages and High-Level Frameworks for High Performance Computing
  (WOLFHPC)}, pages 11--19, Nov 2016.
\newblock \doi{10.1109/WOLFHPC.2016.06}.

\bibitem[{Kundu} et~al.(2017){Kundu}, {Bach}, and
  {Bhattacharyya}]{arXiv171006465K}
A.~{Kundu}, F.~{Bach}, and C.~{Bhattacharyya}.
\newblock {Convex optimization over intersection of simple sets: improved
  convergence rate guarantees via an exact penalty approach}.
\newblock \emph{ArXiv e-prints}, Oct. 2017.

\bibitem[L{\'o}pez and Raydan(2016)]{Lopez2016}
W.~L{\'o}pez and M.~Raydan.
\newblock An acceleration scheme for dykstra's algorithm.
\newblock \emph{Computational Optimization and Applications}, 63\penalty0
  (1):\penalty0 29--44, Jan 2016.
\newblock ISSN 1573-2894.
\newblock \doi{10.1007/s10589-015-9768-y}.
\newblock URL \url{https://doi.org/10.1007/s10589-015-9768-y}.

\bibitem[Louboutin et~al.(2018)Louboutin, Witte, Lange, Kukreja, Luporini,
  Gorman, and Herrmann]{doi:10.1190/tle37010069.1}
M.~Louboutin, P.~Witte, M.~Lange, N.~Kukreja, F.~Luporini, G.~Gorman, and F.~J.
  Herrmann.
\newblock Full-waveform inversion, part 2: Adjoint modeling.
\newblock \emph{The Leading Edge}, 37\penalty0 (1):\penalty0 69--72, 2018.
\newblock \doi{10.1190/tle37010069.1}.
\newblock URL \url{https://doi.org/10.1190/tle37010069.1}.

\bibitem[Lustig et~al.(2007)Lustig, Donoho, and Pauly]{MRM:MRM21391}
M.~Lustig, D.~Donoho, and J.~M. Pauly.
\newblock Sparse mri: The application of compressed sensing for rapid mr
  imaging.
\newblock \emph{Magnetic Resonance in Medicine}, 58\penalty0 (6):\penalty0
  1182--1195, 2007.
\newblock ISSN 1522-2594.
\newblock \doi{10.1002/mrm.21391}.
\newblock URL \url{http://dx.doi.org/10.1002/mrm.21391}.

\bibitem[Macdonald and Ruthotto(2018)]{Macdonald2018}
J.~Macdonald and L.~Ruthotto.
\newblock Improved susceptibility artifact correction of echo-planar mri using
  the alternating direction method of multipliers.
\newblock \emph{Journal of Mathematical Imaging and Vision}, 60\penalty0
  (2):\penalty0 268--282, Feb 2018.
\newblock ISSN 1573-7683.
\newblock \doi{10.1007/s10851-017-0757-x}.
\newblock URL \url{https://doi.org/10.1007/s10851-017-0757-x}.

\bibitem[Mallat and Zhang(1992)]{MallatZhang}
S.~Mallat and Z.~Zhang.
\newblock Adaptive time-frequency decomposition with matching pursuits.
\newblock In \emph{[1992] Proceedings of the IEEE-SP International Symposium on
  Time-Frequency and Time-Scale Analysis}, pages 7--10, Oct 1992.
\newblock \doi{10.1109/TFTSA.1992.274245}.

\bibitem[Mansour et~al.(2010)Mansour, Saab, Nasiopoulos, and
  Ward]{BPDNdeSaturation}
H.~Mansour, R.~Saab, P.~Nasiopoulos, and R.~Ward.
\newblock Color image desaturation using sparse reconstruction.
\newblock In \emph{2010 IEEE International Conference on Acoustics, Speech and
  Signal Processing}, pages 778--781, March 2010.
\newblock \doi{10.1109/ICASSP.2010.5494984}.

\bibitem[Nash(2000)]{doi:10.1080/10556780008805795}
S.~G. Nash.
\newblock A multigrid approach to discretized optimization problems.
\newblock \emph{Optimization Methods and Software}, 14\penalty0 (1-2):\penalty0
  99--116, 2000.
\newblock \doi{10.1080/10556780008805795}.
\newblock URL \url{https://doi.org/10.1080/10556780008805795}.

\bibitem[Nash(2014)]{doi:10.1080/10556788.2012.759571}
S.~G. Nash.
\newblock Properties of a class of multilevel optimization algorithms for
  equality-constrained problems.
\newblock \emph{Optimization Methods and Software}, 29\penalty0 (1):\penalty0
  137--159, 2014.
\newblock \doi{10.1080/10556788.2012.759571}.
\newblock URL \url{https://doi.org/10.1080/10556788.2012.759571}.

\bibitem[Nishihara et~al.(2015)Nishihara, Lessard, Recht, Packard, and
  Jordan]{nishihara2015general}
R.~Nishihara, L.~Lessard, B.~Recht, A.~Packard, and M.~I. Jordan.
\newblock A general analysis of the convergence of admm.
\newblock In \emph{Int. Conf. Mach. Learn.}, volume~37, pages 343--352, 2015.

\bibitem[Nocedal and Wright(2000)]{Nocedal:2000}
J.~Nocedal and S.~J. Wright.
\newblock \emph{{Numerical optimization}}.
\newblock Springer, 2000.

\bibitem[O'Donoghue et~al.(2016)O'Donoghue, Chu, Parikh, and
  Boyd]{ODonoghue2016}
B.~O'Donoghue, E.~Chu, N.~Parikh, and S.~Boyd.
\newblock Conic optimization via operator splitting and homogeneous self-dual
  embedding.
\newblock \emph{Journal of Optimization Theory and Applications}, 169\penalty0
  (3):\penalty0 1042--1068, Jun 2016.
\newblock ISSN 1573-2878.
\newblock \doi{10.1007/s10957-016-0892-3}.
\newblock URL \url{https://doi.org/10.1007/s10957-016-0892-3}.

\bibitem[Paige and Saunders(1982)]{Paige:1982:LAS:355984.355989}
C.~C. Paige and M.~A. Saunders.
\newblock Lsqr: An algorithm for sparse linear equations and sparse least
  squares.
\newblock \emph{ACM Trans. Math. Softw.}, 8\penalty0 (1):\penalty0 43--71, Mar.
  1982.
\newblock ISSN 0098-3500.
\newblock \doi{10.1145/355984.355989}.
\newblock URL \url{http://doi.acm.org/10.1145/355984.355989}.

\bibitem[Pakazad et~al.(2015)Pakazad, Andersen, and
  Hansson]{doi:10.1080/10556788.2014.902056}
S.~K. Pakazad, M.~S. Andersen, and A.~Hansson.
\newblock Distributed solutions for loosely coupled feasibility problems using
  proximal splitting methods.
\newblock \emph{Optimization Methods and Software}, 30\penalty0 (1):\penalty0
  128--161, 2015.
\newblock \doi{10.1080/10556788.2014.902056}.
\newblock URL \url{https://doi.org/10.1080/10556788.2014.902056}.

\bibitem[Parikh and Boyd(2014)]{OPT-003}
N.~Parikh and S.~Boyd.
\newblock Proximal algorithms.
\newblock \emph{Foundations and Trends® in Optimization}, 1\penalty0
  (3):\penalty0 127--239, 2014.
\newblock ISSN 2167-3888.
\newblock \doi{10.1561/2400000003}.
\newblock URL \url{http://dx.doi.org/10.1561/2400000003}.

\bibitem[Parpas(2017)]{doi:10.1137/16M1082299}
P.~Parpas.
\newblock A multilevel proximal gradient algorithm for a class of composite
  optimization problems.
\newblock \emph{SIAM Journal on Scientific Computing}, 39\penalty0
  (5):\penalty0 S681--S701, 2017.
\newblock \doi{10.1137/16M1082299}.
\newblock URL \url{https://doi.org/10.1137/16M1082299}.

\bibitem[Peters and Herrmann(2017)]{doi:10.1190/tle36010094.1}
B.~Peters and F.~J. Herrmann.
\newblock Constraints versus penalties for edge-preserving full-waveform
  inversion.
\newblock \emph{The Leading Edge}, 36\penalty0 (1):\penalty0 94--100, 2017.
\newblock \doi{10.1190/tle36010094.1}.
\newblock URL \url{http://dx.doi.org/10.1190/tle36010094.1}.

\bibitem[Peters et~al.(2018)Peters, Smithyman, and Herrmann]{ournewpreprint}
B.~Peters, B.~R. Smithyman, and F.~J. Herrmann.
\newblock Projection methods and applications for seismic nonlinear inverse
  problems with multiple constraints.
\newblock \emph{GEOPHYSICS}, 0\penalty0 (ja):\penalty0 1--100, 2018.
\newblock \doi{10.1190/geo2018-0192.1}.
\newblock URL \url{https://doi.org/10.1190/geo2018-0192.1}.

\bibitem[Pratt et~al.(1998)Pratt, Shin, and Hicks]{Pratt98}
G.~Pratt, C.~Shin, and G.~Hicks.
\newblock {Gauss-Newton and full Newton methods in frequency-space seismic
  waveform inversion}.
\newblock \emph{Geophysical Journal International}, 133\penalty0 (2):\penalty0
  341--362, May 1998.
\newblock ISSN 0956540X.
\newblock \doi{10.1046/j.1365-246X.1998.00498.x}.
\newblock URL \url{http://doi.wiley.com/10.1046/j.1365-246X.1998.00498.x}.

\bibitem[Ruthotto et~al.(2017)Ruthotto, Treister, and
  Haber]{doi:10.1137/16M1081063}
L.~Ruthotto, E.~Treister, and E.~Haber.
\newblock jinv--a flexible julia package for pde parameter estimation.
\newblock \emph{SIAM Journal on Scientific Computing}, 39\penalty0
  (5):\penalty0 S702--S722, 2017.
\newblock \doi{10.1137/16M1081063}.
\newblock URL \url{https://doi.org/10.1137/16M1081063}.

\bibitem[Saad(1989)]{doi:10.1137/0910073}
Y.~Saad.
\newblock Krylov subspace methods on supercomputers.
\newblock \emph{SIAM Journal on Scientific and Statistical Computing},
  10\penalty0 (6):\penalty0 1200--1232, 1989.
\newblock \doi{10.1137/0910073}.
\newblock URL \url{https://doi.org/10.1137/0910073}.

\bibitem[Schmidt et~al.(2009)Schmidt, Van Den~Berg, Friedlander, and
  Murphy]{schmidt2009optimizing}
M.~Schmidt, E.~Van Den~Berg, M.~P. Friedlander, and K.~Murphy.
\newblock Optimizing costly functions with simple constraints: A limited-memory
  projected quasi-newton algorithm.
\newblock In \emph{Proc. of Conf. on Artificial Intelligence and Statistics},
  2009.

\bibitem[Schmidt et~al.(2012)Schmidt, Kim, and Sra]{pnmethods}
M.~Schmidt, D.~Kim, and S.~Sra.
\newblock \emph{Projected Newton-type Methods in Machine Learning}, volume~35,
  chapter~11, pages 305--327.
\newblock MIT Press, 04 2012.

\bibitem[Serón et~al.(1990)Serón, Sanz, Kindelán, and
  Badal]{CNM:CNM1630060505}
F.~J. Serón, F.~J. Sanz, M.~Kindelán, and J.~I. Badal.
\newblock Finite-element method for elastic wave propagation.
\newblock \emph{Communications in Applied Numerical Methods}, 6\penalty0
  (5):\penalty0 359--368, 1990.
\newblock ISSN 1555-2047.
\newblock \doi{10.1002/cnm.1630060505}.
\newblock URL \url{http://dx.doi.org/10.1002/cnm.1630060505}.

\bibitem[Smithyman et~al.(2015)Smithyman, Peters, and
  Herrmann]{smithyman2015constrained}
B.~Smithyman, B.~Peters, and F.~Herrmann.
\newblock Constrained waveform inversion of colocated vsp and surface seismic
  data.
\newblock In \emph{77th EAGE Conference and Exhibition 2015}, 2015.

\bibitem[Song et~al.(2016)Song, Yoon, and
  Pavlovic]{Song:2016:FAA:3015812.3015924}
C.~Song, S.~Yoon, and V.~Pavlovic.
\newblock Fast admm algorithm for distributed optimization with adaptive
  penalty.
\newblock In \emph{Proceedings of the Thirtieth AAAI Conference on Artificial
  Intelligence}, AAAI'16, pages 753--759. AAAI Press, 2016.
\newblock URL \url{http://dl.acm.org/citation.cfm?id=3015812.3015924}.

\bibitem[Tarantola(1986)]{TarantolaA}
A.~Tarantola.
\newblock A strategy for nonlinear elastic inversion of seismic reflection
  data.
\newblock \emph{GEOPHYSICS}, 51\penalty0 (10):\penalty0 1893--1903, 1986.
\newblock \doi{10.1190/1.1442046}.
\newblock URL \url{http://library.seg.org/doi/abs/10.1190/1.1442046}.

\bibitem[Tibshirani(2017)]{NIPS2017_6655}
R.~J. Tibshirani.
\newblock Dykstra\textquotesingle s algorithm, admm, and coordinate descent:
  Connections, insights, and extensions.
\newblock In I.~Guyon, U.~V. Luxburg, S.~Bengio, H.~Wallach, R.~Fergus,
  S.~Vishwanathan, and R.~Garnett, editors, \emph{Advances in Neural
  Information Processing Systems 30}, pages 517--528. Curran Associates, Inc.,
  2017.

\bibitem[Trussell and Civanlar(1984)]{STE_1}
H.~Trussell and M.~Civanlar.
\newblock The feasible solution in signal restoration.
\newblock \emph{IEEE Transactions on Acoustics, Speech, and Signal Processing},
  32\penalty0 (2):\penalty0 201--212, April 1984.
\newblock ISSN 0096-3518.
\newblock \doi{10.1109/TASSP.1984.1164297}.

\bibitem[Udell et~al.(2014)Udell, Mohan, Zeng, Hong, Diamond, and
  Boyd]{ConvexJL}
M.~Udell, K.~Mohan, D.~Zeng, J.~Hong, S.~Diamond, and S.~Boyd.
\newblock Convex optimization in julia.
\newblock In \emph{2014 First Workshop for High Performance Technical Computing
  in Dynamic Languages}, pages 18--28, Nov 2014.
\newblock \doi{10.1109/HPTCDL.2014.5}.

\bibitem[van~den Berg and Friedlander(2009)]{doi:10.1137/080714488}
E.~van~den Berg and M.~P. Friedlander.
\newblock Probing the pareto frontier for basis pursuit solutions.
\newblock \emph{SIAM Journal on Scientific Computing}, 31\penalty0
  (2):\penalty0 890--912, 2009.
\newblock \doi{10.1137/080714488}.
\newblock URL \url{https://doi.org/10.1137/080714488}.

\bibitem[Vasin(1970)]{Vasin1970}
V.~V. Vasin.
\newblock Relationship of several variational methods for the approximate
  solution of ill-posed problems.
\newblock \emph{Mathematical notes of the Academy of Sciences of the USSR},
  7\penalty0 (3):\penalty0 161--165, Mar 1970.
\newblock ISSN 1573-8876.
\newblock \doi{10.1007/BF01093105}.
\newblock URL \url{https://doi.org/10.1007/BF01093105}.

\bibitem[Venkatakrishnan et~al.(2013)Venkatakrishnan, Bouman, and
  Wohlberg]{PandP_ADMM}
S.~V. Venkatakrishnan, C.~A. Bouman, and B.~Wohlberg.
\newblock Plug-and-play priors for model based reconstruction.
\newblock In \emph{2013 IEEE Global Conference on Signal and Information
  Processing}, pages 945--948, Dec 2013.
\newblock \doi{10.1109/GlobalSIP.2013.6737048}.

\bibitem[Virieux and Operto(2009)]{virieux09}
J.~Virieux and S.~Operto.
\newblock {An overview of full-waveform inversion in exploration geophysics}.
\newblock \emph{Geophysics}, 74\penalty0 (6):\penalty0 WCC1--WCC26, 2009.
\newblock URL \url{http://dx.doi.org/10.1190/1.3238367}.

\bibitem[Wang et~al.(2013)Wang, Zhang, Zhang, and Yi]{openblas_paper}
Q.~Wang, X.~Zhang, Y.~Zhang, and Q.~Yi.
\newblock Augem: Automatically generate high performance dense linear algebra
  kernels on x86 cpus.
\newblock In \emph{2013 SC - International Conference for High Performance
  Computing, Networking, Storage and Analysis (SC)}, pages 1--12, Nov 2013.
\newblock \doi{10.1145/2503210.2503219}.

\bibitem[Witte et~al.(2018)Witte, Louboutin, Lensink, Lange, Kukreja, Luporini,
  Gorman, and Herrmann]{doi:10.1190/tle37020142.1}
P.~Witte, M.~Louboutin, K.~Lensink, M.~Lange, N.~Kukreja, F.~Luporini,
  G.~Gorman, and F.~J. Herrmann.
\newblock Full-waveform inversion, part 3: Optimization.
\newblock \emph{The Leading Edge}, 37\penalty0 (2):\penalty0 142--145, 2018.
\newblock \doi{10.1190/tle37020142.1}.
\newblock URL \url{https://doi.org/10.1190/tle37020142.1}.

\bibitem[{Wytock} et~al.(2015){Wytock}, {Wang}, and {Zico
  Kolter}]{epsilonsoftware}
M.~{Wytock}, P.-W. {Wang}, and J.~{Zico Kolter}.
\newblock {Convex programming with fast proximal and linear operators}.
\newblock \emph{ArXiv e-prints}, Nov. 2015.

\bibitem[Xu et~al.(2016)Xu, De, Figueiredo, Studer, and
  Goldstein]{empiricalncvxadmm}
Z.~Xu, S.~De, M.~Figueiredo, C.~Studer, and T.~Goldstein.
\newblock An empirical study of admm for nonconvex problems.
\newblock In \emph{NIPS workshop on nonconvex optimization}, 2016.

\bibitem[Xu et~al.(2017{\natexlab{a}})Xu, Figueiredo, and
  Goldstein]{pmlr-v54-xu17a}
Z.~Xu, M.~Figueiredo, and T.~Goldstein.
\newblock {Adaptive ADMM with Spectral Penalty Parameter Selection}.
\newblock In A.~Singh and J.~Zhu, editors, \emph{Proceedings of the 20th
  International Conference on Artificial Intelligence and Statistics},
  volume~54 of \emph{Proceedings of Machine Learning Research}, pages 718--727,
  Fort Lauderdale, FL, USA, 20--22 Apr 2017{\natexlab{a}}. PMLR.
\newblock URL \url{http://proceedings.mlr.press/v54/xu17a.html}.

\bibitem[Xu et~al.(2017{\natexlab{b}})Xu, Figueiredo, Yuan, Studer, and
  Goldstein]{Xu_2017_CVPR}
Z.~Xu, M.~A.~T. Figueiredo, X.~Yuan, C.~Studer, and T.~Goldstein.
\newblock Adaptive relaxed admm: Convergence theory and practical
  implementation.
\newblock In \emph{The IEEE Conference on Computer Vision and Pattern
  Recognition (CVPR)}, July 2017{\natexlab{b}}.

\bibitem[Xu et~al.(2017{\natexlab{c}})Xu, Taylor, Li, Figueiredo, Yuan, and
  Goldstein]{pmlr-v70-xu17c}
Z.~Xu, G.~Taylor, H.~Li, M.~A.~T. Figueiredo, X.~Yuan, and T.~Goldstein.
\newblock Adaptive consensus {ADMM} for distributed optimization.
\newblock In D.~Precup and Y.~W. Teh, editors, \emph{Proceedings of the 34th
  International Conference on Machine Learning}, volume~70 of \emph{Proceedings
  of Machine Learning Research}, pages 3841--3850, International Convention
  Centre, Sydney, Australia, 06--11 Aug 2017{\natexlab{c}}. PMLR.
\newblock URL \url{http://proceedings.mlr.press/v70/xu17c.html}.

\bibitem[Ying et~al.(2005)Ying, Demanet, and Candes]{ying20053d}
L.~Ying, L.~Demanet, and E.~Candes.
\newblock 3d discrete curvelet transform.
\newblock In \emph{Wavelets XI}, volume 5914, page 591413. International
  Society for Optics and Photonics, 2005.

\bibitem[Yong et~al.(2018)Yong, Liao, Huang, and Li]{TVWRI2}
P.~Yong, W.~Liao, J.~Huang, and Z.~Li.
\newblock Total variation regularization for seismic waveform inversion using
  an adaptive primal dual hybrid gradient method.
\newblock \emph{Inverse Problems}, 34\penalty0 (4):\penalty0 045006, 2018.
\newblock URL \url{http://stacks.iop.org/0266-5611/34/i=4/a=045006}.

\bibitem[Youla and Webb(1982)]{STE_2}
D.~C. Youla and H.~Webb.
\newblock Image restoration by the method of convex projections: Part 1-theory.
\newblock \emph{IEEE Transactions on Medical Imaging}, 1\penalty0 (2):\penalty0
  81--94, Oct 1982.
\newblock ISSN 0278-0062.
\newblock \doi{10.1109/TMI.1982.4307555}.

\bibitem[Zhang et~al.(2017)Zhang, Zuo, Gu, and Zhang]{CNNdenoiser}
K.~Zhang, W.~Zuo, S.~Gu, and L.~Zhang.
\newblock Learning deep cnn denoiser prior for image restoration.
\newblock In \emph{2017 IEEE Conference on Computer Vision and Pattern
  Recognition (CVPR)}, pages 2808--2817, July 2017.
\newblock \doi{10.1109/CVPR.2017.300}.

\end{thebibliography}

\end{document}